\documentclass[apj,twocolumn]{openjournal}

\usepackage{enumitem}
\usepackage{amsmath,amssymb,amstext}
\usepackage[utf8]{inputenc}
\usepackage[breaklinks,colorlinks,
   urlcolor=blue,citecolor=blue,linkcolor=blue]{hyperref}
\usepackage{url}
\usepackage{lineno}
\usepackage{xcolor}
\usepackage{orcidlink}
\usepackage{txfonts}
\usepackage{comment}
\usepackage{booktabs}
\usepackage{ulem}
\usepackage{algorithm}
\usepackage{algpseudocode}

\newcommand{\widebar}[1]{\overline{#1}}

\definecolor{frcolor}{rgb}{0,0.5,0}

\graphicspath{ {./figures/} }

\journalinfo{The Open Journal of Astrophysics}

\def\HITS{1}
\def\INAF{2}
\def\INFN{3}
\def\ITA{4} 
\def\MATHINF{5} 
\def\ARI{6}
\def\GSI{7}
\def\KERN{8}
\def\LANL{9}
\def\KEELE{10}
\def\KAVLI{11}

\begin{document}

\title{Phlegethon: a fully compressible magnetohydrodynamic code for simulations in stellar astrophysics}

\author{
    G. Leidi\altaffilmark{\HITS},
    A. Holas\altaffilmark{\HITS},
    K. Vitovsky\altaffilmark{\HITS},
    F. Rizzuti\altaffilmark{\HITS,\INAF,\INFN},
    A. Roy\altaffilmark{\HITS,\ITA},
    J. Reichert\altaffilmark{\HITS,\ITA},
    K. Bayer\altaffilmark{\HITS,\MATHINF},
    D. Gagnier\altaffilmark{\ARI,\HITS},
    R. Andrassy\altaffilmark{\ARI,\HITS},
    P. Christians\altaffilmark{\GSI,\KERN}, \\
    P.~V.~F. Edelmann\altaffilmark{\LANL}, 
    V. Varma\altaffilmark{\KEELE},
    R. Hirschi\altaffilmark{\HITS,\KEELE,\KAVLI}, 
    P. S.-H. Cheung\altaffilmark{\ARI,\HITS}, and
    F.~K. Röpke\altaffilmark{\HITS,\ITA}
}

\affiliation{\altaffilmark{\HITS}Heidelberger Institut f{\"u}r Theoretische Studien, Schloss-Wolfsbrunnenweg 35, D-69118 Heidelberg, Germany}

\affiliation{\altaffilmark{\INAF} INAF, Osservatorio Astronomico di Trieste, via G.B. Tiepolo 11, I-34131 Trieste, Italy}

\affiliation{\altaffilmark{\INFN} INFN, Sezione di Trieste, via Valerio 2, I-34134 Trieste, Italy}

\affiliation{\altaffilmark{\ITA} Zentrum f\"ur Astronomie der Universit\"at Heidelberg, Institut f\"ur Theoretische Astrophysik, Philosophenweg 12, D-69120 Heidelberg, Germany}

\affiliation{\altaffilmark{\MATHINF} Universit\"at Heidelberg, Fakult\"at für Mathematik und Informatik, Im Neuenheimer Feld 205, 69120 Heidelberg, Germany}
\affiliation{\altaffilmark{\ARI} Zentrum f\"ur Astronomie der Universit\"at Heidelberg, Astronomisches Rechen-Institut, M\"onchhofstr. 12-14, D-69120 Heidelberg, Germany}

\affiliation{\altaffilmark{\GSI} GSI Helmholtzzentrum für Schwerionenforschung, Planckstraße 1, 64291 Darmstadt, Germany}

\affiliation{\altaffilmark{\KERN} Institut für Kernphysik (Theoriezentrum), Fachbereich Physik, Technische Universität Darmstadt, Schlossgartenstraße 2, 64289 Darmstadt,
Germany}

\affiliation{\altaffilmark{\LANL} Computing and Artificial Intelligence (CAI) Division and Center for Theoretical Astrophysics (CTA), Los Alamos National Laboratory, Los Alamos, PO Box 1663, NM 87545, USA}

\affiliation{\altaffilmark{\KEELE} Astrophysics Group, Lennard-Jones Laboratories, Keele University, Keele ST5 5BG, UK}

\affiliation{\altaffilmark{\KAVLI} Kavli IPMU (WPI), University of Tokyo, 5-1-5 Kashiwanoha, Kashiwa 277-8583, Japan}

\begin{abstract}
We present \texttt{PHLEGETHON}, a fully compressible, Eulerian magnetohydrodynamic (MHD) code designed for multidimensional simulations in stellar astrophysics. The code uses a time-explicit, second-order, finite-volume method optimized to model a wide range of dynamical processes in stars, from very low-Mach-number turbulent convection in the cores of massive stars to supersonic flows in subsurface convection zones. \texttt{PHLEGETHON} employs low-dissipation Riemann solvers and a well-balanced method to accurately capture slow flows arising from strongly stratified media, alongside shock-capturing schemes for handling supersonic regimes. The induction equation is solved using a staggered constrained-transport method to ensure divergence-free evolution of the magnetic field. The MHD equations can be coupled to arbitrary nuclear reaction networks solved in a time-implicit approach, together with super-time-stepping for efficient treatment of thermal diffusion. Equations of state appropriate for stellar plasmas are available, accounting for partial ionization, electron degeneracy, and electron--positron pair production. Multiple grid geometries are supported, including Cartesian, spherical, and cubed-sphere grids. The code is implemented in a compact and user-friendly manner while retaining computational efficiency, and it scales to tens of thousands of CPU cores using MPI-based domain decomposition. We perform several verification tests to demonstrate the accuracy and versatility of the code, and present  simulations of magnetoconvection in a core-collapse supernova progenitor star. The rich variety of physical effects and numerical methods implemented in \texttt{PHLEGETHON} enables the code to model diverse multidimensional processes that play a crucial role in stellar-interior dynamics, such as reactive convection, convective boundary mixing, internal-wave excitation, and magnetic-field amplification mechanisms. Within a single framework, these phenomena can be investigated across a wide range of stellar evolutionary stages, from main-sequence stars to supernova progenitors. \texttt{PHLEGETHON} is publicly accessible online.
\end{abstract}


\maketitle

\begin{section}{Introduction}

The increasing precision of observations from modern 
ground- and space-based telescopes, e.g., VLT \citep{dekker2000}, CoRoT \citep{corot2009}, Gaia \citep{gaia2016}, or TESS \citep{tess2015},
places stringent constraints on theoretical models of stellar
structure and evolution. In particular, asteroseismology provides direct probes of the internal structure of stars with unprecedented accuracy \citep{aerts2010}. Interpreting these observations
requires reliable models of convection, convective boundary
mixing, wave excitation, diffusive energy transport, angular momentum transport, and
magnetic field amplification. Traditional one-dimensional (1D) stellar evolution models cannot capture these inherently
multidimensional processes in a self-consistent way.
Instead, they rely on simplified parameterized
prescriptions, which substantially limit their predictive power \citep[e.g.,][]{davis2019,kaiser2020,temaj2024}.

Fully compressible hydrodynamic and magnetohydrodynamic
(MHD) simulations provide a general framework for studying
stellar interiors, although they are often restricted to the
relatively short dynamical timescale of stellar convection because of their
high computational cost \citep[see, e.g.,][]{brun2017,kupka2017,muller2020,lecoanet2023}. In such simulations,
multidimensional fluid-dynamical processes emerge directly
from the governing equations without the need for
phenomenological parameterizations. At the same time, the diverse
dynamical conditions present in stellar interiors pose
significant challenges for standard numerical methods used
in astrophysical MHD \citep[e.g.,][]{miczek2015,wongwathanarat2016,edelmann2021a}. Convective flows in deep stellar
envelopes and cores typically have velocities $u$ much
smaller than the local sound speed $c_\mathrm{s}$, so that their
Mach number ${\mathcal{M}=|u|/c_\mathrm{s}}$ lies between $10^{-4}$ and
$10^{-2}$ \citep[e.g.,][]{kippenhahn2012,horst2021,jermyn2022}.  In advanced burning shells of massive stars,
the Mach number may reach ${\mathcal{M} \sim 0.1}$ \citep{muller2020}, while convection in subsurface regions can become supersonic \citep{freytag2012,grassitelli2015}. Accurate models of stellar-interior dynamics therefore require numerical methods capable of handling flows spanning several orders of magnitude in Mach number in a wide range of physical regimes.

Standard compressible Godunov-type schemes relying on explicit time stepping and total-variation-diminishing reconstruction can, in
principle, be applied across all flow regimes. In very low Mach number
subsonic flows, however, these schemes often encounter significant numerical
difficulties \citep{guillard1999,fleischmann2020}. Upwinding based on acoustic wave speeds
introduces excessive numerical dissipation when the flow
velocity is small compared to the sound speed, degrading the quality of the solution \citep[e.g.,][]{dumbser2019,minoshima2021,leidi2022}. Explicit time integration is further constrained by the Courant--Friedrichs--Lewy (CFL) stability
criterion, which can lead to prohibitively small
time steps when ${\mathcal{M} \ll 10^{-3}}$. Implicit
methods are unconditionally stable and can relax this time step restriction \citep[e.g.,][]{viallet2011,kifonidis2012,miczek2015}, but they
require iterative solvers that may fail to converge,
particularly when both low- and high-Mach-number flows
are present on the same grid \citep{dumbser2019,fambri2021}. Alternative approaches rely
on modified sets of equations, such as the anelastic \citep[e.g.,][]{brun2005,rogers2013,featherstone2022} or the pseudo-compressibility \citep{almgren2007}
approximations, which remove acoustic waves and thereby
relax the time step constraint.  These formulations,
however, cannot accurately capture shocks or transonic
flows and do not follow compressible pressure modes by
design. Still, retaining such modes is highly desirable because
they broaden the range of numerical results that can be
compared with asteroseismic observations \citep[e.g.,][]{mosser2012,deheuvels2012,montalban2013,coppee2024}.

An additional challenge arises from the fact that, in
layers deep inside stars, slow flows often develop as
perturbations around hydrostatic configurations. Standard finite volume
Godunov-type methods generally do not preserve
hydrostatic solutions exactly at the discrete level \citep[e.g.,][]{edelmann2021a,kappeli2022}.
Fluxes are evaluated at cell faces, while the
gravitational force per unit volume depends on
cell-volume averages stored at cell centers. This
mismatch can drive artificial motions that easily
overwhelm the physical flows. These artifacts become
particularly severe when the subsonic buoyant flows arise from steep stratifications where pressure gradients are large, as is typical in stellar interiors \citep{edelmann2021a}.

The numerical challenges faced by MHD simulations of stellar interiors motivate the development of numerical
methods that remain accurate across both low- and high-Mach regimes
while preserving the hydrostatic equilibrium of strongly stratified backgrounds. A promising strategy is to combine
explicit time-stepping with low-dissipation Riemann
solvers \citep[e.g.,][]{thornber2008,xie2019,minoshima2021,leidi2024} and well-balanced discretizations \citep{berberich2019,edelmann2021a}.
Low-dissipation solvers allow low-Mach, nearly incompressible flows
to be resolved without introducing excessive numerical
diffusion. Well-balanced schemes, in turn, preserve
hydrostatic equilibrium at the discrete level and prevent
spurious accelerations even in strongly stratified
regions with large background pressure gradients.
Together, these techniques allow explicit schemes to
remain computationally competitive with implicit
approaches when the Mach number exceeds roughly
$\,{\sim 10^{-3}}\,$ \citep{brun2017,leidi2022}, while still capturing shocks and strongly
nonlinear flows.
Long-term simulations of stellar MHD flows can therefore benefit from this
combination, achieving both high accuracy in low-Mach regions
and robustness in more dynamic layers.

Realistic simulations of stars also require an equation of state (EoS) that covers a wide range of physical conditions, from partial ionization to electron--positron pair production in advanced burning stages \citep{maeder2009}. Nuclear reaction networks and thermal neutrino cooling are required to accurately track energy generation and composition changes in late evolutionary phases. Radiative and conductive diffusion can strongly affect the evolution of dynamical processes of interest \citep[e.g.,][]{kapyla2019,schwab2020,andrassy2023}, and must therefore also be modeled accurately. 

Moreover, the solenoidal constraint $\,{\boldsymbol{\nabla} \cdot \mathbf{B} = 0}\,$ on the magnetic field  is not automatically satisfied when discretizing the MHD equations on a finite grid, so further care is required for MHD simulations. In fact, if the divergence-free property of the magnetic field is violated, magnetic monopole errors introduce spurious accelerations through the Lorentz force, distort the magnetic-field topology, and potentially lead to numerical instabilities \citep{brackbill1980}. Staggered constrained transport (CT) methods are particularly attractive in this regard because, by construction, they evolve the magnetic fluxes through cell faces in a manner that preserves one discretization of ${\boldsymbol{\nabla} \cdot \mathbf{B}}$ to round-off error \citep[see, e.g.,][for a review]{mignone2021}.

Several existing compressible hydrodynamic and MHD codes, such as \texttt{FLASH} \citep{fryxell2000}, \texttt{CASTRO} \citep{almgren2011}, \texttt{ATHENA++} \citep{stone2020}, \texttt{PPMSTAR} \citep{woodward2015}, \texttt{MUSIC} \citep{viallet2011}, \texttt{SEVEN-LEAGUE HYDRO} \citep[\texttt{SLH},][]{miczek2013}, \texttt{PROMPI} \citep{meakin2007}, \texttt{CO5BOLD} \citep{freytag2012}, or \texttt{PENCIL} \citep{pencil2021}, provide powerful tools for studying astrophysical flows and stellar-interior dynamics. Nevertheless, modeling MHD flows across multiple evolutionary stages and in different types of stars requires a specific combination of low-Mach methods, well-balanced treatment of gravity, robust shock-capturing divergence-free MHD schemes, nuclear networks, and flexible stellar plasma microphysics. At present, such a combination is not found in a single framework. 

In this work, we present \texttt{PHLEGETHON}\footnote{\texttt{PHLEGETHON} is publicly available at \\\url{https://github.com/phlegethon-stellar-hydro/phlegethon}}, a
time-explicit, fully compressible Godunov-type MHD code
developed for simulations of stellar interiors in regimes of Mach numbers ${\mathcal{M} \gtrsim 10^{-3}}$, where explicit
time integration remains computationally efficient. The
 code uses a finite-volume discretization on Eulerian grids. It employs low-dissipation Riemann solvers \citep{minoshima2021}
that recover the correct asymptotic behavior of the underlying governing equations in the limit $\,{\mathcal{M} \rightarrow 0}$, together
with well-balanced discretizations \citep{berberich2019,edelmann2021a} that preserve
hydrostatic equilibrium in strongly stratified
media. The MHD
system of equations is coupled to flexible time-implicit nuclear reaction network solvers
and advanced equations of state that include thermal
radiation, partial ionization effects (Vetter et al., in prep), and arbitrarily
degenerate and relativistic electron--positron plasmas via the Helmholtz EoS \citep{timmes2000}. 
Stiff diffusive processes are treated using super-time-stepping \citep{meyer2012,meyer2014}. Magnetic fields are evolved using the Contact-CT scheme of \cite{gardiner2005}, which preserves a divergence-free evolution of the magnetic field while employing an upwinded discretization of the induction equation. The code supports multiple grid geometries, enabling applications ranging from local
convection studies to simulations of extended stellar
stratifications and a wide range of astrophysical fluid dynamics problems, from convective boundary mixing (CBM) and magnetic dynamos in massive main-sequence stars to the coupling of convection, nuclear burning, and magnetic fields in the late stages of core-collapse supernova progenitors (see Sect.~\ref{sec:ccsnp})

The paper is organized as follows. Section \ref{sec:methods} describes the governing equations and numerical methods, and provides benchmark measurements of code efficiency. Section \ref{sec:verification} presents a series of verification tests. Section \ref{sec:ccsnp} demonstrates the capabilities of the code through simulations of magnetoconvection in a convective shell of a core-collapse supernova progenitor star. Finally, Section \ref{sec:conclusions} summarizes the results.
\end{section}

\begin{section}{Methods}\label{sec:methods}

In this section, we summarize the numerical methods that allow \texttt{PHLEGETHON} to perform multidimensional MHD simulations in stellar astrophysics. We present the governing equations solved by the code, followed by an overview of the finite-volume discretization and time-integration scheme. The implementation of additional components required for stellar-interior applications, including equations of state of stellar plasmas, treatment of gravity, thermal diffusion, and nuclear energy generation, is also outlined. We also discuss the computational performance of the code, including its parallel scaling and the relative cost of the main algorithmic components. The accuracy and robustness of these methods are assessed through a series of benchmark tests presented in Sect.~\ref{sec:verification}.

\subsection{Governing equations}\label{sec:pdes}

\texttt{PHLEGETHON} solves the equations of fully compressible ideal MHD including source terms which account for gravity and nuclear energy generation. The system can be written in conservative form as
\begin{align}
\partial_t \rho + \boldsymbol{\nabla} \cdot (\rho \mathbf{u}) &= 0, \label{eq:mass-conservation} \\ 
\partial_t (\rho \mathbf{u}) + \boldsymbol{\nabla} \cdot \left( \rho \mathbf{u} \otimes \mathbf{u} + P_\mathrm{t} \mathbf{I} - \mathbf{B} \otimes \mathbf{B} \right) &= \rho \mathbf{g}, \\
\partial_t (\rho X_l) + \boldsymbol{\nabla} \cdot (\rho X_l \mathbf{u}) &= \rho \dot{X}_l, \\
\partial_t \mathbf{B} + \boldsymbol{\nabla} \cdot (\mathbf{u} \otimes \mathbf{B} - \mathbf{B} \otimes \mathbf{u}) &= \mathbf{0}, \\
\boldsymbol{\nabla} \cdot \mathbf{B} &= 0, \label{eq:solenoidal} \\
\partial_t (\rho e) + \boldsymbol{\nabla} \cdot \left[ (\rho e + P_\mathrm{t}) \mathbf{u} - \mathbf{B} (\mathbf{B} \cdot \mathbf{u}) - K \boldsymbol{\nabla} T \right] &= \rho \mathbf{g} \cdot \mathbf{u} + \rho \dot{e}.  \label{eq:energy}
\end{align}
Here, \(\rho\) is the mass density, \(\mathbf{u}\) the fluid velocity, \(\mathbf{B}\) the magnetic field\footnote{Throughout this work, we use Heaviside--Lorentz units, $\,{\mathbf{B} = \mathbf{b}\,/\sqrt{4\pi}}\,$.}, and \(P_\mathrm{t} = P + |\mathbf{B}|^2/2\) the total pressure including both thermal and magnetic contributions. The specific energy is ${e = e_\mathrm{int} + |\mathbf{u}|^2/2 + |\mathbf{B}|^2/(2\rho)}$, where $e_\mathrm{int}$ is the specific internal energy. $K$ denotes the thermal conductivity, $T$ the temperature, while $\dot{e}$ and $\dot{X}_l$ denote an energy source (including neutrino cooling) and the rate of change of the mass fraction of species $l$, respectively. $\mathbf{I}$ is the identity tensor and $\otimes$ is the tensor product operator. Gravity is assumed to be Newtonian, $\,{\mathbf{g}=-\boldsymbol{\nabla} \phi\,}$, and the gravitational potential, $\phi$, is obtained from Poisson's equation,
\begin{equation}
\nabla^2 \phi = 4 \pi G \rho,
\end{equation}
where $G$ is the gravitational constant. 

If the gravitational potential is time-independent, redefining the   specific energy as
\begin{equation}
    e \rightarrow\ e + \phi,
\end{equation}
yields the conservative form
\begin{equation}\label{eq:etot}
    \partial_t (\rho e) + \boldsymbol{\nabla} \cdot \left[ (\rho e + P_\mathrm{t}) \mathbf{u} - \mathbf{B} (\mathbf{B} \cdot \mathbf{u}) - K \boldsymbol{\nabla} T \right] = \rho \dot{e}.
\end{equation}
For isolated systems in the absence of energy sources or sinks, a numerical scheme that solves Eq.~(\ref{eq:etot}) preserves the total energy exactly. This property is highly desirable especially in simulations of subsonic flows, where even small energy conservation errors can drive the dynamics of the system \citep{muller2020}.

Finally, \texttt{PHLEGETHON} offers the option of solving Eqs.~(\ref{eq:mass-conservation})--(\ref{eq:energy}) in a rotating frame with a fixed angular frequency vector $\mathbf{\Omega}$. When this option is enabled, the centrifugal and Coriolis forces are added to the right-hand side of the momentum and energy equations as
\begin{align}
    \mathbf{S}_{\rho \mathbf{u}} &= -2 \rho \, \mathbf{\Omega} \times \mathbf{u} - \rho \, \mathbf{\Omega} \times (\mathbf{\Omega} \times \mathbf{r}), \\
    \mathbf{S}_{\rho e} &= - \rho \, \mathbf{u} \cdot \bigl[ \mathbf{\Omega} \times (\mathbf{\Omega} \times \mathbf{r}) \bigr],
\end{align}
where $\mathbf{r}$ is the position vector measured from the rotation axis. 

\subsection{Equation of state}\label{sec:eos}

The system of conservation laws in Eqs.~(\ref{eq:mass-conservation})--(\ref{eq:energy}) is closed by an equation of state that relates temperature and thermal pressure to mass density, composition, and specific internal energy. In \texttt{PHLEGETHON}, the EoS is formulated in terms of the primary variables $(\rho, T, X_l)$, from which all thermodynamic quantities are computed. However, temperature is not directly evolved in the governing equations, so it must be recovered from the conserved variables by finding the root of
\begin{equation}
    e_\mathrm{int} - e_\mathrm{int,EoS}(\rho,T,X_l) = 0,
\end{equation}
where $e_\mathrm{int}$ is the specific internal energy obtained from the conserved variables $(\rho,\rho \mathbf{u},\rho X_l,\mathbf{B},\rho e)$, and $e_\mathrm{int,EoS}$ is the specific internal energy prescribed by the EoS for a given set $(\rho,T,X_l)$. Once the temperature is known, the thermal pressure is computed as
\begin{equation}
    P = P_\mathrm{EoS}(\rho,T,X_l).
\end{equation}
This temperature inversion is performed using a Newton--Raphson iterative scheme. First, the specific internal energy and mass fractions are computed from the set of conserved variables. Then, starting from an initial guess $\tilde{T}$ taken from the previous time step, the residual is computed as
\begin{equation}
    \delta e_\mathrm{int} = e_\mathrm{int} - e_\mathrm{int,EoS}(\rho,\tilde{T},X_l).
\end{equation}
The temperature is then updated iteratively according to
\begin{equation}
    \tilde{T} \rightarrow \tilde{T} + \frac{\delta e_\mathrm{int}}{c_v(\rho,\tilde{T},X_l)},
\end{equation}
where $\,{c_v = \left. \partial e_\mathrm{int,EoS}/\partial T \right|_{\rho,X_l}}\,$ is the specific heat at constant volume evaluated at the current temperature guess. The iteration proceeds until the relative residual satisfies ${|\delta e_\mathrm{int}| / e_\mathrm{int} < \epsilon}$ (${\epsilon = 10^{-11}}$ is the default value in \texttt{PHLEGETHON}). Usually, the convergence criterion is met after a few iterations only. An analogous Newton--Raphson procedure can be applied if temperature must be recovered from density, composition, and thermal pressure (see Sect.~\ref{sec:riemann}).  

Depending on the physical regime of interest, several equations of state are available in the code:
\begin{enumerate}[label=(\roman*)]
\item For simple cases, the fluid can be modeled as a classical ideal gas with an optional contribution from thermal radiation,
\begin{align}
    P & = \frac{\rho \mathcal{R} T}{\mu} + \frac{a T^4}{3}, \label{eq:prad} \\
    e_\mathrm{int} & = \frac{\mathcal{R} T}{(\gamma-1)\mu} + \frac{a T^4}{\rho}, \label{eq:erad}
\end{align}
where $\mathcal{R}$ is the gas constant, $a$ the radiation constant, and $\gamma$ the adiabatic index of the gas. In \texttt{PHLEGETHON}, the mean molecular weight $\mu$ can either be prescribed by the user as a global parameter or computed from mass fraction abundances assuming full ionization,
\begin{equation}
    \mu = \frac{\widebar{A}}{\widebar{Z}+1},
\end{equation}
with
\begin{align}
    \widebar{A} & = \left( \sum_l \frac{X_l}{A_l} \right)^{-1}, \\
    \widebar{Z} & = Y_e \widebar{A}, \\
    Y_e & = \sum_l \frac{Z_l}{A_l} X_l.
\end{align}
$A_l$ and $Z_l$ are the mass and charge numbers of species $l$, respectively. In this notation, $\widebar{A}$ and $\widebar{Z}$ represent the mean mass and charge number of the mixture, while $Y_e$ is the electron fraction. In simulations where individual isotopic abundances are not required, the reduced composition variables $\rho Y_e$ and $\rho \widebar{A}^{\ -1}$ can be advected instead, 
\begin{align}
  \partial_t (\rho Y_e) + \boldsymbol{\nabla} \cdot (\rho Y_e \mathbf{u}) & = 0, \label{eq:rhoye} \\
  \partial_t (\rho \widebar{A}^{\ -1}) + \boldsymbol{\nabla} \cdot (\rho \widebar{A}^{\ -1} \mathbf{u}) & = 0, \label{eq:iabar}
\end{align}
allowing the EoS to still be evaluated.
\item In scenarios where electrons and positrons must be treated as an arbitrarily relativistic and degenerate Fermi--Dirac gas, \texttt{PHLEGETHON} can employ the Helmholtz EoS of \cite{timmes2000}. In this EoS, the gas is assumed to be fully ionized, and electron--positron pair production is included assuming chemical equilibrium for the reactions
\begin{align}
   \gamma + \mathcal{N} \rightarrow & \ e^- + e^+ + \mathcal{N}, \\
   e^- + e^+ \rightarrow &\  \gamma + \gamma  ,   
\end{align}
where $\mathcal{N}$ represents a nucleus. Thermodynamic variables associated with the electron--positron mixture are derived via biquintic Hermite interpolation of the Helmholtz free energy tabulated in $(\rho Y_e, T)$, ensuring that thermodynamic consistency relations are satisfied to round-off error \citep[see][]{timmes2000}. Radiation is included assuming local thermodynamic equilibrium as in Eqs.~(\ref{eq:prad})--(\ref{eq:erad}), while ions are treated as a classical ideal gas with adiabatic index $\gamma = 5/3$.
The combined contributions to the EoS read
\begin{align}
    P & = P_\mathrm{ele,pos} + P_\mathrm{Coul} + \frac{\rho \mathcal{R} T}{\widebar{A}} + \frac{aT^4}{3}, \\
    e_\mathrm{int} & = e_\mathrm{ele,pos} + e_\mathrm{Coul} + \frac{3}{2}\frac{\mathcal{R} T}{\widebar{A}} +  \frac{aT^4}{\rho},
\end{align}
where $P_\mathrm{ele,pos}$ and $e_\mathrm{ele,pos}$ denote the pressure and specific internal energy of the electron-positron gas, resulting from the interpolation of the tabulated Helmholtz free energy. $P_\mathrm{Coul}$ and $e_\mathrm{Coul}$ denote optional Coulomb corrections to the EoS, included following \cite{yakovlev1989}. Also for this EoS, $\rho Y_e$ and $\rho \widebar{A}^{\ -1}$ can be advected according to Eqs.~(\ref{eq:rhoye}) and~(\ref{eq:iabar}) when tracking individual isotopic abundances in the simulation is not desirable.

\item For partially ionized plasmas, \texttt{PHLEGETHON} employs a novel EoS based on biquintic interpolation of the Helmholtz free energy, closely analogous to the fully ionized case. Tables are constructed for arbitrary mixtures of chemical species at fixed mass fractions, ranging from ${}^{1}\mathrm{H}$ to ${}^{56}\mathrm{Fe}$, including molecular hydrogen ($\mathrm{H}_2$). All species are treated as classical particles obeying Maxwell--Boltzmann statistics, and all ionization stages are assumed to be in the ground state. Populations are computed using the Saha equation. Dissociation of $\mathrm{H}_2$ is also considered, including internal vibrational and rotational partition functions following \cite{tomida2013}. The Helmholtz free energy is tabulated in $(\rho, T)$, guaranteeing full thermodynamic consistency while accurately capturing ionization effects (a detailed description of this EoS can be found in Vetter et al., in prep). Thermal radiation can optionally be included, yielding
\begin{align}
    P & = P_\mathrm{gas} + \frac{aT^4}{3}, \\
    e_\mathrm{int} & = e_\mathrm{gas} + \frac{aT^4}{\rho},
\end{align}
where $P_\mathrm{gas}$ and $e_\mathrm{gas}$ are the interpolated values of pressure and specific internal energy for the partially ionized mixture, respectively. This EoS is particularly suited for modeling stellar envelopes, where partial ionization is crucial to determine the stability of the medium against convection \citep{kippenhahn2012}. Moreover, the interpolation routines developed for the fully ionized Helmholtz EoS can be reused directly by setting $Y_e = 1$ during table lookups in $(\rho Y_e, T)$, simplifying code implementation and maintenance. 
\end{enumerate}
\subsection{Solution strategy}

The equations of compressible MHD with source terms presented in Sect.~\ref{sec:pdes} are discretized using the finite-volume framework \citep[e.g.,][]{leveque2002}. This approach ensures exact conservation of mass, momentum, energy, and magnetic flux for an isolated system, as long as gravity is included in conservative form as in Eq.~(\ref{eq:etot}) and no additional non-conservative source terms are present. The computational domain is divided into control volumes (cells) $V_{i,j,k}$, each indexed by $(i,j,k)$. The integral form of the conservation laws is solved over each cell,
\begin{equation}
\frac{\partial}{\partial t} \int_{V_{i,j,k}} \mathbf{U} \, \mathrm{d}V + \oint_{A(V_{i,j,k})} \mathbf{F} \cdot \mathbf{n} \, \mathrm{d}A = \int_{V_{i,j,k}} \mathbf{S} \, \mathrm{d}V,
\end{equation}
where $\mathbf{U}$ is the vector of conserved variables, $\mathbf{F}$ the flux tensor, $\mathbf{n}$ the outward normal at the cell faces, and $\mathbf{S}$ the source term vector (e.g., gravity, fictitious forces, or nuclear energy generation). The surface $A(V_{i,j,k})$ denotes the boundary of the control volume $V_{i,j,k}$.

In \texttt{PHLEGETHON}, both volume and surface integrals are approximated in space to second-order accuracy using the midpoint rule \citep[see, e.g.,][]{leveque2002}, while leaving the system continuous in time. This semi-discrete, directionally-unsplit approach yields a system of ordinary differential equations (ODEs),
\begin{equation}\label{eq:fv}
\frac{\partial}{\partial t} \langle \mathbf{U}\rangle_{i,j,k}
+ \frac{1}{V_{i,j,k}}\sum_f A_f \mathbf{F}_f \cdot \mathbf{n}_f
= \mathbf{S}_{i,j,k},
\end{equation}
where $\langle \mathbf{U}\rangle_{i,j,k}$ is the cell-volume averaged vector of conserved variables. The sum in Eq.~(\ref{eq:fv}) runs over the cell faces $f$ of cell $(i,j,k)$, where $A_f$ denotes the face area and $\mathbf{F}_f$ the flux evaluated at the face center. The source term $\mathbf{S}_{i,j,k}$ is evaluated to second-order accuracy at the cell center.

The code supports multiple grid geometries in two and three dimensions, including Cartesian, polar, and spherical geometries, as well as the cubed-sphere grid of \cite{calhoun2008}\footnote{At present, MHD and radiative diffusion are not supported on the cubed-sphere grid.}. For any geometry, the computational domain is represented in a logically rectangular coordinate system $(x_1,x_2,x_3)$ and discretized into $\,{N_{x_1} \times N_{x_2} \times N_{x_3}}\,$ cells. Parallelization is implemented via domain decomposition of the logical grid, with inter-process communication handled using the Message Passing Interface (MPI) \citep{mpi50} on CPUs. 

For polar and spherical grids, the governing equations are formulated in curvilinear coordinates, which give rise to geometric source terms \citep[see, e.g.,][]{mignone2014,luo2026}. In these geometries the equations are written in differential form, and the divergence of the flux tensor is discretized using conservative flux differences that are algebraically equivalent to the finite-volume formulation. For example, in two-dimensional (2D) polar coordinates $(r,\psi)$ the divergence of the flux tensor,
\begin{equation}
\boldsymbol{\nabla} \cdot \mathbf{F} =
\frac{1}{r}\frac{\partial (r F_r)}{\partial r}
+
\frac{1}{r}\frac{\partial F_\psi}{\partial \psi},
\end{equation}
is approximated as
\begin{align}\label{eq:divf-curv}
(\boldsymbol{\nabla} \cdot \mathbf{F})_{i,j,k} & =
\frac{1}{r_{i,j,k}}
\frac{r_{i+1/2,j,k}F_{r,i+1/2,j,k} - r_{i-1/2,j,k}F_{r,i-1/2,j,k}}{r_{i+1/2,j,k}-r_{i-1/2,j,k}} \\
&\quad + \frac{1}{r_{i,j,k}}
\frac{F_{\psi,i,j+1/2,k}-F_{\psi,i,j-1/2,k}}{\psi_{i,j+1/2,k}-\psi_{i,j-1/2,k}}. \nonumber
\end{align}
Here, face-centered indices such as $(i+1/2,j,k)$ denote the location of the flux on the cell face. Analogous expressions are used for spherical grids. Geometric source terms in $\mathbf{S}_{i,j,k}$ are evaluated to second-order accuracy at cell centers. Grid spacing can be uniform or stretched. Currently, only radial stretching is supported for polar and spherical geometries.

\subsection{Constrained transport}

Magnetic fields are evolved using the Contact-CT method of \cite{gardiner2005}. In this approach, the induction equation is solved in curl form,
\begin{equation}\label{eq:induction-curl}
    \partial_t \mathbf{B} + \boldsymbol{\nabla} \times \mathcal{E} = \mathbf{0},
\end{equation}
where $\mathcal{E}= -\mathbf{u}\times\mathbf{B}$ is the electromotive force (EMF). Integrating Eq.~(\ref{eq:induction-curl}) over the cell surface $A_f$ leads to the evolution equation for the face-averaged magnetic field component normal to a cell interface, $\langle B_f \rangle$,
\begin{equation}\label{eq:induction}
\frac{\partial}{\partial t} \langle B_f \rangle + \frac{1}{A_f}\sum_d l_d \mathcal{E}_d = 0,
\end{equation}
where the sum runs over cell edges $d$, $l_d$ is the edge length, and $\mathcal{E}_d$ the edge-centered EMF. For polar and spherical grids, the constrained-transport update is applied to the curvilinear components of the face-centered magnetic field. 

In Contact-CT, the EMF is computed to second-order accuracy as the arithmetic average of the four neighboring cell faces, with additional upwind terms based on the direction propagation of the contact wave at those faces \citep{gardiner2005}. These upwind terms are essential to maintain stability and prevent numerical artifacts and spurious growth of the magnetic field \citep[e.g.,][]{flock2010,mignone2021}.

By solving the induction equation in curl form using the EMF, the Contact-CT scheme preserves a discrete approximation of the magnetic-field divergence, up to the accumulation of round-off errors. On a Cartesian grid, the preserved divergence is
\begin{equation}
\label{divb}
\begin{split}
(\boldsymbol{\nabla} \cdot \mathbf{B})_{i,j,k} =
& \frac{B_{x,i+1/2,j,k}-B_{x,i-1/2,j,k}}{(\Delta x)_{i,j,k}} \\
+&\frac{B_{y,i,j+1/2,k}-B_{y,i,j-1/2,k}}{(\Delta y)_{i,j,k}} \\
+&\frac{B_{z,i,j,k+1/2}-B_{z,i,j,k-1/2}}{(\Delta z)_{i,j,k}},
\end{split}
\end{equation}
where $(\Delta x)_{i,j,k}$, $(\Delta y)_{i,j,k}$, and $(\Delta z)_{i,j,k}$ denote the grid-cell sizes in the three spatial directions. For curvilinear grids, the divergence expressed in curvilinear components, analogous to Eq.~(\ref{eq:divf-curv}), is preserved. 
\subsection{Spatial reconstruction}\label{sec:reconstruction}

The face-centered fluxes $\mathbf{F}_f$ in Eq.~(\ref{eq:fv}) are computed using the Godunov algorithm \citep{godunov1959}. In this approach, a local 1D Riemann problem is solved at each cell interface to determine the fluxes. In \texttt{PHLEGETHON}, the left and right Riemann states at each cell face are obtained by reconstructing the set of primitive variables ${\mathbf{W} = (\rho,\mathbf{u},X_l,\mathbf{B},P)}$
from cell-centered values\footnote{Because the spatial reconstruction algorithms and Riemann solvers implemented in \texttt{PHLEGETHON} are 1D, for simplicity we drop the $j$ and $k$ indices here.},
\begin{align}
    \mathbf{W}_{i-1/2,\mathrm{R}} &= \mathcal{P}^-(\mathbf{W}_{i-q,\dots,i+q}), \\
    \mathbf{W}_{i+1/2,\mathrm{L}} &= \mathcal{P}^+(\mathbf{W}_{i-q,\dots,i+q}).
\end{align}
Here, $\mathcal{P}$ denotes the reconstruction operator, $2q+1$ is the width of the stencil, and the superscripts $+$ and $-$ indicate reconstruction from the cell center toward the right and left interfaces, respectively.

Three reconstruction schemes are implemented in the code, all operating in logical coordinates: 
\begin{enumerate}[label=(\roman*)]
    \item A second-order accurate linear reconstruction method with the van Leer slope limiter \citep{vanleer1974a}.
    \item The third-order piecewise parabolic hybrid (PPH) method of \cite{leidi2024}, in which all primitive variables except the mass fractions are reconstructed using unlimited parabolas, while mass fractions use limited parabolic reconstruction.
    \item A limited fifth-order polynomial reconstruction following the piecewise quintic hybrid (PQH) method of \cite{leidi2024}. Unlike the PQH method presented in \cite{leidi2024}, here, limited fifth-order  reconstruction is applied to all variables, not just mass fraction abundances.
\end{enumerate}

The van Leer linear reconstruction method is very robust for shocks but is the most diffusive of the available methods. PPH is accurate for smooth flows and it is one of the cheapest methods to achieve a given numerical accuracy in subsonic regimes \citep{leidi2024}, though its partially unlimited nature makes it unsuitable for supersonic flows. Both the van Leer method and PPH use a three-cell-wide stencil (${q=1}$). Our limited fifth-order reconstruction combines high accuracy for smooth flows with good robustness near shocks, but it is approximately twice as expensive as the van Leer and PPH methods on a given grid and it requires a wider stencil (${q=2}$). We stress, however, that the global spatial convergence of the code remains at best second-order even when higher-than-second-order one-dimensional reconstruction schemes are used. This is because the cell-surface flux integrals are evaluated using the midpoint formula, and the point-valued primitive variables at cell centers are obtained directly from cell-volume-averaged conserved quantities and are therefore also limited to second-order accuracy.

\subsection{Approximate Riemann solvers}\label{sec:riemann}

The reconstructed pair of Riemann states, $\mathbf{W}_{i+1/2,\mathrm{L}}$ and $\mathbf{W}_{i+1/2,\mathrm{R}}$, is passed to an approximate Riemann solver to compute the interface flux $\mathbf{F}_{i+1/2}$. The code supports state-of-the-art approximate Riemann solvers for hydrodynamic and magnetohydrodynamic flows:
\begin{enumerate}[label=(\roman*)]
    \item For purely hydrodynamic problems, the three-wave Harten--Lax--van Leer--Contact solver \citep[HLLC,][]{toro1994}  accurately resolves contact discontinuities and shocks.
    \item For magnetized flows, the five-wave Harten--Lax--van Leer--Discontinuities solver \citep[HLLD,][]{miyoshi2005} is employed, which captures fast magnetosonic, Alfv\'en, and contact waves. HLLD  discards slow-magnetosonic waves, which are averaged out in the region of the Riemann fan between the contact and Alfv\'en waves.
\end{enumerate}
In both the HLLC and HLLD solvers, the interface flux is evaluated based on the direction of propagation of the characteristic waves. For instance, on any given cell face, the HLLD flux is determined according to
\begin{align}
\label{flux-hlld}
  \mathbf{F}(\mathbf{W}_\mathrm{L},\mathbf{W}_\mathrm{R}) = \left\{
     \begin{array}{@{}l@{\thinspace}l}
       \mathbf{F}_\mathrm{L}    & \text{if } 0 \le C_\mathrm{L},\\
       \mathbf{F}^*_\mathrm{L}  & \text{if } C_\mathrm{L} < 0 \le C^*_\mathrm{L},\\
       \mathbf{F}^{**}_\mathrm{L}  & \text{if } C^*_\mathrm{L} < 0 \le C^*,\\       
       \mathbf{F}^{**}_\mathrm{R}  & \text{if } C^* < 0 \le C^*_\mathrm{R},\\
       \mathbf{F}^*_\mathrm{R}  & \text{if } C^*_\mathrm{R} < 0 \le C_\mathrm{R},\\       
       \mathbf{F}_\mathrm{R}    & \text{if } C_\mathrm{R} < 0,
     \end{array}
   \right.
\end{align}
where the fluxes are computed in the different regions of the Riemann fan delimited by the fast-magnetosonic waves, $C_\mathrm{L}$ and $C_\mathrm{R}$, the Alfvén waves, $C^*_\mathrm{L}$ and $C^*_\mathrm{R}$, and the contact discontinuity, $C^*$. 

To approximate the speeds $C_\mathrm{L/R}$ of the fastest waves (either sound waves or fast-magnetosonic waves), the solver must compute the sound speed from both the left and right states. Moreover, the specific internal energy is not directly reconstructed to cell faces, so its value must be recovered from the reconstructed primitive variables. Computing $c_\mathrm{s}$ and $e_\mathrm{int}$ requires an iterative procedure (see Sect.~\ref{sec:eos}) to first determine the temperature from the set $(\rho, P, X_l)$. Because this operation must be repeated for every cell face on the computational grid multiple times per time step, it can become relatively expensive, especially for a tabulated EoS.

Although \texttt{PHLEGETHON} provides this option, a significantly cheaper alternative is to reconstruct the auxiliary indices
\begin{align}
    \gamma_e &= \frac{P}{\rho e_\mathrm{int}} + 1, \\
    \gamma_c &= \frac{\rho c_\mathrm{s}^2}{P},
\end{align}
in addition to the set of primitive quantities \citep[e.g.][]{colella1985,reinecke1999a}. We verified that reconstructing $\gamma_e$ and $\gamma_c$ does not introduce appreciable numerical errors, consistent with expectations given that these indices are of order unity and vary smoothly in most relevant thermodynamic regimes. $\gamma_e$ and $\gamma_c$ can be computed from the EoS using cell-centered quantities once per grid cell before reconstructing the primitive variables to the six faces of the cell. The reconstructed indices can then be combined with the reconstructed $\rho$ and $P$ states to compute
\begin{align}
    e_{\mathrm{int},\mathrm{L/R}} &= \frac{1}{\rho_\mathrm{L/R}}\frac{P_{\mathrm{L/R}}}{\gamma_{e,\mathrm{L/R}} - 1}, \\
    c_{\mathrm{s},\mathrm{L/R}} &= \left( \gamma_{c,\mathrm{L/R}} \frac{P_{\mathrm{L/R}}}{\rho_{\mathrm{L/R}}} \right)^{1/2},
\end{align}
without requiring an iterative procedure. We find that this approach reduces the computational cost of the Riemann solve by a factor up to six without affecting the accuracy of the numerical scheme when the Helmholtz EoS is employed (see also Sect.~\ref{sec:perf}). It is therefore chosen as the default method in \texttt{PHLEGETHON}.

In the standard HLLC and HLLD flux functions, the expression of the thermal pressure in the star regions is typically derived as the sum of a physical and a dissipative contribution. For HLLC, the interface pressure can be obtained from the linearized gas-dynamics solver of \cite{toro1991},
\begin{equation}\label{eq:pstar}
P^* = \frac{1}{2}(P_\mathrm{L}+P_\mathrm{R})-\frac{1}{2}\bar{\rho}\bar{c}_\mathrm{s}(u_{x_1,\mathrm{R}}-u_{x_1,\mathrm{L}}),
\end{equation}
where the first term on the right-hand side corresponds to the physical central-flux estimate, while the second term introduces upwinding. Here, $\bar{\rho}$ and $\bar{c}_s$ denote suitable averages of the mass density and sound speed \citep[see, e.g.,][]{leidi2024}. Analogous expressions can be derived for HLLD \citep[e.g.,][]{miyoshi2005,minoshima2021}. The numerical upwind term in Eq.~(\ref{eq:pstar}) scales linearly with the local Mach number \citep{minoshima2021}. This scaling guarantees that sufficient dissipation is provided to stabilize the numerical solution in transonic and supersonic regimes, but it is inconsistent with the asymptotic behavior of thermal pressure fluctuations as $\,{\mathcal{M} \rightarrow 0}$, whose amplitude scales as $\mathcal{O}(\mathcal{M}^2)$ \citep{guillard1999,miczek2015}. As a result, in low-Mach-number regimes the numerical dissipation can dominate the physical evolution of the flow.

To alleviate this issue, low-dissipation extensions of HLLC and HLLD, known as LHLLC and LHLLD \citep{minoshima2021,leidi2024}, are implemented in \texttt{PHLEGETHON}. In these solvers, the numerical dissipation term in $P^*$ is multiplied by an additional factor $\xi$ as 
\begin{equation}
P^* = \frac{1}{2}(P_\mathrm{L}+P_\mathrm{R})-\xi \frac{1}{2}\bar{\rho}\bar{c}_\mathrm{s}(u_{x_1,\mathrm{R}}-u_{x_1,\mathrm{L}}),
\end{equation}
where 
\begin{equation}\label{eq:low-mach-fix}
    \xi = \zeta(2-\zeta), 
\end{equation}
with
\begin{equation}
    \zeta = \max \Bigg( \frac{|\mathbf{u}_\mathrm{L}|}{c_{s,\mathrm{L}}},\frac{|\mathbf{u}_\mathrm{R}|}{c_{s,\mathrm{R}}} \Bigg).
\end{equation}
The factor $\xi$ scales as ${\mathcal{O}(\mathcal{M})}$, thus restoring the correct low-Mach-number scaling of the numerical dissipation term in $P^*$. In the code, we apply this correction only when the local Mach  number falls below a conservative cut-off value, i.e. ${\zeta<\mathcal{M}_\mathrm{cut}=0.6}$, ensuring that the character of the Riemann solution remains unchanged as the flow approaches the transonic regime. Despite its simplicity, this modification has been shown to significantly improve the quality of numerical solutions in studies of subsonic convection and turbulent dynamos \citep[e.g.,][]{leidi2023,leidi2024}, yielding a solver that remains robust and accurate from very low-Mach to mildly supersonic regimes.  In all verification tests discussed in Sect.~\ref{sec:verification}, unless explicitly stated otherwise, we use the LHLLC/D Riemann solvers.

\subsection{Shock-flattening procedure}\label{sec:shock-flat}

To ensure robustness in the presence of strong shocks and to avoid the carbuncle instability \citep{quirk1994}, a `shock-flattening' procedure can be used \citep[e.g.,][]{marti2003,stone2008}. After performing the reconstruction step in Sect.~\ref{sec:reconstruction}, regions satisfying
\begin{equation}
\frac{|P_{i+1,j,k} - P_{i-1,j,k}| + |P_{i,j+1,k} - P_{i,j-1,k}| + |P_{i,j,k+1} - P_{i,j,k-1}|}{\epsilon_\mathrm{SF} P_{i,j,k}} > 1
\end{equation}
are flagged as shock regions (${\epsilon_\mathrm{SF}=0.1}$ is chosen as the default value in \texttt{PHLEGETHON}). All cells within a 3×3×3 neighborhood of a flagged cell are also marked, ensuring that the shock-flattener smoothly covers the shock and its immediate surroundings. In these cells, the primitive variables are reconstructed using minmod slope-limited linear reconstruction \citep[see, e.g.,][]{leveque2002}. Additionally, the Riemann solver is switched to the more diffusive two-wave HLL solver of \cite{harten1983}, which provides additional damping of non-physical oscillations near shocks (see also Sect.~\ref{sec:bw}). This approach preserves high-order accuracy in smooth regions of the flow while still capturing strong-shock dynamics \citep{mignone2005,stone2008}. 
\subsection{Well-balancing}\label{sec:wb}

To maintain hydrostatic solutions and reduce discretization errors arising from the imbalance between volumetric gravitational accelerations and pressure gradients, \texttt{PHLEGETHON} employs the deviation method \citep[][]{berberich2019,edelmann2021a}. This method splits the set of conserved quantities into a hydrostatic background state $\hat{\mathbf{U}}$ and a finite-amplitude deviation $\Delta\mathbf{U}$,
\begin{equation}
    \mathbf{U} = \hat{\mathbf{U}} + \Delta \mathbf{U}.
\end{equation}
The background state $\hat{\mathbf{U}}$ needs to be known a priori and is held fixed in time. 

In the deviation method, the system of conservation laws applied to the background state is subtracted from the system applied to the full state, yielding
\begin{equation}\label{eq:wb}
    \partial_t \mathbf{U} + \boldsymbol{\nabla} \cdot \left[ \mathbf{F}(\mathbf{U} ) - \mathbf{F}(\hat{\mathbf{U}}) \right]
    = \mathbf{S}(\mathbf{U})-\mathbf{S}(\hat{\mathbf{U}}).
\end{equation}
At the level of the partial differential equations, subtracting the system of conservation laws applied to the background state is equivalent to subtracting 0. However, in the discrete sense, this procedure preserves hydrostatic solutions to round-off error. In fact, although Godunov-type schemes generally result in
\begin{equation}
    \boldsymbol{\nabla} \cdot \mathbf{F}(\hat{\mathbf{U}}) \neq \mathbf{S}(\hat{\mathbf{U}}),
\end{equation}
substituting $\mathbf{U}=\hat{\mathbf{U}}$ into Eq.~(\ref{eq:wb}) immediately yields
\begin{equation}
    \partial_t \mathbf{U} = \mathbf{0}.
\end{equation}
For this scheme to work, the cell-centered equilibrium source terms $\hat{\mathbf{S}} :=\mathbf{S}(\hat{\mathbf{U}})$, as well as the face-centered equilibrium fluxes $\hat{\mathbf{F}} := \mathbf{F}(\hat{\mathbf{U}})$, must be known a priori. Practically, the subtraction of the equilibrium fluxes and source terms in Eq.~(\ref{eq:wb}) is performed in two stages:
\begin{enumerate}[label=(\roman*)]
\item After the Riemann solve step described in Sect.~\ref{sec:riemann}, the equilibrium fluxes are subtracted from the interface flux,
\begin{equation}\label{wb1}
    \mathbf{F}_{i+1/2} \rightarrow \mathbf{F}_{i+1/2}-\hat{\mathbf{F}}_{i+1/2}.
\end{equation}
Analogous expressions can be derived for the fluxes in the $x_2$- and the $x_3$-direction.
\item Equilibrium source terms are subtracted at cell centers,
\begin{equation}\label{wb2}
    \mathbf{S}_{i,j,k} \rightarrow \mathbf{S}_{i,j,k} - \hat{\mathbf{S}}_{i,j,k}.
\end{equation}
\end{enumerate}

In addition to subtracting numerical residuals that arise from discretization errors, the deviation method also modifies the spatial reconstruction step described in Sect.~\ref{sec:reconstruction} by reconstructing only the deviation from the hydrostatic background state to the cell interfaces. Within the Godunov algorithm implemented in \texttt{PHLEGETHON}, the deviation of the primitive variables is first computed at cell centers,
\begin{equation}
    \Delta \mathbf{W}_{i} = \mathbf{W}_i-\hat{\mathbf{W}}_i.
\end{equation}
Next, the deviation is reconstructed to cell interfaces using the same methods introduced in Sect.~\ref{sec:reconstruction}, yielding
\begin{align}
    \Delta \mathbf{W}_{i-1/2,\mathrm{R}} &= \mathcal{P}^-(\Delta\mathbf{W}_{i-q,\dots,i+q}), \\
    \Delta\mathbf{W}_{i+1/2,\mathrm{L}} &= \mathcal{P}^+(\Delta\mathbf{W}_{i-q,\dots,i+q}).
\end{align}
The equilibrium state is then added back at cell interfaces, e.g.,
\begin{equation}
 \mathbf{W}_{i+1/2,\mathrm{L/R}} =
 \Delta \mathbf{W}_{i+1/2,\mathrm{L/R}} + \hat{\mathbf{W}}_{i+1/2}.
\end{equation}
Finally, the resulting pair of Riemann states is passed to the Riemann solver. Riemann solvers are numerically consistent with the underlying system of governing equations, so when $\Delta \mathbf{W}_{i+1/2,\mathrm{L/R}}=\mathbf{0}$ everywhere, as is the case for hydrostatic solutions, the interface flux reads 
\begin{equation}
    \mathbf{F}_{i+1/2}=\mathbf{F}(\hat{\mathbf{W}}_{i+1/2},\hat{\mathbf{W}}_{i+1/2})=\hat{\mathbf{F}}_{i+1/2}.
\end{equation}
Using this flux in Eq.~(\ref{wb1}), alongside Eq.~(\ref{wb2}), yields a well-balanced evolution. To use the deviation method, the equilibrium primitive variables must be known a priori both at cell centers, $\hat{\mathbf{W}}_{i,j,k}$, and at cell faces, e.g., $\hat{\mathbf{W}}_{i+1/2,j,k}$.

Reconstructing the deviation is particularly important when modeling low-Mach-number flows arising from steep background pressure gradients, typical of layers deep inside stars. In such cases, reconstructing the full state can introduce spatial discretization errors that overwhelm the small physical pressure fluctuations that arise in low-Mach-number flows, which are $\,{\Delta P / \hat{P} \approx \mathcal{M}^2\,}$ \citep[see, e.g.,][]{miczek2015}. By contrast, reconstructing the deviation typically produces reconstruction errors that remain smaller than the dynamical fluctuations. This aspect of the deviation method allows a Godunov-like scheme to simulate Mach numbers as low as permitted by the finite precision of floating-point arithmetic \citep[e.g.,][]{edelmann2021a}. Additionally, when the auxiliary indices $\gamma_e$ and $\gamma_c$ are used in the simulation (see Sect.~\ref{sec:riemann}), we find that reconstructing the deviation of $\gamma_e$ from its background state helps suppress numerical artifacts that can arise in strongly stratified flows whose Mach number is $\,{\mathcal{M}\lesssim 10^{-3}}$.

One design flaw of the deviation method is that \textit{any} time-independent background state $\hat{\mathbf{U}}$ is balanced by construction, even if it is not a formal hydrostatic solution of the governing equations. Using  background stratifications that formally are not hydrostatic leads to a numerical discretization that is inconsistent with the governing equations and may over-stabilize the time evolution of the conserved quantities $\mathbf{U}$. Therefore, when only numerical solutions are available for $\hat{\mathbf{U}}$, it is necessary to ensure that the residual ${|\boldsymbol{\nabla} \hat{P} - \hat{\rho}\mathbf{g}|}$ vanishes up to discretization errors.

\subsection{Time discretization}\label{sec:time-disc}

The system of ODEs resulting from the spatial discretization of fluxes and source terms is advanced in time using explicit strong stability preserving (SSP) Runge--Kutta (RK) methods. \texttt{PHLEGETHON} incorporates the SSP-RK2 and SSP-RK3 methods of \cite{shu1988}, which are second- and third-order accurate in time, respectively.  

In particular, the SSP-RK2 update from $t^n$ to $\,{t^{n+1}=t^n+\Delta t}\,$  is performed in two stages,
\begin{align}
\langle \mathbf{U}^{(*)} \rangle_{i,j,k} &= \langle \mathbf{U}^{(n)} \rangle_{i,j,k} -  \mathbf{R}_{i,j,k}\Big(\langle \mathbf{U}^{(n)} \rangle\Big)\Delta t, \\
\langle\mathbf{U}^{(n+1)}\rangle_{i,j,k} &= \frac{1}{2}\langle\mathbf{U}^{(n)}\rangle_{i,j,k} + \frac{1}{2}\langle\mathbf{U}^{(*)}\rangle_{i,j,k} -  \frac{1}{2}\mathbf{R}_{i,j,k}\Big(\langle\mathbf{U}^{(*)}\rangle\Big)\Delta t.
\end{align}
The SSP-RK3 update requires three stages,
\begin{align}
\langle \mathbf{U}^{(*)} \rangle_{i,j,k} &= \langle \mathbf{U}^{(n)} \rangle_{i,j,k} -  \mathbf{R}_{i,j,k}\Big(\langle \mathbf{U}^{(n)} \rangle\Big)\Delta t, \\
\langle \mathbf{U}^{(**)}\rangle_{i,j,k} &= \frac{3}{4}\langle \mathbf{U}^{(n)} \rangle_{i,j,k} + \frac{1}{4} \langle\mathbf{U}^{(*)}\rangle_{i,j,k} -  \frac{1}{4}\mathbf{R}_{i,j,k}\Big(\langle\mathbf{U}^{(*)}\rangle\Big)\Delta t, \\
\langle\mathbf{U}^{(n+1)}\rangle_{i,j,k} &= \frac{1}{3}\langle\mathbf{U}^{(n)}\rangle_{i,j,k} + \frac{2}{3}\langle\mathbf{U}^{(**)}\rangle_{i,j,k} -  \frac{2}{3}\mathbf{R}_{i,j,k}\Big(\langle\mathbf{U}^{(**)}\rangle\Big)\Delta t.
\end{align}

Here, $\mathbf{R}(\mathbf{U})$ denotes the spatial residual vector, computed using the methods described in the previous sections,
\begin{equation}\label{eq:spatres}
    \mathbf{R}_{i,j,k}(\mathbf{U}) =  \frac{1}{V_{i,j,k}}\sum_f A_f \mathbf{F}_f(\mathbf{U}) \cdot \mathbf{n}_f
- \mathbf{S}_{i,j,k}(\mathbf{U}).
\end{equation}
Magnetic-field components at cell faces are updated concurrently with the other conserved quantities. The corresponding residuals, for example at the face $i+1/2,j,k$, are given by
\begin{equation}
    \mathbf{R}_{i+1/2,j,k}(\mathbf{U}) = \frac{1}{A_f}\sum_d l_d \mathcal{E}_d(\mathbf{U}).
\end{equation}
Source terms, including gravity, rotation, and thermal diffusion, are incorporated directly in $\mathbf{R}_{i,j,k}(\mathbf{U})$ and are advanced together with the fluxes in each SSP-RK stage. Stiff source terms that require special treatment are discussed in Sects.~\ref{sec:sts} and~\ref{sec:nuclear-network}.

\texttt{PHLEGETHON} evaluates the time step $\Delta t$ according to the CFL stability criterion that constraints the hyperbolic part of the governing equations,
\begin{equation}\label{eq:cfl}
    \Delta t = \Delta t_\mathrm{h} = \mathrm{CFL} \left[ \max_{i,j,k} \left( \sum_{s=1}^3 \frac{|u_{s,i,j,k}|+C_{x_s,i,j,k}}{\Delta x_{s,i,j,k}}\right) \right]^{-1},
\end{equation}
where $s$ denotes the spatial direction and $C_{x_s}$ is the local fastest characteristic speed along that direction (sound speed for hydrodynamics and fast-magnetosonic speed for MHD). In all verification tests presented in Sect.~\ref{sec:verification}, we use SSP-RK3 with ${\mathrm{CFL}=0.8}$, unless explicitly stated otherwise. 

To ensure that the mass fraction abundances sum to unity after each SSP-RK substep update, the consistent multi-fluid advection method of \cite{plewa1999} is used. Before entering the Riemann solver subroutine (see Sect.~\ref{sec:riemann}), the mass fractions are rescaled as
\begin{equation}
    X_l \rightarrow \frac{X_l}{\sum_m X_m}. 
\end{equation}
These rescaled abundances are then used to compute the advective fluxes $\rho X_l \mathbf{u}$. This procedure guarantees that the sum of the individual species mass fluxes is consistent with the total mass flux,
\begin{equation}
  \rho \mathbf{u} = \sum_l \rho X_l \mathbf{u},
\end{equation}
even at the discrete level.

We note that the CFL constraint associated with explicit time integration in Eq.~(\ref{eq:cfl}) becomes increasingly stringent at low Mach numbers. This constraint can, in principle, be alleviated using semi-implicit or fully implicit time-integration schemes \citep[e.g.,][]{chacon2008,lutjens2010,fambri2021,leidi2022,boscheri2024}. Such methods, however, involve important trade-offs. Although they permit time steps exceeding the explicit CFL limit without losing numerical stability, their computational cost per time step is generally substantially higher than that of explicit methods, and the resulting performance advantage becomes significant only at sufficiently low Mach numbers, approximately $\mathcal{M}\lesssim10^{-3}$ \citep{leidi2022}. Since the primary applications targeted by \texttt{PHLEGETHON} typically involve Mach numbers higher than $10^{-3}$, implicit time integration is not generally advantageous for these applications. Moreover, implicit methods require considerably more numerical machinery than explicit schemes, including the evaluation and use of Jacobian matrices for the governing equations, making the implementation of additional physical processes and numerical features substantially more involved. Finally, taking time steps based on the much longer advective or Alfvénic cell-crossing times can lead to poor temporal resolution of acoustic and fast-magnetosonic waves in low-Mach-number flows. Accurately resolving these compressible modes is important for stellar simulations aimed at producing asteroseismic synthetic observables that can be directly compared with observations \citep[e.g.,][]{mosser2012,deheuvels2012,montalban2013,coppee2024}. 
 
\subsection{Super-time-stepping for stiff diffusive heat transport}\label{sec:sts}

Transport of thermal energy is accounted for in the diffusive limit. The thermal flux in Eq.~(\ref{eq:energy}) is first discretized at cell faces using the second-order finite-difference formula (here for simplicity given for a Cartesian grid geometry) 
\begin{equation}
    F_{\mathrm{th},i+1/2} = - K_{i+1/2} \frac{T_{i+1}-T_{i}}{x_{i+1}-x_{i}},
\end{equation}
and then added to the spatial residuals in Eq.~(\ref{eq:spatres}). 
The thermal conductivity at the cell interface, $K_{i+1/2}$, is computed as the harmonic mean of the values in the two cells neighboring the interface, 
\begin{equation}
    K_{i+1/2} = \frac{4ac}{3}\frac{2\chi_{i} \chi_{i+1}}{\chi_{i}+\chi_{i+1}},
\end{equation}
where
\begin{equation}
    \chi_i = \frac{T_i^3}{\kappa_i\rho_i},
\end{equation}
and $c$ is the speed of light. The total opacity $\kappa$ is derived from the radiative, $\kappa_\mathrm{rad}$, and conductive, $\kappa_\mathrm{cond}$, opacities as
\begin{equation}
    \kappa = \frac{\kappa_\mathrm{rad}\kappa_\mathrm{cond}}{\kappa_\mathrm{rad}+\kappa_\mathrm{cond}}.
\end{equation}
In \texttt{PHLEGETHON}, radiative and conductive opacities can either be mapped onto the grid and held fixed in time, or computed at runtime following the analytical prescriptions of \cite{timmes2000a}, which are functions of $(\rho,T,\widebar{A},Y_e)$\footnote{The implemenation of conductive and radiative opacities in \texttt{PHLEGETHON} is based on \texttt{kapfxt.tar.xz} available at \url{https://cococubed.com/code_pages/kap.shtml} (last accessed 8 August 2026).}.

The operator discretizing thermal diffusion is parabolic, so it can only be integrated in a stable manner using time-explicit methods if the time step satisfies the parabolic CFL stability constraint \citep[see, e.g.,][]{leveque2007},
\begin{equation}
    \Delta t < \Delta t_\mathrm{p} = \frac{1}{2}
    \left\{
    \max_{i,j,k}
    \left[
    \sum_{s=1}^3
    \frac{D_{i,j,k}}{(\Delta x_{s,i,j,k})^2}
    \right]
    \right\}^{-1},
\end{equation}
where
\begin{equation}
    D = \frac{4acT^3}{3\kappa\rho^2 c_v}.
\end{equation}
On fine grids or in regions of low opacity, this stability constraint can become prohibitively restrictive. One way to overcome this limitation is to employ super-time-stepping (STS) techniques, which relax the stability constraint for parabolic operators at moderate computational cost via a sequence of explicit substeps designed to suppress unstable modes. These methods are particularly suitable for non-symmetric operators with purely negative real eigenvalues, as it is the case for thermal diffusion.

\texttt{PHLEGETHON} employs the second-order Runge--Kutta--Legendre (RKL2) super time-stepper scheme of \cite{meyer2012}. The method is implemented through operator splitting and coupled to the system of MHD equations using second-order Strang splitting \citep{strang1968}. During the STS substeps, the mass density, velocity, magnetic field, and mass fractions are held fixed, while only the specific internal energy is evolved. Under these conditions, Eq.~(\ref{eq:energy}) reduces to
\begin{equation}\label{eq:diffusion_eint}
    \rho \partial_t e_\mathrm{int} = \boldsymbol{\nabla}\cdot(K\boldsymbol{\nabla} T),
\end{equation}
which can be written in operator form as
\begin{equation}
     \partial_t e_\mathrm{int} = \mathbf{M}\big[T(e_\mathrm{int})\big].
\end{equation}
The RKL2 method advances the solution over a super time step $\Delta t_{\mathrm{STS}}$ through a sequence of $N_\mathrm{STS}$ substeps. Denoting the internal energy at stage $m$ by $e_\mathrm{int}^{(m)}$, the algorithm proceeds as follows:

\begin{enumerate}[label=(\roman*)]

\item The initial state $e_\mathrm{int}^{(0)}=e_\mathrm{int}^{(n)}$ is set.

\item The first stage is performed using a forward Euler step,
\begin{equation}
e_\mathrm{int}^{(1)} =
e_\mathrm{int}^{(0)} +
\tilde{\mu}_1 \Delta t_{\mathrm{STS}}\,
\mathbf{M}\left[T\left(e_\mathrm{int}^{(0)}\right)\right].
\end{equation}

\item For stages $m=2,\dots,N_\mathrm{STS}$, the solution is updated according to
\begin{equation}
\begin{split}
e_\mathrm{int}^{(m)} &= \mu_m e_\mathrm{int}^{(m-1)} + \nu_m e_\mathrm{int}^{(m-2)} \\
& + (1-\mu_m-\nu_m)e^{(0)}_\mathrm{int} \\
& + \tilde{\mu}_m \Delta t_{\mathrm{STS}}\,
\mathbf{M}\left[T\left(e_\mathrm{int}^{(m-1)}\right)\right] \\
& + \tilde{\gamma}_m \Delta t_{\mathrm{STS}}\,
\mathbf{M}\left[T\left(e_\mathrm{int}^{(0)}\right)\right].
\end{split}
\end{equation}
Explicit expressions for the coefficients $\mu_m$, $\tilde{\mu}_m$, $\nu_m$, and $\tilde{\gamma}_m$ are given in \cite{meyer2012} and are chosen to maximize the permitted time step $\Delta t$ for a given $N_\mathrm{STS}$ while maintaining stability in the integration of parabolic terms.

\item After completing the $N_\mathrm{STS}$ substeps, the solution is updated as
\begin{equation}
e_\mathrm{int}^{(n+1)} = e_\mathrm{int}^{(N_\mathrm{STS})}.
\end{equation}

\end{enumerate}

The number of substeps $N_\mathrm{STS}$ determines the effective time step that can be taken relative to the explicit parabolic stability limit $\Delta t_{\mathrm{p}}$. For the RKL2 scheme, the maximum stable super time step is 
\begin{equation}
    \Delta t_\mathrm{STS} = \frac{N_\mathrm{STS}^2+N_\mathrm{STS}-2}{4}\Delta t_\mathrm{p},
\end{equation}
which for large $N_\mathrm{STS}$ can be approximated as
\begin{equation}
    \Delta t_\mathrm{STS} \approx \frac{N_\mathrm{STS}^2}{4}\Delta t_\mathrm{p}.
\end{equation}
Consequently, increasing the number of substeps allows the diffusion operator to be advanced over time intervals much larger than the explicit stability limit while preserving second-order accuracy in time. In \texttt{PHLEGETHON}, both steps of Strang-splitting determine the number of substeps as if the thermal diffusion operator were to be advanced over the full hyperbolic CFL time step in Eq.~(\ref{eq:cfl}),
\begin{equation}
    N_\mathrm{STS} = \max \left\{
    1 + \Bigg\lfloor
    \frac{1}{2}\left(
    \sqrt{9+16\frac{\Delta t_\mathrm{h}}{\Delta t_\mathrm{p}}}-1
    \right)
    \Bigg\rfloor,
    3
    \right\},
\end{equation}
where $\lfloor \cdot \rfloor$ denotes the floor function\footnote{This implementation is based on that in the \texttt{ATHENA++} code \citep{stone2020}.}. A minimum of three substeps is required to ensure that the scheme retains second-order accuracy in time \citep{meyer2012}. 

One limitation of the RKL2 method is that the equation of state must be evaluated in every substep to determine the temperature corresponding to the current specific internal energy, while holding density and composition fixed (see Sect.~\ref{sec:eos}). When the EoS evaluation is computationally expensive, as in the case of a tabulated EoS, the overall cost of the simulation can become prohibitively high for large numbers of substeps. One possible solution to this problem is to evolve the equivalent temperature equation
\begin{equation}
    \rho c_v \partial_t T = \boldsymbol{\nabla} \cdot (K \boldsymbol{\nabla} T).
\end{equation}
The STS update formulas are equivalent to those used when evolving the specific internal energy, only the operator $\mathbf{M}$ has a different form,
\begin{equation}
    \mathbf{M}\big[T\big]\rightarrow \frac{1}{c_v}\mathbf{M}\big[T\big].
\end{equation}
By fixing the value of $c_v$ at the beginning of the time step, the governing equation can be solved without repeatedly invoking the EoS routine, except for updating the internal energy at the end of the STS step,
\begin{equation}
    e^{(n+1)}_\mathrm{int} =
    e_{\mathrm{int,EoS}}(\rho^{(n+1)},T^{(n+1)},X_l^{(n+1)}).
\end{equation}
Although this choice formally reduces the temporal accuracy of the scheme to first order, the leading first-order error terms are subdominant on grids of practical resolution provided that the specific heat $c_v$ does not vary significantly over a single time step (see also Sect.~\ref{sec:diff-pulse}). In \texttt{PHLEGETHON}, either the method that directly updates the specific internal energy or the method that evolves the temperature equation can be used. 

Moreover, when deviation well-balancing is enabled (see Sect.~\ref{sec:wb}), the code also offers the option of applying the thermal diffusion operator only to the deviation from the background hydrostatic temperature, $\Delta T$, rather than to the full state,
\begin{equation}
\rho \partial_t e_\mathrm{int} = \boldsymbol{\nabla} \cdot (K \boldsymbol{\nabla} \Delta T),
\end{equation}
or, equivalently,
\begin{equation}\label{eq:diffusion_temp}
\rho c_v \partial_t T = \boldsymbol{\nabla} \cdot (K \boldsymbol{\nabla} \Delta T).
\end{equation}
This option is useful whenever thermal excess relative to the background state needs to be diffused in stratified setups, thereby avoiding the generation of large thermal fluxes associated with background temperature gradients (see Sect.~\ref{sec:bb}).

\subsection{Boundary conditions}\label{sec:bcs}

\texttt{PHLEGETHON} employs layers of ghost cells surrounding each local computational domain block to reconstruct pairs of Riemann states at the domain boundaries. MPI send-receive communication calls are used to pass the primitive-variable values from one domain to fill the ghost cells of a neighboring block. In contrast, the ghost cells at the edges of the global grid are filled according to the selected boundary conditions.

The number of ghost cells required to run a given numerical algorithm depends on the reconstruction method. Two ghost cells are sufficient for van Leer and PPH reconstructions, while the limited fifth-order reconstruction method requires three ghost cells. To guarantee that the shock-flattening procedure (Sect.~\ref{sec:shock-flat}) produces consistent results across different domain decompositions, at least three ghost cells are required.

Several boundary conditions are implemented in the code, including periodic, constant-ghost, reflecting, outflow, diode, a purely-normal magnetic field, and fixed-temperature conditions (the latter only for thermal diffusion; see Table~\ref{table:bcs}). In addition, \texttt{PHLEGETHON} supports inscribed solid boundaries, in which solid geometries are embedded within the computational domain by marking grid faces that coincide with the solid surface. Reflecting boundary conditions are enforced on the primitive variables at these interfaces. This capability is currently available only on Cartesian grids (see also Sect.~\ref{sec:ccsnp}).

For our constrained-transport implementation, boundary conditions must also be imposed on the electromotive force to ensure a consistent evolution of the face-centered magnetic field. For instance, for reflecting boundaries, the analytical boundary condition listed in Table~\ref{table:bcs} imposes a vanishing magnetic-field component normal to the boundary, corresponding to a purely tangential magnetic field at the boundary. In the constrained-transport implementation, this condition is enforced using ghost cells. The reflection properties of the electromotive force follow from those of the velocity and magnetic-field components entering ${\mathcal{E}= - \mathbf{u}\times\mathbf{B}}$. For a reflecting boundary, these properties imply that the tangential electromotive-force components change sign across the boundary, whereas the normal component does not. When applying the Contact-CT algorithm, this results in a vanishing tangential electromotive force at the grid boundary to round-off accuracy. Consequently, the constrained-transport update leaves the normal magnetic field at the boundary face unchanged, so that it remains fixed at its initial value to round-off accuracy, which is zero for the reflecting boundary condition. For inscribed solid boundaries, we explicitly mark the grid faces coinciding with the solid surface and apply the same ghost-cell procedure. This treatment ensures that the constrained-transport update preserves the divergence-free condition also in the vicinity of the solid boundary.

\begin{table*} \caption{Boundary conditions available in \texttt{PHLEGETHON}.} \label{table:bcs} \centering \begin{tabular}{ll} \toprule \\ Type & Description \\ \midrule Periodic & Ghost-cell values are copied from the opposite side of the computational domain. \\ Constant ghost & Primitive variables fixed to user-prescribed values in the ghost cells. \\ Reflecting & $u_n = 0$, $B_n = 0$, and $\partial_n W = 0$ for remaining primitive variables. \\ Outflow & $\partial_n W = 0$ for all primitive variables. \\ Diode & Outflow with $u_n \ge 0$. \\ Fixed-temperature (only for thermal diffusion) & $T = T_{\mathrm{bc}}$ at the last cell face; ghost cells filled via 2nd-order extrapolation. \\ Purely normal magnetic field & $\partial_n B_n = 0$ and zero transverse magnetic field components at the boundary, $B_t = 0$. \\ Inscribed solid & Reflecting conditions applied at embedded solid interfaces. \\ \bottomrule \end{tabular} \end{table*}

\newpage

\subsection{Poisson solver}\label{sec:poisson}

In simulations where the gravitational potential must be updated in time, Poisson's equation
\begin{equation}
    \nabla^2 \phi = 4 \pi G \rho,
\end{equation}
can either be solved at the beginning of every Runge--Kutta stage, or before the hyperbolic update using first-order Godunov time splitting. Currently, the Poisson solver is only supported for Cartesian grids with the option to apply an arbitrary stretching function to each coordinate axis separately. Boundary conditions for the gravitational potential are imposed using a multipole expansion of the mass distribution inside the computational domain. The monopole term is always included, while quadrupole and octupole corrections can optionally be added.

A finite-difference discretization of the Laplacian operator for non-uniform Cartesian grids is employed,
\begin{align}
\label{eq:poisson-FD}
    L_{1,x}\phi_{i-1,j,k} + L_{2,x}\phi_{i,j,k} + L_{3,x}\phi_{i+1,j,k}  \ + &\nonumber \\
     L_{1,y}\phi_{i,j-1,k}  + L_{2,y}\phi_{i,j,k} + L_{3,y}\phi_{i,j+1,k} \  + & \\
    L_{1,z}\phi_{i,j,k-1} + L_{2,z}\phi_{i,j,k} + L_{3,z}\phi_{i,j,k+1} \    = & \ 4\pi G\rho_{i,j,k} \nonumber
\end{align}
with
\begin{align}
    L_{1,x} =\ & \frac{2}{\Delta x_-(\Delta x_- + \Delta x_+)}, \\
    L_{2,x} =\ & -\frac{2}{\Delta x_- \Delta x_+}, \\
    L_{3,x} =\ & \frac{2}{\Delta x_+(\Delta x_- + \Delta x_+)}, \\
    L_{1,y} =\ &\frac{2}{\Delta y_-(\Delta y_- + \Delta y_+)}, \\
    L_{2,y} =\ & -\frac{2}{\Delta y_- \Delta y_+}, \\
    L_{3,y} =\ & \frac{2}{\Delta y_+(\Delta y_- + \Delta y_+)}, \\
    L_{1,z} =\ & \frac{2}{\Delta z_-(\Delta z_- + \Delta z_+)}, \\
    L_{2,z} =\ & -\frac{2}{\Delta z_- \Delta z_+}, \\
    L_{3,z} =\ & \frac{2}{\Delta z_+(\Delta z_- + \Delta z_+)},
\end{align}
and 
\begin{align}
    \Delta x_- =\ & x_{i,j,k}-x_{i-1,j,k}, \\
    \Delta x_+ =\ & x_{i+1,j,k}-x_{i,j,k}, \\
    \Delta y_- =\ & y_{i,j,k}-y_{i,j-1,k}, \\
    \Delta y_+ =\ & y_{i,j+1,k}-y_{i,j,k}, \\
    \Delta z_- =\ & z_{i,j,k}-z_{i,j,k-1}, \\
    \Delta z_+ =\ & z_{i,j,k+1}-z_{i,j,k}.
    \label{eq:poisson_last}
\end{align}
This finite-difference discretization is formally only first-order accurate on non-uniform grids, with the leading first-order error term being proportional to $|\Delta \xi_+ - \Delta \xi_-|$ ($\xi = x, y, z$). Although first-order convergence is expected on sufficiently fine grids, in practice second-order convergence is achieved on grids of practical resolution with smoothly varying cell sizes (see Sect.~\ref{sec:collapse}).

The resulting linear system
\begin{equation}\label{eq:Lphi}
L  \boldsymbol{\phi} = 4\pi G \boldsymbol{\rho}
\end{equation}
is solved iteratively using a matrix-free implementation of the bi-conjugate gradient stabilized method (BiCGSTAB) of \cite{vandervorst1992}, preconditioned by a geometric multigrid method. The multigrid hierarchy employs volume-weighted averaging for restriction, trilinear interpolation for prolongation, and weighted red-black Gauss--Seidel smoothing,
\begin{algorithmic}
\For{each color $c\in\{\mathrm{red},\mathrm{black}\}$}
    \For{each cell $(i,j,k)$ of color $c$}
        \State $\phi_{i,j,k} \rightarrow
        (1-\gamma_{\mathrm{GS}})\phi_{i,j,k}
        +\gamma_{\mathrm{GS}}\phi_{i,j,k}^{\mathrm{GS}}$
    \EndFor
\EndFor
\end{algorithmic}
Here, $\gamma_{\mathrm{GS}}$ is the relaxation parameter and
$\phi_{\mathrm{GS}}$ is the value obtained from the standard Gauss--Seidel
update using the neighboring cell values. ${\gamma_{\mathrm{GS}}=1}$ recovers the
standard red-black Gauss--Seidel algorithm. Each application of the multigrid preconditioner performs one V-cycle through the multigrid hierarchy. The numbers of pre- and post-smoothing iterations, the number of smoothing iterations on the coarsest grid, and the maximum depth of the multigrid hierarchy can be specified by the user at compile time. In Eq.~(\ref{eq:Lphi}), $L$ is the discrete Laplacian operator, $\boldsymbol{\phi}$ is the discrete gravitational potential vector, and $\boldsymbol{\rho}$ is the discrete mass density vector.

The iteration is terminated once the root-mean-square relative residual
\begin{equation}\label{eq:poisson-eps}
\sqrt{
\dfrac{1}{N_x N_y N_z}
\sum_{i=1}^{N_x}
\sum_{j=1}^{N_y}
\sum_{k=1}^{N_z}
\left(
\dfrac{(L\boldsymbol{\phi})_{i,j,k} - 4 \pi G \rho_{i,j,k}}{4 \pi G \rho_{i,j,k}}
\right)^2
}
\end{equation}
falls below a prescribed tolerance $\epsilon$ (a default value of ${\epsilon = 10^{-10}}$ is used in \texttt{PHLEGETHON}\footnote{We note that the optimal choice of the tolerance $\epsilon$ is problem-dependent and may require fine-tuning for each setup.}). We choose this relative error form because it provides a dimensionless measure of the residual normalized to the local magnitude of the source term, thereby allowing the convergence criterion to reflect the relative accuracy of the solution across regions with different density scales. The gravity vector ${\mathbf{g} = -\boldsymbol{\nabla} \phi}$, needed to compute source terms in the momentum and energy equations (see Sect.~\ref{sec:pdes}), is discretized using a second-order central-derivative approximation for non-uniform grids,
\begin{align}
g_x = &- \frac{\Delta x_-}{\Delta x_+ (\Delta x_+ + \Delta x_-)} \phi_{i+1,j,k} - \frac{\Delta x_+ - \Delta x_-}{\Delta x_+ \Delta x_-} \phi_{i,j,k}\nonumber\\
&+\frac{\Delta x_+}{\Delta x_- (\Delta x_+ + \Delta x_-)} \phi_{i-1,j,k}, \label{eq:grav_x} \\ 
g_y = &- \frac{\Delta y_-}{\Delta y_+ (\Delta y_+ + \Delta y_-)} \phi_{i,j+1,k} - \frac{\Delta y_+ - \Delta y_-}{\Delta y_+ \Delta y_-} \phi_{i,j,k}\nonumber\\
&+\frac{\Delta y_+}{\Delta y_- (\Delta y_+ + \Delta y_-)} \phi_{i,j-1,k}, \label{eq:grav_y} \\ 
g_z = &- \frac{\Delta z_-}{\Delta z_+ (\Delta z_+ + \Delta z_-)} \phi_{i,j,k+1} - \frac{\Delta z_+ - \Delta z_-}{\Delta z_+ \Delta z_-} \phi_{i,j,k}\nonumber\\
&+\frac{\Delta z_+}{\Delta z_- (\Delta z_+ + \Delta z_-)} \phi_{i,j,k-1} \label{eq:grav_z}.
\end{align}

\subsection{Nuclear reaction network solvers}\label{sec:nuclear-network}

\texttt{PHLEGETHON} incorporates arbitrary nuclear reaction networks that can be coupled to the MHD equations. The nuclear network is treated as a source term for the mass fractions and specific internal energy of the gas. Below, we present the governing equations solved by the code and outline the numerical methods used.

The first law of thermodynamics for a multi-species reacting fluid reads \citep[e.g.,][]{cox2004}
\begin{equation}\label{eq:first-law}
   T D_t s + N_\mathrm{A} \sum_l \Phi_l D_t Y_l = D_t e_\mathrm{int,t} - \frac{P}{\rho^2}D_t \rho = \delta \dot{q},
\end{equation}
where $s$ is the specific entropy, $Y_l = X_l/A_l$ are the molar abundances, $N_\mathrm{A}$ is Avogadro's number, $\Phi_l$ is the chemical potential of species $l$, and $D_t = \partial_t + \mathbf{u} \cdot \boldsymbol{\nabla}$ is the Lagrangian derivative. The total specific internal energy, including the rest mass of each species, $m_l$, is
\begin{equation}
    e_\mathrm{int,t} = e_\mathrm{int} + N_\mathrm{A}\sum_l m_l c^2 Y_l,
\end{equation}
and $\delta \dot{q}$ represents the net heat gained or lost by the mixture. The thermal component, $e_\mathrm{int}$, is the quantity directly tracked in the simulation (see Sect.~\ref{sec:pdes}). Equation~(\ref{eq:first-law}) can be rearranged to yield the evolution equation for $e_\mathrm{int}$,
\begin{equation}
    D_t e_\mathrm{int} = \delta \dot{q} + \frac{P}{\rho^2} D_t \rho - N_\mathrm{A} \sum_l m_l c^2 D_t Y_l.
\end{equation}

At the same time, the evolution of the molar abundance of a species involved in one-, two-, and three-body reactions is governed by \citep[see, e.g.,][]{reichert2023}
\begin{equation}
\begin{split}
    D_t Y_l & = \sum_{r_1} \alpha_{r_1,l} Y_l \lambda_{r_1} \\
    & + \sum_{r_2} \sum_p  \alpha_{r_2,l,p}\frac{Y_l Y_p}{1+\delta_{p,l}} \rho N_\mathrm{A}\langle \sigma u\rangle_{r_2} \\
    & + \sum_{r_3} \sum_p \sum_q  \alpha_{r_3,l,p,q}\frac{Y_l Y_p Y_q}{1+\Delta_{q,p,l}} \rho^2 N_\mathrm{A}^2 \langle l,p,q\rangle_{r_3},
\end{split}
\end{equation}
where $\alpha_{r_1,l}$, $\alpha_{r_2,l,p}$, $\alpha_{r_3,l,p,q}$ are the stoichiometric coefficients for species $l$ in each type of reaction. The sum is carried over all reactions participating in the network. The Kronecker delta $\delta_{p,l}$ ensures correct counting for identical reactants, and
\begin{equation}
    \Delta_{q,p,l} = \delta_{q,p} + \delta_{q,l} + \delta_{p,l} + 2\delta_{p,q,l}.
\end{equation}
Here, $\lambda_{r_1}$ is the one-body reaction rate, $\langle \sigma u \rangle_{r_2}$ is the thermally averaged reaction rate for two-body reactions, and $\langle l,p,q \rangle_{r_3}$ is the rate coefficient for three-body reactions. 

\texttt{PHLEGETHON} uses reaction rates from the JINA REACLIB database \citep{cyburt2010}, which are given as exponentials of temperature-dependent functions. Weak reaction rates for medium-mass nuclei are interpolated from precomputed tables \citep{fuller1982,oda1994,langanke2001} in $(\rho Y_e,T)$. Electron screening corrections can optionally be enabled, following \cite{wallace1982}. Reverse reactions are multiplied by partition functions from \cite{rauscher2000} to account for thermally populated excited nuclear states in high-temperature plasmas.

Nuclear reactions can have characteristic timescales much shorter than the hyperbolic CFL time step in Eq.~(\ref{eq:cfl}) that governs the time integration algorithm. As a result, implicit time integration is generally required. In \texttt{PHLEGETHON}, implicit nuclear network solvers are coupled to the MHD system of equations via either Godunov or Strang splitting. During the time-split network step, the mass density is held fixed and species advection is neglected because it is already accounted for in the MHD update. Under these assumptions, the Lagrangian derivative reduces to a simple Eulerian derivative,
\begin{equation}
    D_t \rightarrow \partial_t,
\end{equation}
and the network step requires solving the coupled, nonlinear system of ODEs
\begin{equation}\label{eq:nuclear}
\begin{aligned}
& \partial_t e_\mathrm{int} = \delta \dot{q} - N_\mathrm{A} \sum_l m_l c^2 \, \partial_t Y_l, \\
&\partial_t Y_l = \sum_{r_1} \alpha_{r_1,l} Y_l \lambda_{r_1} 
+ \sum_{r_2} \sum_p  \alpha_{r_2,l,p}\frac{Y_l Y_p}{1+\delta_{p,l}} \rho N_\mathrm{A}\langle \sigma u\rangle_{r_2} \\
& \quad \quad + \sum_{r_3} \sum_p \sum_q  \alpha_{r_3,l,p,q}\frac{Y_l Y_p Y_q}{1+\Delta_{q,p,l}} \rho^2 N_\mathrm{A}^2 \langle l,p,q\rangle_{r_3}.
\end{aligned}
\end{equation}
The medium is assumed to be transparent for neutrinos, so $\delta \dot{q}<0$ accounts for mean neutrino losses, which have both thermal and nuclear contributions. Thermal neutrino losses are computed following \cite{itoh1996}, which includes pair, photo-, plasma, Bremsstrahlung, and recombination neutrinos\footnote{The implementation of thermal neutrino energy loss rates in \texttt{PHLEGETHON} is based on \texttt{sneut5.tbz} available at \url{https://cococubed.com/code_pages/nuloss.shtml} (last accessed 10 April 2026).}, while nuclear neutrino energy losses are included via weak rates tabulated in ($\rho Y_e$,$T$) \citep{fuller1982,oda1994,langanke2001} .

\texttt{PHLEGETHON} solves Eq.~(\ref{eq:nuclear}) using either the first-order backward Euler (BE) method or the trapezoidal backward-differentiation second-order (TR-BDF2) method of \cite{hosea1996}, the latter being an explicit-first-stage, singly diagonally implicit Runge--Kutta scheme. Both methods are A- and L-stable \citep[see, e.g.,][]{hosea1996}, which makes them suited for solving stiff ODE systems. To reduce implementation complexity, reaction rates are evaluated at the beginning of the nuclear network step, allowing the thermal energy equation to be decoupled from the species evolution. It should be noted that holding the value of the reaction rates fixed throughout the implicit solve can reduce numerical accuracy in vigorous burning regimes, such as in strong detonations \citep[see, e.g.,][]{zingale2021}, but such extreme regimes are outside the intended scope of \texttt{PHLEGETHON}. 

When casting the governing equation for the molar abundances $Y_l$ as
\begin{equation}\label{eq:yl}
    \partial_t Y_l - R_l(\mathbf{Y}) = 0,
\end{equation}
where $R_l(\mathbf{Y})$ denotes the net production rate of species $l$, the BE method reads
\begin{equation}
    Y_l^{n+1} - Y_l^n - R_l(\mathbf{Y}^{n+1})\Delta t = 0.
\end{equation}
The discrete form of Eq.~(\ref{eq:yl}) in the TR-BDF2 method can be obtained from the Butcher tableau \citep[see][]{hosea1996}
{\renewcommand{\arraystretch}{1.5}
\[
\begin{array}{c|ccc}
0 & 0 & 0 & 0 \\
\theta & \frac{\theta}{2} & \frac{\theta}{2} & 0 \\
1 & \frac{3\theta-\theta^2-1}{2 \theta} & \frac{1-\theta}{2\theta} & \frac{\theta}{2} \\
\hline
 & \frac{3\theta-\theta^2-1}{2 \theta} & \frac{1-\theta}{2\theta} & \frac{\theta}{2}
\end{array}
\]
}
with $\,{\theta = 2 - \sqrt{2}}$. Both the BE and TR-BDF2 implementations employ a Newton--Raphson root-finding algorithm, using $\,{Y_l^* = Y_l^n}\,$ as an initial guess. In \texttt{PHLEGETHON}, the Jacobian of the system is computed and stored as a dense matrix. For small reaction networks ($N_\mathrm{species}\, {\lesssim}\, 400$), this approach remains competitive with sparse-matrix solvers while substantially reducing implementation complexity \citep{reichert2023}.

Focusing on BE for simplicity, the residuals in each substep of the root-finding algorithm are obtained as
\begin{equation}
    E_l(\mathbf{Y}^*) =  Y_l^{*} - Y_l^n - R_l(\mathbf{Y}^{*})\Delta t,
\end{equation}
where $Y^*$ is the current guess. Therefore, the Jacobian is
\begin{equation}
    J_{lb} = \frac{\partial E_l}{\partial Y^*_b} = \delta_{lb} - \frac{\partial R_l(\mathbf{Y}^{*})}{\partial Y_b}\Bigg|_{\mathbf{Y}^*}\Delta t.
\end{equation}
The increment $\Delta Y^*_l$ on the guess $Y^*$ is obtained by solving a ${N_\mathrm{species} \times N_\mathrm{species}}$ linear system of equations,
\begin{equation}
    \mathbf{J} \Delta \mathbf{Y}^* = - \mathbf{E}.
\end{equation}
This linear system is solved iteratively using BiCGSTAB. When the solution to the linear system is obtained, the Newton--Raphson iteration proceeds until the root mean square relative residual 
\begin{equation}\label{eq:nuclear-thr}
    \sqrt{\frac{1}{N_\mathrm{species}} \sum_l E_l^2(\mathbf{Y}^*)}
\end{equation}
falls below a prescribed threshold (the default value is $10^{-13}$ in \texttt{PHLEGETHON}).

Once convergence is reached for the abundances, the internal energy is updated. In BE,
\begin{equation}
    e_\mathrm{int}^{\,n+1} =
    e_\mathrm{int}^{\,n}
    + \delta \dot{q}^{n+1} \Delta t
    - N_\mathrm{A}\sum_l  m_l c^2 \big( Y_l^{\,n+1} - Y_l^n \big).
\end{equation}
Analogous expressions follow straightforwardly for the TR-BDF2 scheme. The temperature is then reconstructed from the updated internal energy and composition, using the iterative approach described in Sect.~\ref{sec:eos}. Because the nuclear network solver implemented in \texttt{PHLEGETHON} does not generally preserve the normalization condition $\sum_l X_l=1$, we employ the same renormalization procedure as that used in the consistent multi-fluid advection method of \cite{plewa1999}. Specifically, after the nuclear network solve the mass fractions are rescaled according to
\begin{equation}
X_l \rightarrow \frac{X_l}{\sum_m X_m}.
\end{equation}

The nuclear reaction network solver implemented in \texttt{PHLEGETHON} does not currently apply an explicit positivity correction to the mass fractions. The verification tests presented in Sect.~\ref{sec:verification} that employ a nuclear reaction network indicate that, for the time steps encountered in the simulations, any negative mass fractions produced by the nuclear-network solve remain at levels comparable to the nonlinear-solver tolerance, which is $10^{-13}$ by default in \texttt{PHLEGETHON}. The nonlinear implicit solvers used by the nuclear network do not, however, enforce a strict positivity constraint, and larger negative abundances could in principle arise for sufficiently large time steps or in reaction networks exhibiting particularly stiff transient behavior. No such behavior is encountered in the calculations presented in this study. If it were to occur, the time step would need to be reduced, for example by lowering the CFL factor, until the nonlinear solve remains within the physically admissible region.
\newline

\subsection{Summary of the time integration algorithm}

The governing equations presented in Sect.~\ref{sec:pdes} are integrated in time using an operator-split approach in which the hyperbolic MHD equations, thermal diffusion, nuclear reactions, gravity, and other source terms are advanced sequentially within each global time step. Depending on the problem and the numerical method employed, some of these operators can be combined within a single update of the Godunov scheme, or symmetric operator splitting. Therefore, the algorithm described below represents a typical implementation adopted in \texttt{PHLEGETHON}, in which diffusive and reactive source terms are coupled to the MHD system of equations using second-order Strang splitting. 

The update from time $t^n$ to $\,{t^{n+1}=t^n+\Delta t}\,$ on the set of conserved quantities $\mathbf{U}$ proceeds through a sequence of intermediate stages $\mathbf{U}^{(m)}$ as follows:

\begin{enumerate}[label=(\roman*)]

\item At time $t^n$, the global time step $\Delta t$ is evaluated based on the hyperbolic wave speeds as in Eq.~(\ref{eq:cfl}).

\item Poisson's equation is solved at the mass density state $\rho^{(n)}$, yielding the gravitational potential $\phi^{(n)}$ (see Sect.~\ref{sec:poisson}).

\item The nuclear reaction network is advanced over a half time step, $\Delta t/2$, updating mass fractions and specific internal energy, $\mathbf{U}^{(n)}\rightarrow\mathbf{U}^{(1)}$. During this substep, mass density, velocity, and magnetic field are held fixed (Sect.~\ref{sec:nuclear-network}). 

\item Thermal diffusion is advanced over a half time step, $\Delta t/2$, updating the specific internal energy, $\mathbf{U}^{(1)}\rightarrow\mathbf{U}^{(2)}$. During this substep, mass density, velocity, magnetic field, and mass fractions are held fixed (see Sect.~\ref{sec:sts}).

\item The ideal MHD system of equations, including source terms such as gravity or rotation, is advanced over the full time step $\Delta t$, updating all conserved quantities, $\mathbf{U}^{(2)}\rightarrow\mathbf{U}^{(3)}$ (see Sects.~\ref{sec:reconstruction}--\ref{sec:time-disc}).

\item Thermal diffusion is advanced over the remaining half time step, $\Delta t/2$, updating the specific internal energy, $\mathbf{U}^{(3)}\rightarrow\mathbf{U}^{(4)}$. 

\item The nuclear reaction network is advanced over the remaining half time step, $\Delta t/2$, updating mass fractions and specific internal energy, $\mathbf{U}^{(4)}\rightarrow\mathbf{U}^{(5)}$. 

\item The global time is updated, $t^{n+1} = t^n + \Delta t$, as well as the set of conserved variables, $\mathbf{U}^{(n+1)}=\mathbf{U}^{(5)}$, completing the integration cycle.

\end{enumerate}
The cycle is repeated until a stopping criterion is met. In \texttt{PHLEGETHON}, the stopping criterion may be a maximum simulation time, a maximum number of time steps, or a maximum wall-clock time. During the time integration, the code outputs grid snapshots and restart files at user-specified intervals, using the HDF5 library \citep{hdf5} for parallel I/O. The flowchart of a typical time integration loop is shown in Fig.~\ref{fig:flow-chart}.

\begin{figure*}
\centering
\includegraphics[width=0.9\textwidth]{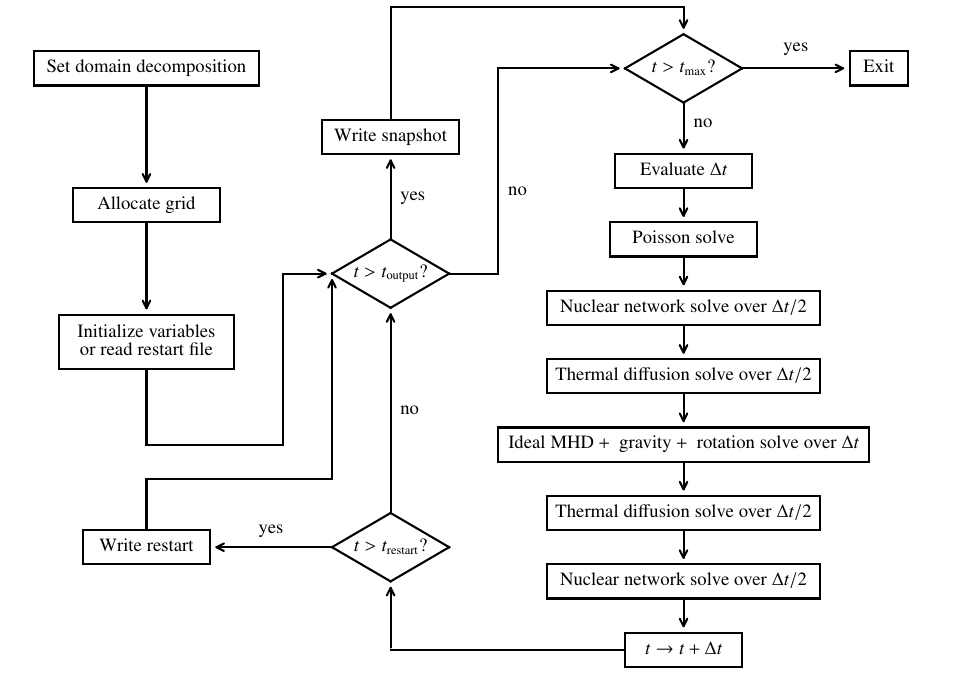}
\caption{Flowchart illustrating a typical full simulation execution loop in \texttt{PHLEGETHON}.}
\label{fig:flow-chart}
\end{figure*}

\subsection{Code scaling and relative cost of the numerical algorithms}\label{sec:perf}

Figure~\ref{fig:scaling} shows the weak scaling efficiency of \texttt{PHLEGETHON}, measured on the Tier-2 HoreKa HPC system\footnote{\url{https://www.scc.kit.edu/en/services/horeka.php}}, using  Intel Xeon `Ice Lake' 2.4~GHz processors. To avoid load imbalances, we consider a three-dimensional (3D) Cartesian grid with spatially homogeneous thermodynamic variables. Specifically, we adopt a temperature of $\,{T = 2.06 \times 10^9\ \mathrm{K}}\,$ and a density of $\,{\rho = 4.83 \times 10^5\ \mathrm{g\,cm^{-3}}}$, together with a 12-species oxygen-burning network. The chosen temperature and density correspond to conditions at the base of the convective shell in the core-collapse supernova progenitor model described in Sect.~\ref{sec:ccsnp}. The same 12-species oxygen-burning network is used there. All magnetic, velocity, and gravitational field components are set to zero but are retained in the computation.
The nuclear network is coupled to the MHD equations via Godunov splitting, and the nuclear network step is solved with the time-implicit backward-Euler method. We include thermal neutrinos and use the Helmholtz EoS. The base Godunov algorithm employs the PPH reconstruction scheme with two ghost cells (including the reconstruction of $\gamma_e$ and $\gamma_c$), the LHLLD Riemann solver, and the three-stage SSP-RK3 time stepper. To ensure that all tests perform the same number of operations per Runge--Kutta stage, the time step is fixed at $\,{\Delta t = 0.01}\,$ s. The setup is evolved for 100 time steps.

Although this setup does not correspond to a physically realistic scenario, it engages all algorithmic components. It performs the full set of operations typical of a production-quality simulation and is designed to stress-test the code’s scaling. In fact, the computation involves communicating a total of 20 collocated quantities and six staggered (non-collocated) quantities\footnote{Cell-centered quantities include 12 species, $\rho$, $\,{\mathbf{u}=(u_{x_1},u_{x_2},u_{x_3})}$, $P$, $T$, $\gamma_e$, and $\gamma_c$, while face-centered quantities include two components of the electromotive force per cell face.} three times per time step. For these tests, we use a 64-core simulation as a reference, nearly filling a HoreKa node (76 cores), so that the per-core memory bandwidth is close to that of production-quality simulations at scale. \texttt{PHLEGETHON} demonstrates good weak scaling: on HoreKa, increasing the number of cores by a factor of 512, from 64 to 32768, reduces the parallel efficiency by $\,{\approx 24\%}\,$ for $32^3$ local grid sizes and by only $\,{\approx 15\%}\,$ for $64^3$ local grid sizes. 

For reference, Table~\ref{tab:costs} summarizes the typical computational cost of the main algorithmic components relative to a single Godunov MHD update obtained on a 3.8 GHz AMD EPYC 9554 64-Core Processor. Thermal diffusion is included, solved both explicitly and using super-time-stepping. The value of the opacity is varied so that the super time-stepper performs between 3 and 43 substeps. Several equations of state are also considered, from ideal gas law to the Helmholtz EoS. A Poisson solver is coupled to the MHD system of equations via Godunov splitting, and both the gravitational potential and the gravity vector are reset to zero before calling the Godunov algorithm. Overall, the computational cost of the numerical methods implemented in \texttt{PHLEGETHON} is in most cases smaller than, or at most comparable to, the cost of the core Godunov algorithm. Not surprisingly, the most expensive components are the super time-stepper and the nuclear reaction network solver, the latter being a factor of 2.6 more expensive than the Godunov algorithm itself. Moreover, the reconstruction of the auxiliary thermodynamic indices $\gamma_e$ and $\gamma_c$ (see Sect.~\ref{sec:reconstruction}) is essential for reducing the computational cost of the simulation. In setups that are partially electron-degenerate, this approach can lead to a speed-up of up to a factor of six. 

\begin{figure}
\includegraphics[width=\linewidth]{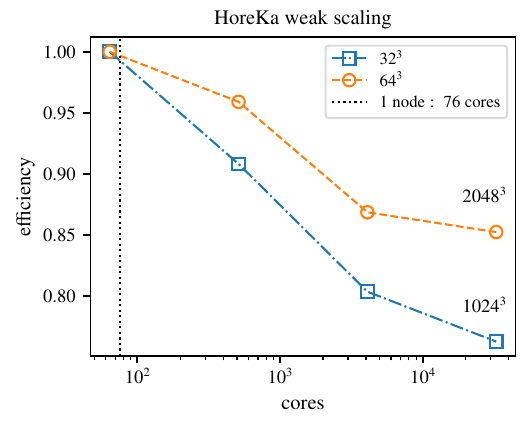}
\caption{Weak scaling of \texttt{PHLEGETHON} for the MHD setup described in Sect.~\ref{sec:perf}. Local grids of $32^3$ and $64^3$ cells are considered. The panel shows the efficiency measured on the Tier-2 HoreKa HPC system, which uses Intel Xeon `Ice Lake' 2.4~GHz processors with 38 cores per socket and two sockets per node. The maximum global grid size reached in each scaling is indicated. To ensure that the per-core memory bandwidth is representative of typical usage in production-quality simulations at scale, the reference run for the weak scaling on HoreKa is taken at 64 cores, nearly filling a node.  For these performance measurements, \texttt{PHLEGETHON} is compiled with \texttt{gcc 13.3.0} and \texttt{Open MPI 5.0} through the \texttt{mpifort} MPI compiler wrapper, using the \texttt{-O3}, \texttt{-mtune=native}, \texttt{-march=native}, \texttt{-ftree-vectorize}, \texttt{-funroll-loops}, and \texttt{-fforce-addr} optimization flags.}
\label{fig:scaling}
\end{figure}

\begin{table}[ht]
\centering
\caption{Typical computational cost of key algorithmic components in \texttt{PHLEGETHON} on a 3.8 GHz AMD EPYC 9554 64-Core Processor. The cost is given relative to the cost of a single Godunov MHD update. Values are averaged over three runs of the setup described in Sect.~\ref{sec:perf}. Unless specified, all tests are run on a $32^3$ Cartesian grid on a single processor. 1 $\sigma$ errors in the averages are smaller than $1\%$.}
\label{tab:costs}
\begin{tabular}{p{0.5\columnwidth} c}
\hline
Component & Relative cost per time step$^{*}$ \\
\hline
Godunov MHD, 12 species$^{1}$  & 1 \\
Contact-CT & 0.37 \\
Well balancing & 0.05 \\
EoS (ideal gas + thermal radiation) & 0.35 \\
EoS (ideal gas + thermal radiation, \\
 $\gamma_{e,c}$ reconstruction) & 0.08 \\
EoS (Helmholtz) & 8.9 \\
EoS (Helmholtz, $\gamma_{e,c}$ reconstruction) & 1.4 \\
Explicit thermal diffusion & 0.05 \\
STS$^{2}$ (ideal gas + thermal radiation, \\
 $\gamma_{e,c}$ reconstruction) & 0.18--0.72--2.2 \\
Poisson solver$^{3}$ & 0.8 \\
Nuclear network$^{4}$  & 2.6 \\
Thermal neutrino losses & 0.39 \\
\hline
\end{tabular}

\vspace{4pt}

\footnotesize
$^{1}$ The wall-clock time required to perform a Godunov update for one cell over a single time step, excluding EoS calls, is 1.03 $\mu$s. \\
$^{2}$ For 3, 14, and 43 STS RKL2 substeps, respectively.\\
$^{3}$ Coupled to the MHD equations via Godunov splitting.\\
$^{4}$ Coupled to the MHD equations via Godunov splitting and solved using the time-implicit backward Euler method. \\
${^*}$ For these performance measurements, \texttt{PHLEGETHON} is compiled with \texttt{gcc 13.3.0} and \texttt{Open MPI 4.1.6} through the \texttt{mpifort} MPI compiler wrapper, using the \texttt{-O3}, \texttt{-mtune=native}, \texttt{-march=native}, \texttt{-ftree-vectorize}, \texttt{-funroll-loops}, and \texttt{-fforce-addr} optimization flags.
\end{table}

\end{section}

\begin{section}{Code verification}\label{sec:verification}

In this section, we present a set of verification tests to assess the stability and accuracy of the numerical methods described in Sect.~\ref{sec:methods} on 1D, 2D, and 3D grids. The benchmarks are chosen to probe different aspects of \texttt{PHLEGETHON}, including its shock-capturing capability, its asymptotic-preserving behavior at low Mach numbers, and its ability to handle stiff source terms such as fast thermal diffusion and nuclear energy generation.  

The tests include standard problems such as the Brio--Wu shock tube and the Balsara vortex, which verify wave propagation and evaluate the MHD solver's performance across both high- and low-Mach-number regimes. A 2D MHD blast-wave problem on curvilinear grids further tests the MHD implementation across different grid geometries and verifies the preservation of the divergence-free condition of the magnetic field. Fast thermal diffusion of a 2D temperature pulse is used to verify the implementation of the super time-stepper described in Sect.~\ref{sec:sts}. The accuracy of our Poisson gravity solver is quantified using a problem with a known analytical solution. The 3D Evrard collapse test is used to assess energy conservation in the presence of a time-dependent gravitational potential. A one-zone hot-CNO cycle model tests our nuclear network solver, validated against the established \texttt{PYNUCASTRO} code \citep{pynucastro2}. Simulations of a non-magnetic burning-radiating buoyant bubble verify the code's ability to evolve slow flows in a strongly stratified medium, as well as the integration of nuclear source terms and their coupling to hydrodynamics and diffusive heat transport. The code-comparison setup of \cite{andrassy2022}, which includes heat-driven turbulent convection, CBM, and wave excitation on the same grid, provides a realistic cross-code benchmark for stellar hydrodynamics, confirming that our results are consistent with established codes such as \texttt{SLH}, \texttt{PROMPI}, \texttt{FLASH}, \texttt{PPMSTAR}, and \texttt{MUSIC}.

These tests provide a comprehensive view of \texttt{PHLEGETHON}’s capabilities across physical regimes relevant to stellar MHD. A full application  is presented in Sect.~\ref{sec:ccsnp}, for which we ran an MHD simulation of a core-collapse supernova progenitor star, thus demonstrating the code’s reliability in a production-quality astrophysical scenario. All tests were performed using version \texttt{v2026.4.1} of the code \citep{leidi_2026_19554676}.

\subsection{Brio--Wu shock tube}\label{sec:bw}

To assess the shock-capturing capabilities of \texttt{PHLEGETHON}, we consider the Brio--Wu shock-tube test problem \citep{brio1988}. This is a 1D Riemann problem for ideal MHD that gives rise to a rich family of waves, including a compound structure in which the left-going slow-magnetosonic and Alfv\'en waves propagate together, posing a challenge for many MHD schemes.

The computational domain spans $x \in (0,1)$, and the primitive variables exhibit a discontinuity at $x=0.5$. The left and right states at the discontinuity are given by
\begin{align}
    (\rho,u_x,u_y,B_y,P)_\mathrm{L} & = (1,0,0,1,1), \\
    (\rho,u_x,u_y,B_y,P)_\mathrm{R} & = (0.125,0,0,-1,0.1),
\end{align}
with $B_x=0.75$. The equation of state is that of an ideal gas with adiabatic index $\gamma=2$. At both boundaries of the domain, outflow boundary conditions are imposed.

Higher-order limited reconstruction methods, such as our limited fifth-order scheme (see Sect.~\ref{sec:reconstruction}), are known to suffer from spurious oscillations near discontinuities \citep[e.g.,][]{stone2008}. In this test, we therefore explore the robustness of our limited fifth-order reconstruction method when combined with the shock-flattener procedure described in Sect.~\ref{sec:shock-flat}. For all simulations presented in this section, we set the hyperbolic CFL number to 0.4 to improve the stability of the numerical schemes in capturing MHD shocks. The shock-flattening parameter $\epsilon_\mathrm{SF}$  (see Sect.~\ref{sec:shock-flat}) is set to 0.1 in all tests.

As commonly done for this test problem in the literature, we examine the solution at $\,{t=0.08}\,$ (in code units) using a grid with 500 cells in the $x$-direction. For comparison, we also show results obtained with van Leer reconstruction on the same grid, used without shock flattening. This second-order reconstruction method is generally more robust and more accurate than higher-order reconstruction methods in the presence of large nonlinear discontinuities \citep{greenough2004}. Additionally, we compute a reference solution using van Leer reconstruction on a grid with 4000 cells. The results are shown in Fig.~\ref{fig:bw}. The solution consists of a sequence of MHD waves, including a left-going fast rarefaction, a compound slow--Alfv\'en wave, a contact discontinuity, a slow shock, and a right-going fast rarefaction.

Using the fifth-order method to reconstruct pairs of Riemann states at cell faces leads to large-amplitude spurious oscillations in the numerical solution, as expected. These artifacts are most clearly visible in the region between the right-going fast rarefaction and the slow shock, as well as across the contact discontinuity. Despite the presence of oscillations, the positions of the discontinuities agree closely with those in the reference solution.

When the higher-order method is combined with the shock-flattening procedure, the amplitude of the spurious oscillations is substantially reduced, yielding results comparable to those obtained with van Leer reconstruction. At the same time, the jumps across the MHD waves, including the compound wave, remain well resolved and occur at the correct locations, although the use of the minmod slope limiter in the shock-flattening procedure broadens the discontinuities by approximately 3 to 7 additional cells compared with the other tested methods\footnote{We find that using van Leer reconstruction as a substitute for the limited fifth-order method in the shock-flattening procedure does not sufficiently reduce the amplitude of the spurious waves.}. These results demonstrate that the shock-flattening procedure improves the robustness of the fifth-order reconstruction scheme while still allowing discontinuities and rarefaction waves to be captured with good accuracy. We also find that varying the value of the cut-off Mach number $\mathcal{M}_\mathrm{cut}$ in the LHLLD solver (see Sect.~\ref{sec:riemann}) over the range ${0.1 < \mathcal{M}_\mathrm{cut} < 0.9}$ does not generate appreciable differences in the numerical solution for this test problem. \\

\begin{figure*}
\centering
\includegraphics[width=\textwidth]{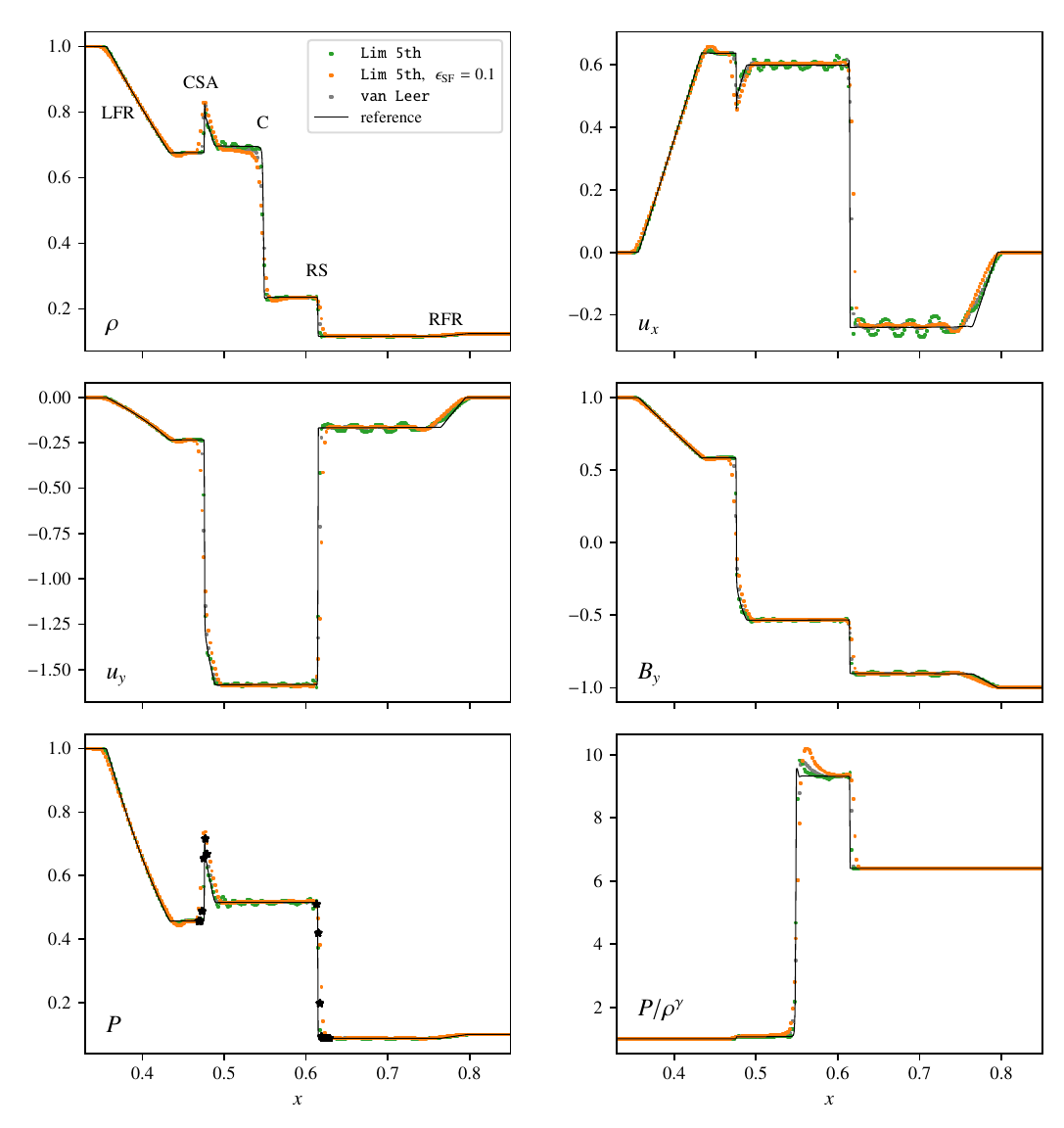}
\caption{Brio--Wu shock-tube test computed on a grid with 500 cells. The panels show the solution at ${t=0.08}$ (in code units) for the mass density, $x$- and $y$-velocity components, transverse magnetic field, pressure, and pseudo-entropy $P/\rho^\gamma$. Results obtained with the limited fifth-order reconstruction method (green), the same method combined with the shock-flattening procedure (orange), and second-order van Leer limited reconstruction (gray) are compared. The black solid lines represent the reference solution computed using van Leer reconstruction on a grid with 4000 cells. The solution consists of a left-going fast rarefaction (LFR), a compound slow--Alfv\'en wave (CSA), a contact discontinuity (C), a right-going slow shock (RS), and a right-going fast rarefaction (RFR). The black-starred points indicate the locations where the shock-flattening procedure was activated during the last time step.}
\label{fig:bw}
\end{figure*}

\subsection{Balsara vortex}

We consider here the 2D MHD vortex of \cite{balsara2004}, a smooth, stationary solution of the homogeneous MHD equations in which centrifugal, Lorentz, and pressure-gradient forces are in balance. The grid is Cartesian, covering the spatial domain $\,{(x,y) \in (-5,5)\times(-5,5)}$, with periodic boundary conditions on all sides. For this test, we employ the PPH reconstruction scheme. The EoS is that of a classical ideal gas. The primitive states are mapped onto the Cartesian grid as
\begin{equation}
\begin{split}
    &(u_x,u_y)  =   \tilde{u} e^{\frac{1-r^2}{2}}(-y,x), \\
    &(B_x,B_y)  =   \tilde{B} e^{\frac{1-r^2}{2}}(-y,x), \\
    &P          =  1 + \left[ \frac{\tilde{B}^2}{2}(1-r^2) - \frac{\tilde{u}^2}{2}\right] e^{1-r^2}, \\
    &\rho       = 1,
\end{split}
\end{equation}
where $r = \sqrt{x^2+y^2}$, and $\tilde{u}$ and $\tilde{B}$ are the maximum rotational speed and largest Alfv\'en velocity on the grid, respectively. Following \cite{leidi2022}, the vortex is advected along the diagonal of the grid with an advective speed $u_\mathrm{adv} = \tilde{u}$,
\begin{align}
        u_x & \rightarrow u_x + \frac{u_\mathrm{adv}}{\sqrt{2}}, \\
        u_y & \rightarrow u_y + \frac{u_\mathrm{adv}}{\sqrt{2}}.
\end{align}
The advection is followed for one grid-crossing time, ${\tau = 10\sqrt{2}/\tilde{u}}$, during which the vortex rotates approximately 2.25 times around its barycenter. 

To test the low-Mach capabilities of \texttt{PHLEGETHON}, we run a grid of models spanning $\tilde{u} \in \{10^{-4}, 10^{-3}, 10^{-2}, 10^{-1}\}$ and $\beta_\mathrm{K} \in \{10^{-2}, 10^{-1}, 1, 10^{1}, 10^2\}$, where 
\begin{equation}
\beta_\mathrm{K} = \left(\frac{\tilde{B}}{\tilde{u}}\right)^2
\end{equation}
is the initial magnetic-to-kinetic energy ratio. Therefore, this setup explores Mach numbers from $\,{\sim 10^{-4}}\,$ to $\,{\sim 0.1}\,$ (the sound speed is $\,{\approx 1}\,$) in both weakly and strongly magnetized regimes.  
Figure~\ref{fig:balsara-2d} shows the volumetric magnetic energy, $|\mathbf{B}|^2/2$, normalized by $\tilde{B}^2/2$, for the chosen parameter space on $64^2$ grids. Despite the wide range of Mach numbers, after one grid-crossing time both the fraction of magnetic energy converted into heat by numerical resistive processes and the shape of the vortex are essentially independent of $\tilde{u}$, reflecting the asymptotic-preserving property of the LHLLD scheme as $\mathcal{M} \to 0$ (see Sect.~\ref{sec:riemann}). The numerical dissipation, however, increases with the magnetic-to-kinetic energy ratio, ranging from $2\%$ for $\,{\beta_\mathrm{K} = 10^{-2}}\,$ up to $16\%$ for $\,{\beta_\mathrm{K} = 10^2}$. This is expected because the upwind terms in LHLLD scale linearly with the Alfv\'enic Mach number $\mathcal{M}_\mathrm{a}$ \citep[see, e.g.,][]{leidi2022},
\begin{equation}
    \mathcal{M}_\mathrm{a} = \frac{|\mathbf{u}|}{C_\mathrm{a}}, \qquad C_\mathrm{a} = \frac{|\mathbf{B}|}{\sqrt{\rho}},
\end{equation}
where $C_\mathrm{a}$ is the Alfv\'en velocity. Therefore, LHLLD is not strictly asymptotic-preserving in the limit $\,{\mathcal{M}_\mathrm{a} \rightarrow 0}$. Fully asymptotic-preserving schemes for subsonic and sub-Alfv\'enic flows exist \citep[e.g.,][]{boscheri2024}, but they have so far only been implemented with semi-implicit time discretizations. Notably, for $\,{\beta_\mathrm{K} = 10}$, where the medium is already significantly magnetized, only about $6\%$ of the total magnetic energy is dissipated after $\,{t = \tau}$. Moreover, the symmetry of the vortex is well preserved in all cases, suggesting that dispersion errors play a subdominant role even in strongly magnetized regimes.

To examine the scaling of global truncation errors with grid resolution, we repeat the Balsara vortex test for a case representative of stellar-interior MHD systems targeted by \texttt{PHLEGETHON}, i.e., $\tilde{u} = 10^{-2}$ and $\beta_\mathrm{K} = 1$\footnote{Such dynamical conditions are expected in subsonic stellar convection zones in the presence of a large-scale dynamo \citep[e.g.,][]{brun2017}.}, on grids ranging from $64^2$ to $1024^2$ cells, doubling the linear resolution $N$ each time. Figure~\ref{fig:balsara-l1} shows the resolution dependence of the $L_1$ error norm in the primitive variables at $t=\tau$ relative to their distribution at $t=0$,
\begin{equation}
   L_1(N) = \frac{1}{N^2} \sum_{i=1}^{N} \sum_{j=1}^{N} \big| q_{i,j} - q_{0,i,j} \big|,
\end{equation}
where $q$ denotes a primitive variable. The errors are normalized by $\tilde{u}$ for velocity and magnetic field, and by the ram pressure $\tilde{u}^2$ for mass density\footnote{In low-Mach-number flows, density fluctuations  scale as $\mathcal{M}^2$ \citep{meakin2007}.} and thermal pressure. As expected, all primitive variables exhibit second-order convergence on fine grids, with error magnitudes comparable across variables, within a factor of two.

\begin{figure*}
\centering
\includegraphics[width=0.88\textwidth]{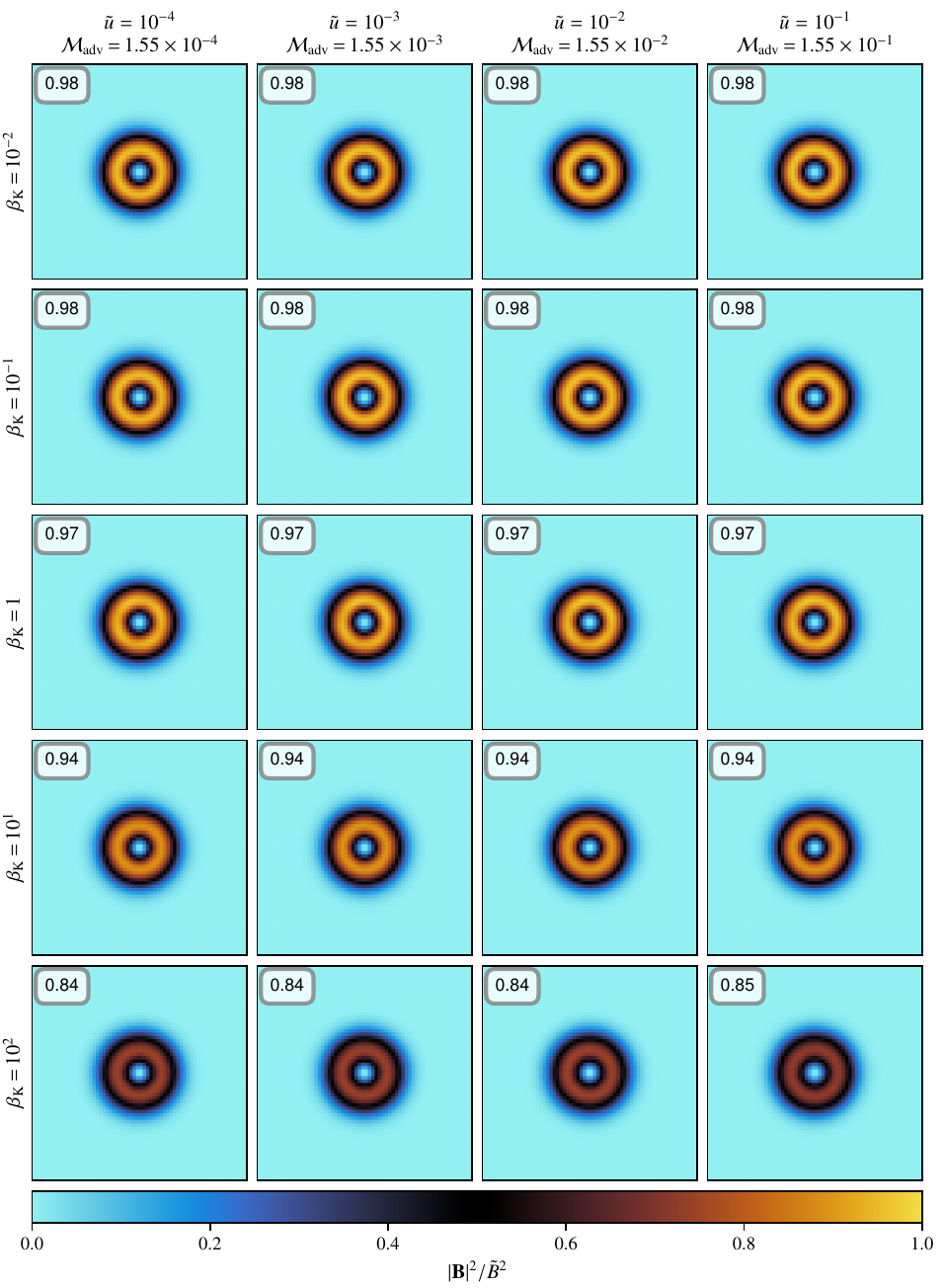}
\caption{Magnetic energy density normalized by its maximum value at $\,{t=0}\,$ for the 2D Balsara vortex test, shown for all combinations of the initial magnetic-to-kinetic energy ratio $\beta_\mathrm{K}$ and advective speed $\tilde{u}$. Here, the grid is $64^2$. The vortex is advected along the grid diagonal with advective speed $\,{u_\mathrm{adv}=\tilde{u}}$. The corresponding advective Mach number, $\mathcal{M}_\mathrm{adv}$, is also indicated. All snapshots are taken after one advective crossing time of the vortex. Numbers in the insets show the fraction of magnetic energy remaining relative to the initial value.}
\label{fig:balsara-2d}
\end{figure*}

\begin{figure}
\includegraphics[width=\linewidth]{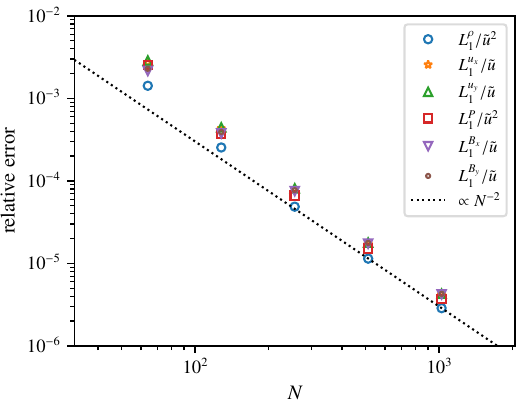}
\caption{$L_1$ global truncation error norms for the primitive variables as a function of linear grid resolution $N$ in the Balsara vortex test. Here, the advective speed and magnetic-to-kinetic energy ratio are $\,{\tilde{u} = 10^{-2}}\,$ and \mbox{$\beta_\mathrm{K} = 1$}, respectively. Errors in velocity and magnetic field are normalized by $\tilde{u}$, which is appropriate since $\,{\tilde{u} = \tilde{B}}\,$ in this case. Errors in mass density and thermal pressure are rescaled by the ram pressure $\tilde{u}^2$. The black  dotted line indicates second-order scaling.}
\label{fig:balsara-l1}
\end{figure}

\subsection{MHD blast on 2D curvilinear grids}
\begin{table*}[t]
\caption{Summary of the parameters used in the 2D MHD blast tests.} 
\label{table:mhd-blast} 
\centering 
\begin{tabular}{lllll} 
\toprule \\ 
Name of the test & Domain & $(x_0,y_0)$ & $(r_0,\psi_0)$ & Riemann Solver \\ 
\midrule 
Cartesian & $(x,y)\in(-1,1)\times(-1,1)$ & $(0,0)$ & - & LHLLD\\ 
$r_0=10$ & $(r,\psi)\in(10,12)\times(\frac{\pi}{4}-\frac{1}{11},\frac{\pi}{4}+\frac{1}{11})$ & $(\frac{11}{\sqrt{2}},\frac{11}{\sqrt{2}})$ & $(10,\frac{\pi}{4})$ & LHLLD\\ 
$r_0=3$ & $(r,\psi)\in(3,5)\times(\frac{\pi}{4}-\frac{1}{4},\frac{\pi}{4}+\frac{1}{4})$ & $(2\sqrt{2},2\sqrt{2})$ & $(3,\frac{\pi}{4})$ & LHLLD\\ 
$r_0=1$ & $(r,\psi)\in(1,3)\times(\frac{\pi}{4}-\frac{1}{2},\frac{\pi}{4}+\frac{1}{2})$ & $(\sqrt{2},\sqrt{2})$ & $(1,\frac{\pi}{4})$ & LHLLD\\ 
$2\pi$ & $(r,\psi)\in(0.001,1)\times(0,2\pi)$ & $(0,0)$ & - & LHLLD\\ 
$2\pi~\mathrm{offset}$ & $(r,\psi)\in(0.001,1)\times(0,2\pi)$ & $(0.1,0.1)$ & - & LHLLD\\ 
$r_0=1,~\mathrm{Q~III}$ & $(r,\psi)\in(1,3)\times(\frac{5\pi}{4}-\frac{1}{2},\frac{5\pi}{4}+\frac{1}{2})$ & $(-\sqrt{2},-\sqrt{2})$ & $(1,\frac{5\pi}{4})$ & LHLLD\\ 
$2\pi~\mathrm{offset,~HLLD}$ & $(r,\psi)\in(0.001,1)\times(0,2\pi)$ & $(0.1,0.1)$ & - & HLLD\\ 
\bottomrule 
\end{tabular} 
\end{table*}

To test the MHD module in \texttt{PHLEGETHON} on a curvilinear grid, we perform 2D MHD blast simulations \citep{ZMC1994} on various settings of a polar grid.
The initial conditions of the blast problem are defined by
\begin{align}
    \rho&=1,\\
    v_x&=0,\\
    v_y&=0,\\
    \label{eq:mhd_blast_p}
    P&=0.1 + \frac{9}{20}\left(1 -\tanh\left[\frac{\sqrt{(x-x_0)^2+(y-y_0)^2}-0.1}{0.01}\right]\right),\\
    \mathbf{A}&=\frac{\hat{z}}{\sqrt{2}}(y-x),
\end{align}
where $(x_0,y_0)$ is the center of the over-pressurized circular region and $\mathbf{A}$ is the magnetic vector potential, from which the magnetic field components are computed as ${\mathbf{B}=\nabla\times\mathbf{A}}$.
We initialize the pressure with a smooth radial transition (Eq.~\eqref{eq:mhd_blast_p}) rather than a discontinuous step, since a sharp initial jump is discretized differently depending on the grid's local cell geometry, introducing a resolution-dependent initial condition error independent of the subsequent shock evolution. 
The finite-width transition mitigates this artifact and ensures a consistent initial representation across coordinate systems.
The adiabatic index $\gamma$ is $5/3$.
The test uses second-order, van Leer limited spatial reconstruction.
All the tests stop at ${t=0.18}$. 

The benchmark test is a $512^2$ uniform Cartesian grid with a spatial domain $(x,y)\in(-1,1)\times(-1,1)$ and $(x_0,y_0)=(0,0)$. 
The explosion is also performed on four $512^2$ uniform polar wedge grids \citep{Zhang_2019} and three $512^2$ uniform $2\pi$ full-circle grids \citep{ZHANG2019276}.
The domain of the polar wedge grid coordinates and the over-pressurized centers are defined by
\begin{align}
    r&\in(r_0, r_0+l),\\
    \psi&\in\left(\psi_0-\frac{1}{r_0+l/2},\psi_0+\frac{1}{r_0+l/2}\right),\\
    (x_0,y_0)&=\left[\left(r_0+\frac{l}{2}\right)\cos(\psi_0),\left(r_0+\frac{l}{2}\right)\sin(\psi_0)\right],
\end{align}
where $l$ is the wedge half-length, $r_0$ are the radial coordinates of the inner radial boundaries of the grids, and $\psi_0$ is the center angle for the wedge orientation. 
By decreasing $r_0$, we increase the curvature effect of the grids.
Outflow boundary conditions are applied at all sides for both the uniform Cartesian grids and the uniform polar wedge grids (see Table \ref{table:bcs}).
We present three tests in Quadrant I with ${l=2}$, and ${\psi_0=\pi/4}$, scanning ${r_0\in\{10,3,1\}}$, and one test in Quadrant III with ${(l,r_0,\psi_0)=(2,1,5\pi/4)}$ to demonstrate robustness across angular placements.
These tests are performed with LHLLD.

For the three uniform $2\pi$ full-circle grids, the spatial domains are ${(r,\psi)\in(0.001,1)\times(0,2\pi)}$, with outflow boundary conditions applied on both radial sides and periodic boundary conditions applied on the angular sides.
Purely-normal magnetic field boundary conditions are also applied at the inner radial sides (see Table \ref{table:bcs}).
In one of the tests, the over-pressurized circular region is centered at ${(x_0,y_0)=(0,0)}$ and one other test is centered at ${(x_0,y_0)=(0.1,0.1)}$.
Both the above tests are performed with LHLLD. 
The final test has its over-pressurized circular region centering at ${(x_0,y_0)=(0.1,0.1)}$, but with HLLD.
Summary of all eight tests can be found in Table~\ref{table:mhd-blast}

We present the results in Fig.~\ref{fig:mhd-blast}. 
The morphology of the explosion (shown by the density contours at ${t=0.18}$) for all the grid settings are consistent with each other qualitatively.
We also present the density profiles extracted along the diagonal lines (${y=x}$, illustrated by the black dashed line on Fig.~\ref{fig:mhd-blast}) of the blast.
To compare the numerical value across different grids, we first align the grids by their initial over-pressurized centers $(x_0,y_0)$, and then use a bilinear interpolator to interpolate the density distribution onto the Cartesian grid.
The results show good agreement between different grids.
The percentage differences are usually less than $5\%$ within the blast region.
The positions of the shock fronts and their thickness are consistent across grid configurations and treatment in the Riemann solvers.

To assess the preservation of the divergence-free condition of the magnetic field, we calculate the maximum and mean values of the normalized magnetic monopole error $(g_{\mathrm{m},i,j})$ as
\begin{align}\label{gmon}
    g_\mathrm{m,i,j}=\frac{| \nabla \cdot \mathbf{B} |_{i,j}}{ |\mathbf{B}|_{i,j}} (\Delta \Omega)^{1/2}_{i,j},
\end{align}
where $(\Delta \Omega)_{i,j}$ is the cell area.
The terms on the right-hand side of Eq.~(\ref{gmon}) are then computed, for the 2D Cartesian grid, as
\begin{equation}
\begin{split} 
|\nabla\cdot\mathbf{B}|_{i,j}=&\left|\frac{B_{x,i+1/2,j}-B_{x,i-1/2,j}}{x_{i+1/2,j}-x_{i-1/2,j}}+\right.\\
&\left.\frac{B_{y,i,j+1/2}-B_{y,i,j-1/2}}{y_{i,j+1/2}-y_{i,j-1/2}}\right|, \\
|\mathbf{B}|^2_{i,j}=&B^2_{x,i,j}+B^2_{y,i,j}, \\
(\Delta \Omega)_{i,j}=&(x_{i+1/2,j}-x_{i-1/2,j})(y_{i,j+1/2}-y_{i,j-1/2}),
\end{split}
\end{equation}
and for the 2D polar grids as
\begin{equation}
\begin{split} 
|\nabla\cdot\mathbf{B}|_{i,j}=&\frac{1}{r_{i,j}}\left|\frac{r_{i+1/2,j}B_{r,i+1/2,j}-r_{i-1/2,j}B_{r,i-1/2,j}}{r_{i+1/2,j}-r_{i-1/2,j}}\right.\\
&\left.+\frac{B_{\psi,i,j+1/2}-B_{\psi,i,j-1/2}}{\psi_{i,j+1/2}-\psi_{i,j-1/2}}\right|, \\
|\mathbf{B}|^2_{i,j} = &B^2_{r,i,j}+B^2_{\psi,i,j}, \\
(\Delta \Omega)_{i,j}= & r_{i,j}(r_{i+1/2,j}-r_{i-1/2,j})(\psi_{i,j+1/2}-\psi_{i,j-1/2}).
\end{split}
\end{equation}

The time evolution of the mean and maximum values of the normalized magnetic monopole error is shown in Fig.~\ref{fig:mhd-blast-divB}. The mean magnetic monopole error remains at ${\sim10^{-21}}$--${10^{-19}}$, consistent with errors at the level of double-precision round-off after accounting for the normalization and spatial averaging used in its definition of Eq.~(\ref{gmon}). The maximum value across all cells is ${\sim10^{-16}}$--${10^{-13}}$.
The three tests performed on the uniform $2\pi$ angular grids exhibit larger errors than those performed on the wedge-shaped grids, owing to their coarser angular resolution. Overall, these 2D MHD blast tests demonstrate the ability of the code to maintain a nearly divergence-free magnetic field across different grid geometries.
\begin{figure}
\includegraphics[width=\linewidth]{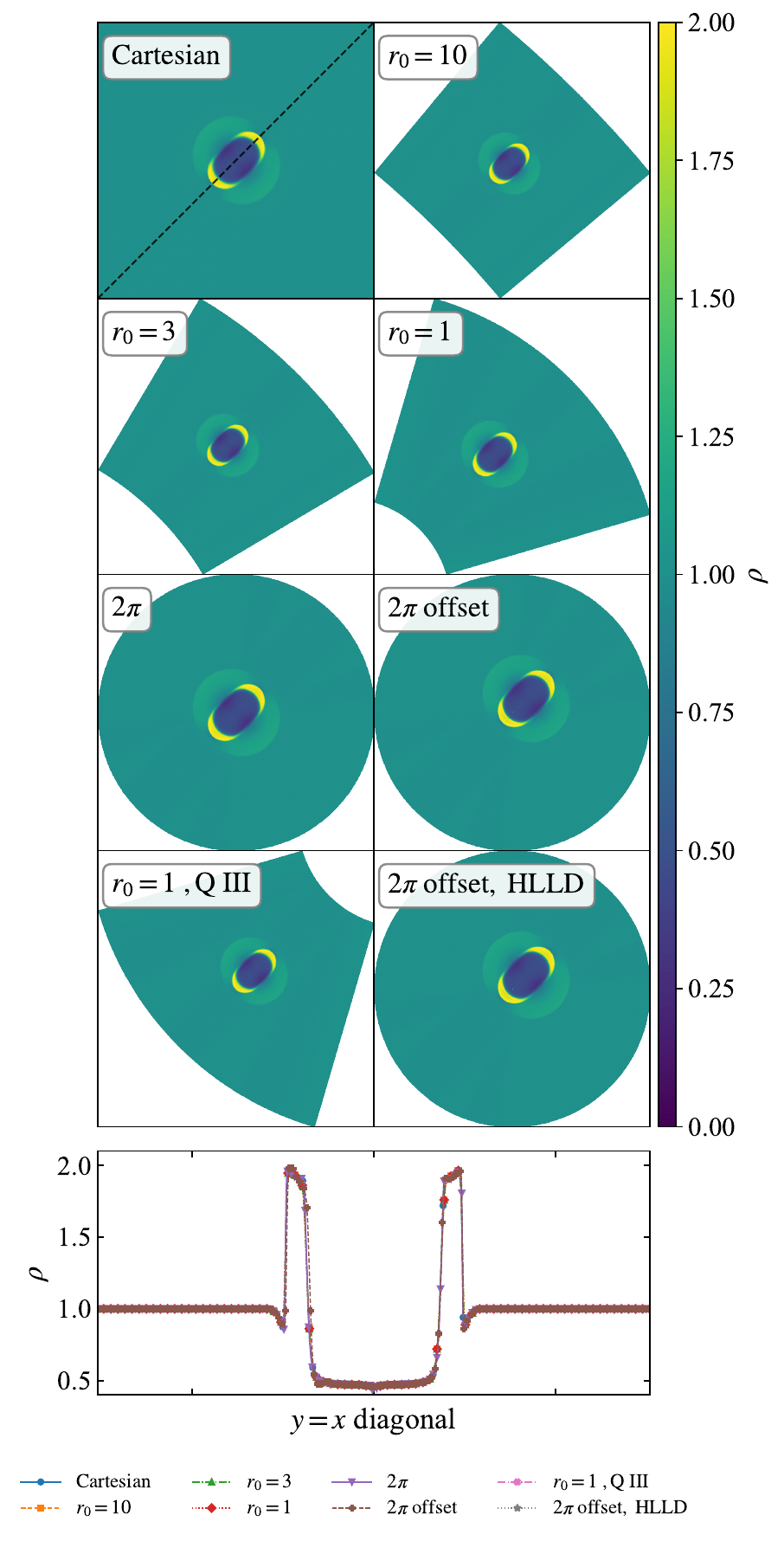}
\caption{{\it Top panels}: Density contours of the 2D MHD blast tests on $8$ different grids at ${t=0.18}$. Each grid is $512^2$. The black dashed diagonal line indicates the line extraction for the analysis. {\it Bottom panels}: Density line profiles extracted from the diagonal lines. The markers on the lines are indicating the numerical value for every 5 grid points.}
\label{fig:mhd-blast}
\end{figure}
\begin{figure}
\includegraphics[width=\linewidth]{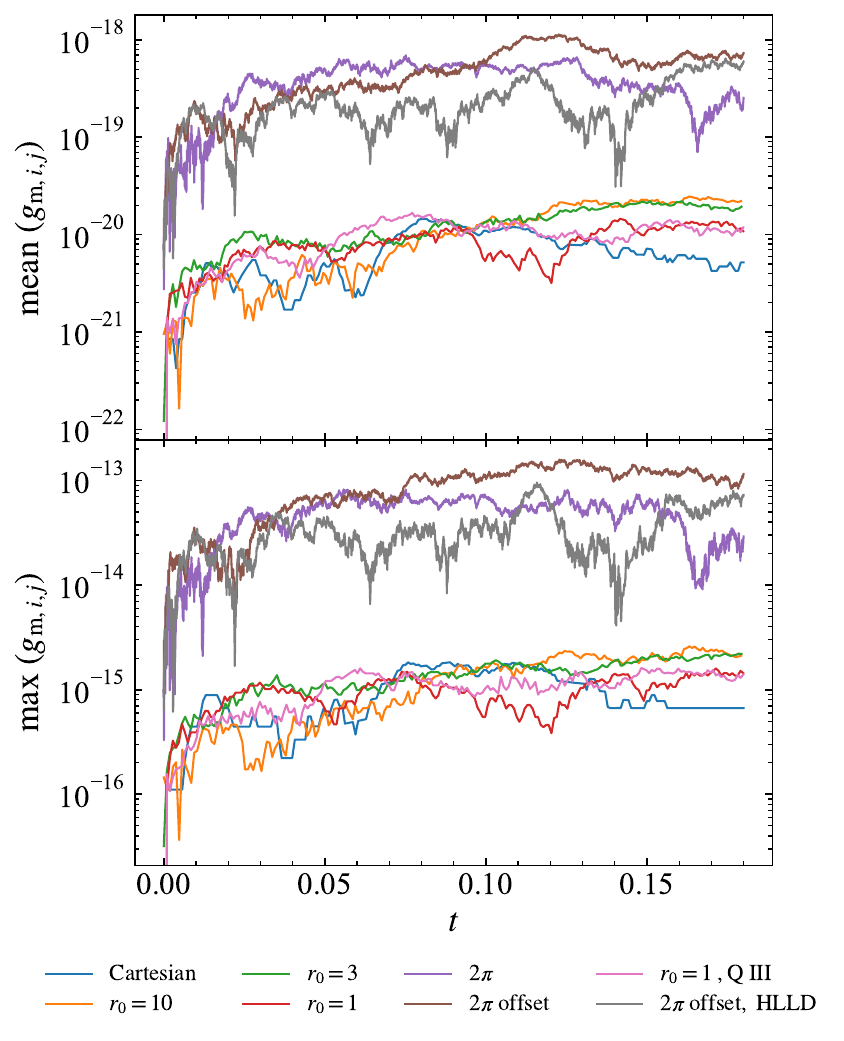}
\caption{Time evolution of the mean ({\it Top}) and maximum ({\it Bottom}) of the normalized magnetic monopole error for the 2D MHD blast tests.}
\label{fig:mhd-blast-divB}
\end{figure}
\subsection{Thermal diffusion of a 2D temperature pulse}\label{sec:diff-pulse}

To benchmark the super time-stepper solver for treating stiff thermal diffusion, described in Sect.~\ref{sec:sts}, we consider the 2D diffusion of a temperature pulse. The grid geometry is Cartesian, and periodic boundary conditions are applied on all sides. The spatial domain spans $\,{(x,y) \in (-5 R_\odot,5R_\odot) \times (-5 R_\odot,5R_\odot)}$, where $R_\odot$ is the solar radius. The density and opacity are homogeneous in space and take values ${\rho=1.0\ \mathrm{g\,cm^{-3}}}$ and $\,{\kappa=1\ \mathrm{cm^{2}\, g^{-1}}}$, respectively. The temperature is initialized as the sum of a homogeneous base state and a compact-support smooth perturbation,
\begin{equation}
T =
\begin{cases}
T_\mathrm{b} +\delta T\cos^2\left( \frac{\pi r}{2R_\odot} \right), & r \leq R_\odot, \\
T_\mathrm{b}, & r > R_\odot, 
\end{cases}
\end{equation}
where $\,{T_\mathrm{b}=10^7\ \mathrm{K}}\,$ is the base temperature, $\,{\delta T = T_\mathrm{b}/10}$, and $\,{r=\sqrt{x^2+y^2}}$. These thermodynamic conditions are chosen to be representative of those found in the interior of massive main sequence stars \citep{kippenhahn2012}. The equation of state is that of a classical ideal gas with mean molecular weight $\,{\mu=1}\,$ and includes thermal radiation (see Sect.~\ref{sec:eos}).

In this test, the Godunov update is disabled and only thermal diffusion is allowed to operate. The value of the super time step $\Delta t_\mathrm{STS}$ is fixed as a multiple of the parabolic CFL time step,
\begin{equation}
    \Delta t_\mathrm{STS} = \mathrm{CFL}_\mathrm{p} \frac{3}{4}\frac{(\Delta x)^2}{c},
\end{equation}
where $\Delta x$ is the grid spacing. The system is evolved until the initial pulse diffuses over a distance $\,{l=2 R_\odot}\,$
\begin{equation}
    \tau = \frac{3 l^2}{c}.
\end{equation}
This choice yields ${\tau = 64570\ \mathrm{yr}}$.

As a reference, Fig.~\ref{fig:diffusion-xprofs} shows the time evolution of the temperature profile taken along the $x$-direction in a test run with $\,{\mathrm{CFL}_\mathrm{p}=1}$. As expected, as time progresses, the amplitude of the temperature pulse decreases while its width increases.

To quantify the convergence properties of the RKL2 STS method implemented in \texttt{PHLEGETHON}, Fig.~\ref{fig:diffusion-l1} shows $L_1$ error norms as functions of the linear grid resolution $N$, ranging from $64^2$ to $2048^2$ with successive factors of two. The time step is varied such that ${\mathrm{CFL}_\mathrm{p} \in \left\{ 1,10,100,1000\right\}}$, corresponding to numbers of STS substeps $N_\mathrm{STS}$ of 4, 8, 22, and 65, respectively. The $L_1$  error norms are computed as
\begin{equation}
    L_1(N) = \frac{4}{\delta T N^2}\sum_{i=1}^{N/2}\sum_{j=1}^{N/2} |T_{N}^{i,j}-T_{N/2}^{i,j}|,
\end{equation}
where $T_{N}$ and $T_{N/2}$ are the temperature distributions on grids with linear resolution $N$ and $N/2$, respectively. The finer $T_{N}$ grid is restricted to the coarser grid by averaging blocks of $\,{2\times2}\,$ cells.

At a given resolution, the error increases monotonically with $\mathrm{CFL}_\mathrm{p}$. Remarkably, the errors of simulations run up to $\,{\mathrm{CFL}_\mathrm{p}=100}\,$ are comparable within $20\%$ even on a relatively coarse $128^2$ grid. The error dispersion between different cases diminishes as the grid is refined. All cases except $\,{\mathrm{CFL}_\mathrm{p}=1000}\,$ show second-order convergence on the finer grids. The highest CFL case exhibits an apparent third- to fourth-order scaling, but produces errors approximately two orders of magnitude larger than the other cases on coarse grids. On the finest grid, the error of the $\,{\mathrm{CFL}_\mathrm{p}=1000}\,$ run is about $40\%$ larger than that of the reference case with $\mathrm{CFL}_\mathrm{p}=1$.

In Fig.~\ref{fig:diffusion-comp}, we show the ratio of the error obtained from simulations evolving the temperature according to Eq.~(\ref{eq:diffusion_temp}) to the error produced by the more accurate standard mode that evolves the specific internal energy in Eq.~(\ref{eq:diffusion_eint}). For simulations with $\mathrm{CFL}_\mathrm{p}$ up to $10$, the errors remain comparable within a few percent. For the case with $\,{\mathrm{CFL}_\mathrm{p}=100}$, the error in the temperature-evolution mode increases to values up to $40\%$ larger than those of the standard mode but it still scales second-order, while for $\,{\mathrm{CFL}_\mathrm{p}=1000}\,$ the ratio of the two errors approaches one dex. This behavior arises because, on sufficiently fine grids, the magnitude of spatial discretization errors becomes small enough that time-discretization errors are dominant. In particular, the temperature-evolution formulation introduces a first-order contribution associated with holding the specific heat $c_v$ fixed during the update.  For $\,{\mathrm{CFL}_\mathrm{p}=1000}$, the time step is sufficiently large that this first-order contribution dominates the local-truncation error, causing the method to exhibit first-order scaling.

\begin{figure}
\includegraphics[width=\linewidth]{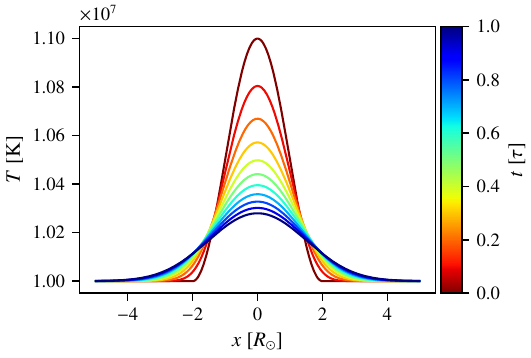}
\caption{Temporal evolution of the temperature profile along the $x$-direction in the 2D thermal diffusion test.}
\label{fig:diffusion-xprofs}
\end{figure}
\begin{figure}

\includegraphics[width=\linewidth]{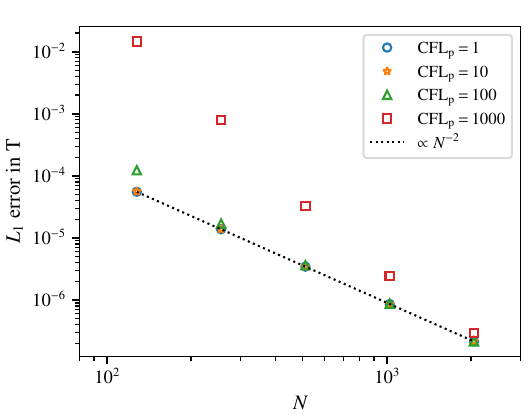}
\caption{$L_1$ global truncation error norms in temperature as functions of the linear grid resolution in the 2D thermal diffusion test. Different colors correspond to different values of the parabolic CFL number $\mathrm{CFL}_\mathrm{p}$, while the black dotted line indicates second-order scaling.}
\label{fig:diffusion-l1}
\end{figure}

\begin{figure}
\includegraphics[width=\linewidth]{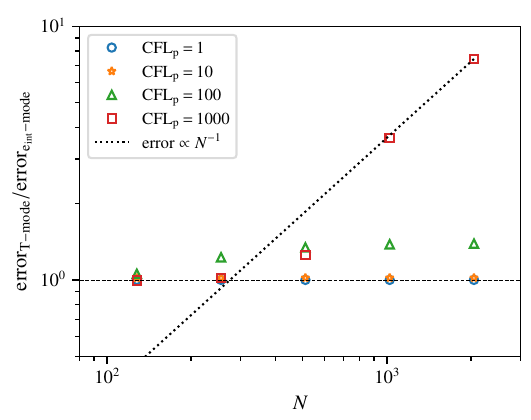}
\caption{Ratio of the $L_1$ error norms in temperature as a function of the linear grid resolution for the temperature-based and specific internal energy-based formulations of the thermal diffusion operator in the 2D thermal diffusion test. Different colors indicate different parabolic CFL numbers, $\mathrm{CFL}_\mathrm{p}$, while the black dotted line shows first-order scaling. The horizontal  black dashed line serves to guide the eye. At high resolution for ${\mathrm{CFL}_\mathrm{p}=1000}$, the apparent first-order scaling arises because the specific heat at constant volume is held fixed throughout the super-time-stepping solve.}
\label{fig:diffusion-comp}
\end{figure}

\subsection{Accuracy of the Poisson solver}\label{sec:collapse}

In this section, we investigate the accuracy of our gravitational Poisson solver using a 3D test problem run on a Cartesian grid. Here, the Godunov algorithm is disabled, and only the Poisson solver is used to compute the numerical solution for a spherical mass density distribution,
\begin{equation}
\rho =
\begin{cases}
\rho_0\left( 1 - \frac{r^2}{r_0^2} \right)^2, & r \leq r_0,\\
0, & r > r_0,
\end{cases}
\end{equation}
with $\,{\rho_0=1}\,$ and $\,{r_0=0.25}$. For this test, $\,{G=1}$. For this mass density distribution, Poisson's equation has an analytical solution \citep[see, e.g.,][]{mandal2023},
\begin{equation}
\phi(r) =
\begin{cases}\label{eq:coll1}
-\frac{2}{3}\pi \rho_0r_0^2 + 4 \pi\rho_0\left(\frac{r^2}{6}-\frac{1}{10}\frac{r^4}{r_0^2}+\frac{1}{42}\frac{r^6}{r_0^4} \right), & r \leq r_0,\\
-\frac{M}{r}, & r > r_0,
\end{cases}
\end{equation}
where the mass of the sphere is $M=32\pi\rho_0r_0^3/105$. The radial gravitational acceleration is
\begin{equation}
g_r(r) =
\begin{cases}\label{eq:coll2}
-4\pi\rho_0\left(\frac{r}{3}-\frac{2}{5}\frac{r^3}{r_0^2}+\frac{1}{7}\frac{r^5}{r_0^4} \right), & r \leq r_0,\\
-\frac{M}{r^2}, & r > r_0.
\end{cases}
\end{equation}

We run a series of tests where the linear resolution of the grid $N$ spans $16$ to $512$, doubling the resolution each time. The spatial domain spans $\,{(x,y,z) \in (-0.5,0.5)\times (-0.5,0.5) \times (-0.5,0.5)}$. The grid is non-uniform, with physical coordinates mapped from logical coordinates as, e.g.,
\begin{equation}\label{eq:nucart}
x_i = \frac{1}{4}\left(\eta_i + \eta_i^5\right),
\end{equation}
where
\begin{equation}
\eta_i = -1 + \Bigg(i-\frac{1}{2}\Bigg)\frac{2}{N}.
\end{equation}
Analogous expressions are used to compute the $y-$ and $z-$coordinates. The chosen mapping yields a linear resolution at the origin that is twice that of a uniform Cartesian grid with the same number of grid cells (see Fig.~\ref{fig:collapse-phi}).
Monopole boundary conditions are used for the gravitational potential, and the tolerance for the BiCGSTAB convergence criterion is set to $10^{-10}$. The initial gravitational potential guess used by the linear solver is set to 0 everywhere.

Since the mass density vanishes outside the sphere, the relative error normalized by density given in Eq.~(\ref{eq:poisson-eps}) is not meaningful. Therefore, during the BiCGSTAB solve, we monitor the residual as
\begin{equation}\label{eq:conv-poisson-evr}
||\mathbf{r}|| = \sqrt{
\dfrac{1}{N^3}
\sum_{i=1}^N
\sum_{j=1}^N
\sum_{k=1}^N
\left[
(L\boldsymbol{\phi})_{i,j,k} - 4 \pi \rho_{i,j,k}
\right]^2
}.
\end{equation}
After solving Poisson's equation, we compute the $L_1$ error norm for the gravitational potential using the numerical solution, $\phi_\mathrm{n}$, and the analytical solution, $\phi_\mathrm{a}$, provided in  Eq.~(\ref{eq:coll1}), as
\begin{equation}
L_1(N) = \frac{1}{N^3}\sum_{i=1}^N\sum_{j=1}^N\sum_{k=1}^N |\phi_{\mathrm{a},i,j,k} - \phi_{\mathrm{n},i,j,k}|.
\end{equation}
A similar procedure is applied to the radial component of gravity.

The error scaling is shown in Fig.~\ref{fig:collapse-l1}. Both the global truncation errors for the gravitational potential and radial gravitational acceleration exhibit second-order convergence from the coarsest to finest of the tested grids. Although the discretization of Poisson's equation used in \texttt{PHLEGETHON} is formally only first-order accurate, on smoothly varying grids, like the one employed in this test, the leading first-order error term becomes subdominant compared to the second-order error term for the grid resolutions considered.

Finally, in Fig.~\ref{fig:collapse-bicgstab}, we show the residual normalized by its initial value as a function of the wall-clock time per cell spent solving Poisson's equation during the BiCGSTAB iteration. For the grid stretching considered here, we find that the best-performing combination of multigrid parameters consists of one pre-smoothing iteration, three post-smoothing iterations, three smoothing iterations on the coarsest grid with $2^3$ cells, and ${\gamma_\mathrm{GS}=1.5}$. As a reference, we repeat the setup on uniform Cartesian grids, using one pre-smoothing iteration, one post-smoothing iteration, one smoothing iteration on the coarsest grid with ${\gamma_\mathrm{GS}=1}$, and a $2^3$ coarsest grid for all resolutions. For these reference runs, the prescribed tolerance of $10^{-10}$ is reached at an approximately resolution-independent rate, with the residual decreasing, on average, by roughly two orders of magnitude per BiCGSTAB iteration. This behavior is consistent with the optimality of multigrid methods, for which the number of iterations required to reduce the residual by a prescribed factor is, ideally, independent of the grid resolution \citep{leveque2007}. In practice, the tolerance is reached in approximately $0.8\ \mu\mathrm{s}$ per cell on a 3.8~GHz AMD EPYC 9554 64-Core Processor. On the nonuniform Cartesian grid, by comparison, the number of iterations required to reach the prescribed tolerance increases approximately logarithmically with resolution. This behavior can be attributed to the reduced effectiveness of the pointwise weighted Gauss--Seidel smoother for error components associated with anisotropic grid spacings. In contrast, the smoother is highly effective at damping error components associated with relatively isotropic modes. Since the geometric multigrid hierarchy is coarsened by a factor of two at each level, a standard V-cycle contains $\mathcal{O}(\log_2 N)$ levels. For the stretched grids considered here, the grid anisotropy persists throughout the multigrid hierarchy, and the pointwise smoother therefore remains less effective at damping the corresponding error components on each level. As the resolution increases, the hierarchy contains a correspondingly larger number of levels over which this reduced smoothing effectiveness is encountered. Consequently, the convergence factor gradually deteriorates, leading to the approximately logarithmic increase in the number of iterations observed here. The convergence rate of the gravity solver is therefore no longer strictly resolution-independent on the stretched grids. Nevertheless, the absolute performance of the gravity solver remains good: the cost is below $3\ \mu\mathrm{s}$ per cell for the highest-resolution grid considered ($512^3$), which is still comparable to the cost of the Godunov algorithm (see Table~\ref{tab:costs}). Moreover, the present benchmark represents a conservative estimate of the cost of the Poisson solve in typical production simulations. In the latter, the gravitational potential from the previous time step provides an initial guess for the iterative solver, which is generally much closer to the converged solution than the initial guess used in this test. Consequently, fewer BiCGSTAB iterations are expected to be required in production simulations.

\begin{figure}
\includegraphics[width=\linewidth]{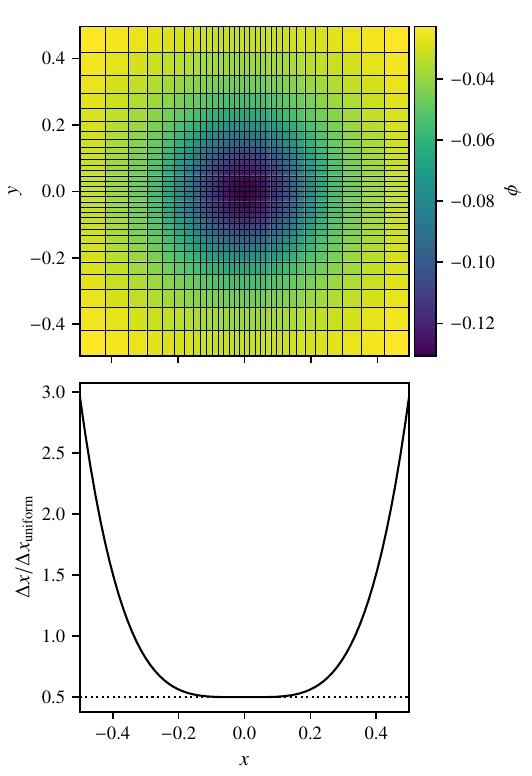}
\caption{{\it Top}: slice in the $\,{z=0}\,$ plane of the numerical solution for the gravitational potential $\phi$ from the Poisson solver accuracy test on a $32^3$ grid. Solid lines delineate the grid cell boundaries. {\it Bottom}: ratio of the grid cell size in the $x$-direction on the non-uniform grid used in this test problem to that of a uniform grid with the same number of cells.}
\label{fig:collapse-phi}
\end{figure}

\begin{figure}
\includegraphics[width=\linewidth]{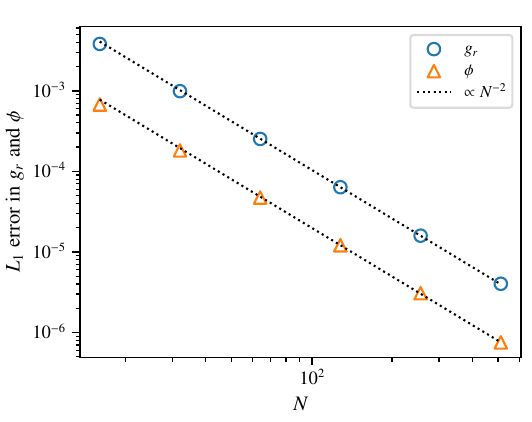}
\caption{$L_1$ global truncation error norm as a function of the linear grid resolution for the radial gravitational acceleration (circles) and  gravitational potential (triangles) in the Poisson solver accuracy test. The black dotted  lines indicate the second-order scaling.}
\label{fig:collapse-l1}
\end{figure}

\begin{figure}
\includegraphics[width=\linewidth]{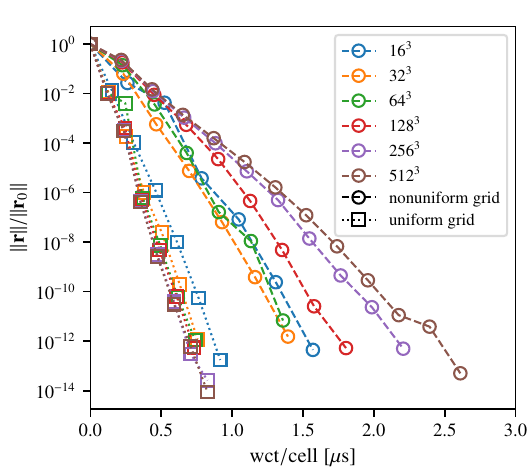}
\caption{Normalized BiCGSTAB residual, $||\mathbf{r}||/||\mathbf{r}_0||$, as a function of the wall-clock time required to update a single cell for the Poisson solver accuracy test. Different colors indicate the grid resolutions shown above, ranging from $16^3$ to $512^3$. Results from the same test performed on a uniform Cartesian grid are shown with square markers. All tests were performed using a single core of an AMD EPYC 9554 64-core processor clocked at 3.8 GHz.}
\label{fig:collapse-bicgstab}
\end{figure}

\subsection{Evrard collapse}

To test the energy conservation properties of \texttt{PHLEGETHON} in simulations with a time-dependent gravitational potential, we consider the purely hydrodynamical, 3D Evrard collapse problem \citep{evrard1988}, which involves the gravitational collapse of an initially cold gas cloud. The initial conditions closely follow those of \cite{mandal2023}. The spatial domain is ${(x,y,z)\in(-2,2)\times(-2,2)\times(-2,2)}$. The grid geometry is analogous to the nonuniform Cartesian grid used in Sect.~\ref{sec:collapse}, with the physical coordinates mapped from logical space as 
\begin{equation}\label{eq:nucart}
x_i = \eta_i + \eta_i^5,
\end{equation}
where
\begin{equation}
\eta_i = -1 + \Bigg(i-\frac{1}{2}\Bigg)\frac{2}{N}.
\end{equation}
Analogous expressions are used to compute the $y-$ and $z-$coordinates. Monopole boundary conditions are used for the gravitational potential. A convergence tolerance of $10^{-10}$ is used for the BiCGSTAB solver. At each iteration, the residual is computed according to Eq.~(\ref{eq:conv-poisson-evr}). The numerical solution of Poisson's equation is coupled to the gas-dynamics equations using Godunov splitting. We use the van Leer slope-limited reconstruction scheme and a CFL factor of 0.4 to improve the robustness of the method in the presence of strong shocks.

The density distribution at $t=0$ is given by 
\begin{equation}
\rho(r) =
\begin{cases}\label{eq:evr1}
\frac{1}{2 \pi r}, & r \leq 1,\\
10^{-4}, & r > 1,
\end{cases}
\end{equation}
where ${r = \sqrt{x^2+y^2+z^2}}$. The gas is described by an ideal-gas equation of state with adiabatic index ${\gamma=5/3}$. The initial value of the specific internal energy is set to ${e_\mathrm{int}=0.05}$. As in \cite{mandal2023}, we set the gravitational constant to ${G=1}$, which results in an initial gravitational binding energy per unit mass of 0.6. The fluid is initially at rest, and reflecting boundary conditions are imposed on all domain boundaries so that the system remains isolated. The subsequent evolution of the system is followed until ${t=3}$.

The time evolution of the internal, kinetic, and gravitational potential energy reservoirs, together with the total energy, is shown in Fig.~\ref{fig:evrard-comps} for a simulation performed on a $512^3$ grid. Because the gravitational binding energy initially dominates over the internal energy of the gas, the cloud undergoes gravitational collapse. The resulting compression increases the internal energy and pressure in the central regions. At approximately ${t\approx 1.1}$, the collapse produces a bounce shock that propagates outward, while the outer layers continue to fall inward. At later times, the system  virializes, with the total kinetic energy decreasing as the cloud settles toward a quasi-hydrostatic configuration.

To quantify the energy-conservation properties of the numerical scheme, we repeat the test on grids with $128^3$ and $256^3$ cells. The time evolution of the relative energy-conservation error is shown in Fig.~\ref{fig:evrard-cons}. The largest contribution to the accumulated error occurs during the late stages of the gravitational collapse, around the time of shock formation. The absolute relative error decreases from approximately $14\%$ on the $128^3$ grid to approximately $2\%$ on the $512^3$ grid, confirming that the energy-conservation error is primarily due to spatial and temporal discretization errors. The use of slope limiters and Godunov splitting to couple Poisson's equation to the system of gas-dynamics equations introduces additional first-order truncation errors, reducing the convergence rate to approximately 1.4 for the grids tested, compared with the expected second-order scaling of the base scheme.

\begin{figure}
\includegraphics[width=\linewidth]{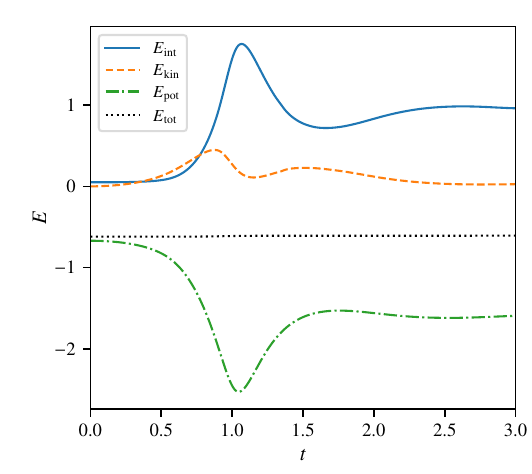}
\caption{Time evolution of the internal, kinetic, and gravitational potential energy reservoirs in the Evrard collapse test on a grid with $512^3$ cells. The black dotted line denotes the evolution of the total energy.}
\label{fig:evrard-comps}
\end{figure}

\begin{figure}
\includegraphics[width=\linewidth]{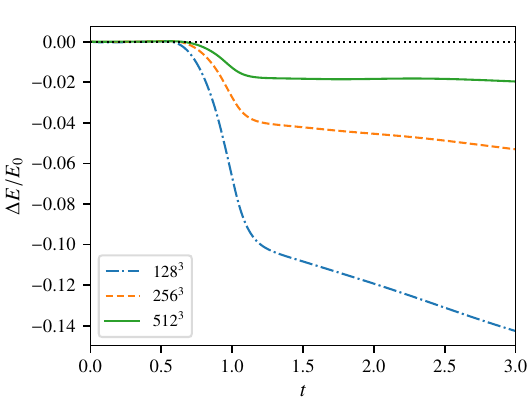}
\label{fig:evrard-cons}
\caption{Time evolution of the energy-conservation error normalized by the initial total energy. Different curves correspond to grid resolutions ranging from $128^3$ to $512^3$ cells. The horizontal black dotted line at zero serves as a guide to the eye.}
\label{fig:evrard-cons}
\end{figure}

\subsection{One-zone network calculation}

Here, we perform one-zone nuclear reaction network calculations. The test is carried out on a logical grid consisting of a single cell, with the Godunov solver disabled. The mass density and temperature are initialized at $\,{t=0}\,$ with values $\,{\rho = 10^4\ \mathrm{g\,cm^{-3}}}\,$ and $\,{T = 2\times10^8\ \mathrm{K}}\,$ and are held fixed throughout the calculation. These thermodynamic conditions are typical of the convective phase preceding a classical nova explosion on a carbon--oxygen white dwarf star \citep[e.g.,][]{casanova2016,jose2020}.

We track nine isotopes, $\mathrm{p}$, $^{4}\mathrm{He}$, $^{12}\mathrm{C}$, $^{13}\mathrm{C}$, $^{13}\mathrm{N}$, $^{14}\mathrm{N}$, $^{15}\mathrm{N}$, $^{14}\mathrm{O}$, and $^{15}\mathrm{O}$, which are connected through a network of eight reactions involved in the hot-CNO cycle \citep[see, e.g.,][]{iliadis2007},
\begin{equation}
\begin{array}{ll}
^{13}\mathrm{N}(\beta^+)^{13}\mathrm{C}, &
^{13}\mathrm{N}(\mathrm{p},\gamma)^{14}\mathrm{O}, \\
^{14}\mathrm{O}(\beta^+)^{14}\mathrm{N}, &
^{13}\mathrm{C}(\mathrm{p},\gamma)^{14}\mathrm{N}, \\
^{15}\mathrm{O}(\beta^+)^{15}\mathrm{N}, &
^{14}\mathrm{N}(\mathrm{p},\gamma)^{15}\mathrm{O}, \\
^{12}\mathrm{C}(\mathrm{p},\gamma)^{13}\mathrm{N}, &
^{15}\mathrm{N}(\mathrm{p},\alpha)^{12}\mathrm{C}.
\end{array}
\end{equation}
The initial composition is $\,{X_\mathrm{p} = 0.5}$, $\,{X_{^{4}\mathrm{He}} = 0.25}$, and $\,{X_{^{12}\mathrm{C}} = 0.25}$. The corresponding molar fractions are computed as $\,{Y_l=X_l/A_l}$. 

As a reference solution, we use results obtained with the established \texttt{PYNUCASTRO} code \citep{pynucastro2}. The same thermodynamic conditions, reaction network, initial composition, and time step sizes are used in \texttt{PYNUCASTRO}, but the governing system of ODEs is solved with the fifth-order accurate Radau solver for stiff systems \citep{hairer1996}, using a relative tolerance of $10^{-14}$. In the \texttt{PHLEGETHON} run, the second-order TR-BDF2 solver is employed (see Sect.~\ref{sec:nuclear-network}). In both calculations, electron screening is disabled so that the comparison isolates differences arising purely from the time integration of the reaction network. Reaction rates are taken from the JINA REACLIB library \citep[][version \texttt{2021-06-24}]{cyburt2010}.

In these one-zone models, the time step is not constrained by the hyperbolic CFL condition. Instead, it is adaptively updated according to
\begin{equation}\label{eq:time-step-nuc0}
\Delta t \rightarrow \Delta t \min_{ \left\{ Y_l>10^{-10} \right\} } \left[ \frac{Y_l}{\max(10^{-15},|\Delta Y_l|)} \right],
\end{equation}
where $\Delta Y_l = Y_l^{n+1}-Y_l^{n}$ denotes the change in the molar fraction of species $l$ during a time step. The conservative factor $10^{-15}$ in Eq.~(\ref{eq:time-step-nuc0}) prevents division by zero when $\,{\Delta Y_l = 0}$. The initial time step is set to $\,{\Delta t = 10^{-9}\,\mathrm{s}}$, which is about five times smaller than what predicted by Eq.~(\ref{eq:time-step-nuc0}) after the first time step integration. This procedure ensures that even the fastest reactions are accurately resolved in time and allows the results from the two codes to be compared over a wide range of temporal scales.

Figure~\ref{fig:onezone-cno} shows the temporal evolution of the mass fractions until $t=10^4\ \mathrm{s}$, which proceeds as follows:

\begin{enumerate}[label=(\roman*)]

\item At early times ($t \lesssim 10^{-1}\,\mathrm{s}$), the network is dominated by rapid proton-capture reactions. The initial $^{12}\mathrm{C}$ seed is quickly processed through $^{12}\mathrm{C}(\mathrm{p},\gamma)^{13}\mathrm{N}$, leading to a rapid increase in the abundance of $^{13}\mathrm{N}$. Subsequent proton capture produces $^{14}\mathrm{O}$ through $^{13}\mathrm{N}(\mathrm{p},\gamma)^{14}\mathrm{O}$. The much slower $\beta^+$ decay of $^{13}\mathrm{N}$ (which occurs on a timescale of $\,{\sim 10^{3}\,\mathrm{s}}\,$) generates only a $\,{\sim10^{-6}\,}$ mass fraction abundance of $^{13}\mathrm{C}$. During this phase, the abundance of $^{12}\mathrm{C}$ decreases rapidly as it is converted into heavier CNO isotopes.

\item At intermediate times ($\,{10^{-1}\,\mathrm{s} \,{\lesssim}\, t \, {\lesssim} 1\,\mathrm{s}}\,$) as the abundance of $^{12}\mathrm{C}$ drops below $10^{-2}$, proton captures on $^{12}\mathrm{C}$ become inefficient and the reaction flow becomes regulated by proton captures on $^{13}\mathrm{N}$. As a result, $^{14}\mathrm{O}$ reaches an abundance of ${\approx}\, 0.3$, even exceeding that of $^{4}\mathrm{He}$.

\item At later times ($\,{1\,\mathrm{s} \lesssim t \lesssim 10^3\,\mathrm{s}}\,$), the system approaches a quasi-steady state in which CNO nuclei act as catalysts that mediate the conversion of hydrogen into helium. The abundances of the intermediate isotopes vary only slowly as a balance is established between proton captures and $\beta^+$ decays along the cycle. The dominant reservoirs within the CNO group are $^{14}\mathrm{O}$ and 
$^{15}\mathrm{O}$.

\item Finally, at late times ($t \gtrsim 10^{3}\,\mathrm{s}$), the decreasing proton abundance leads to the gradual termination of efficient hydrogen burning. As hydrogen is converted into $^{4}\mathrm{He}$, the helium mass fraction increases, while the reaction flow through the CNO cycle diminishes and unstable isotopes decay away exponentially.

\end{enumerate}
Both codes produce solutions that agree to very high accuracy at all times. The maximum relative difference in the mass fractions between $\,{t=1}\,$ and $\,{t=100\,\mathrm{s}}\,$, when none of the abundances are negligibly small, is approximately $10^{-5}$.

We also compare the time-integrated nuclear energy release per unit mass, one of the key quantities in reacting-convective flows,
\begin{equation}
\Delta e_\mathrm{N}(t) = N_\mathrm{A} \sum_l m_l c^2
  \left[ Y_l(t) - Y_l(0) \right].
\end{equation}
In this test, however, it is not used to update the specific internal energy. The evolution of ${\Delta e_\mathrm{N}}$ is shown in Fig.~\ref{fig:onezone-enuc}. Because $\Delta e_\mathrm{N}$ increases monotonically with time, it provides a convenient integrated diagnostic of the network evolution. Over the course of the calculation, the time-integrated nuclear energy release increases by approximately three orders of magnitude, while the relative difference between the results of the two codes remains below $\,{\approx 10^{-4}}\,$ 

In production-quality simulations of stellar interiors, fast transients in a nuclear network often occur on timescales much shorter than the hyperbolic CFL time step. Updating the isotopes during the network solve using the time-step criterion based on Eq.~(\ref{eq:time-step-nuc0}) is therefore computationally infeasible. To test the robustness of our nuclear network solver when dealing with stiff systems and large time steps, we re-run the one-zone hot-CNO cycle model using a fixed time step of $\Delta t = 1\ \mathrm{s}$, which is significantly longer than the initial fast transient dominated by proton captures on $^{12}\mathrm{C}$ and $^{13}\mathrm{N}$. The result is shown in Fig.~\ref{fig:onezone-dt-1s}. Although the implicit solver overshoots the reference solution during the first few time steps for some abundances, the solution subsequently recovers the correct asymptotic values on longer timescales.

To better assess the scaling of global temporal truncation errors generated by our implicit methods, we compute the $L_1$ relative error norm in the time-integrated nuclear energy release at $t=10^2\ \mathrm{s}$ between adjacent time step resolutions as
\begin{equation}
L_1(\Delta t) = \frac{|\Delta e_\mathrm{N,\Delta t}-\Delta e_\mathrm{N,\Delta t/2}|}{\Delta e_\mathrm{N,\Delta t/2}}.
\end{equation}
For this test, the nonlinear convergence tolerance in Eq.~(\ref{eq:nuclear-thr}) is set to a conservative value of $10^{-15}$. The results shown in Fig.~\ref{fig:onezone-L1} confirm the expected first and second-order scaling for BE and TR-BDF2, respectively, as the time step is decreased. The error deviates from the predicted scaling only at the smallest tested time-step values, likely because the global truncation error becomes comparable to the accumulated numerical residuals associated with the finite convergence tolerance of the Newton--Raphson solver. These tests therefore demonstrate that the nuclear network solvers in \texttt{PHLEGETHON} can robustly and accurately integrate stiff reaction systems relevant to stellar-interior applications.

\begin{figure}
\includegraphics[width=\linewidth]{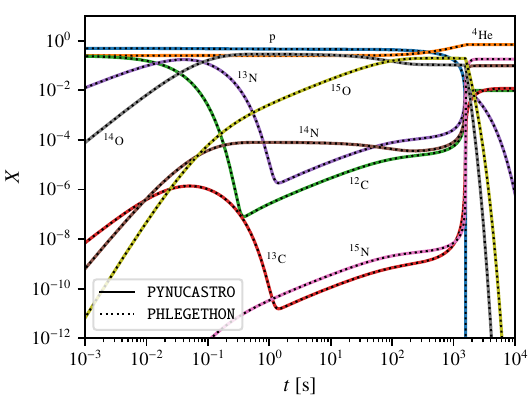}
\caption{
Time evolution of the isotopic mass fractions in the one-zone hot-CNO cycle calculation.
The colored solid lines show the results obtained with the \texttt{PYNUCASTRO} code,
while the black dotted lines correspond to the results from \texttt{PHLEGETHON}.
}
\label{fig:onezone-cno}
\end{figure}

\begin{figure} 
\includegraphics[width=\linewidth]{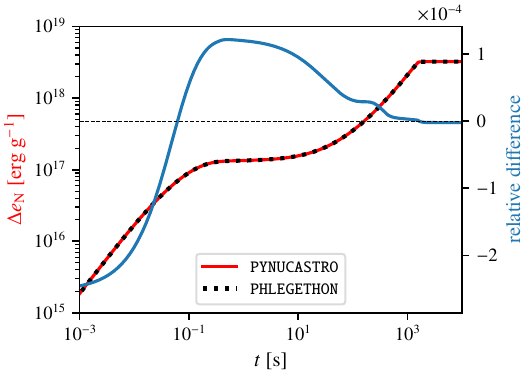}
\caption{
Time evolution of the time-integrated nuclear energy release per unit mass in the one-zone hot-CNO cycle calculation.
The red solid line shows the result obtained with \texttt{PYNUCASTRO}, while the black dotted line shows the result from \texttt{PHLEGETHON}. 
The blue solid curve represents the relative difference between the two codes.
}
\label{fig:onezone-enuc}
\end{figure}

\begin{figure}
\includegraphics[width=\linewidth]{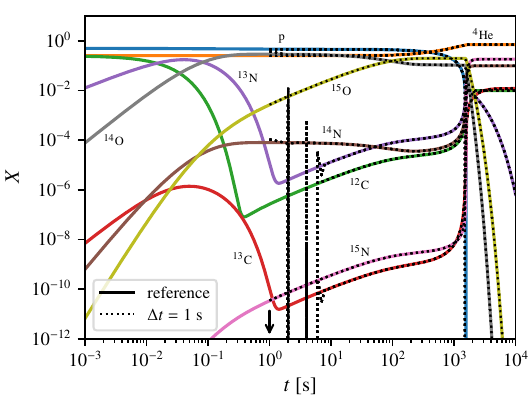}
\caption{Time evolution of the isotopic mass fractions in the one-zone hot-CNO cycle calculation.
The colored solid lines show a reference solution obtained with \texttt{PHLEGETHON} using the adaptive time step criterion in Eq.~(\ref{eq:time-step-nuc0}),
while the dotted black lines correspond to results obtained with a fixed time step of $\Delta t = 1\,\mathrm{s}$. The downward-pointing arrow indicates the end of the first integration step.}
\label{fig:onezone-dt-1s}
\end{figure}

\begin{figure}
\includegraphics[width=\linewidth]{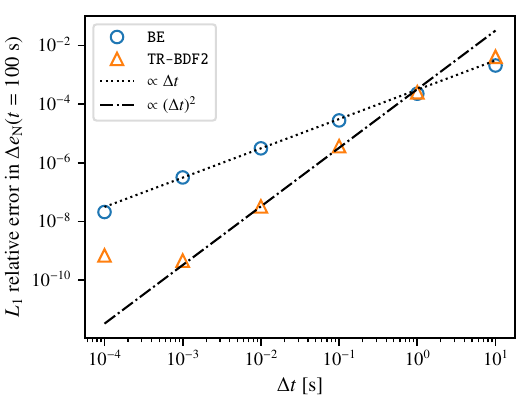}
\caption{$L_1$ relative error norm in the time-integrated nuclear energy release up to $t=100$ s. These errors are functions of the time step, measured between simulations with adjacent time step resolutions in the one-zone hot-CNO cycle model. Results are shown for the BE (circles) and TR-BDF2 (triangles) methods. The black dotted and dash-dotted lines indicate first- and second-order scaling, respectively.}
\label{fig:onezone-L1}
\end{figure}
\subsection{Burning-radiating buoyant bubble}\label{sec:bb}

We design this test problem to verify multiple numerical schemes and physics modules simultaneously, including well-balancing, the nuclear reaction network solver, the thermal diffusion solver, and a tabulated equation of state. In particular, we present simulations of a low-Mach buoyant bubble rising in an isentropic and hydrostatic background. Similar tests have been presented in the literature \citep[see, e.g.,][]{zingale2019,edelmann2021a}. Here, the problem is set up so that the bubble undergoes nuclear burning, while the energy excess generated within it is allowed to diffuse away. The balance between these two processes regulates the bubble’s buoyant acceleration.

The background hydrostatic stratification is assumed to be isentropic, i.e., $\,{\nabla=\nabla_\mathrm{ad}}$, where $\,{\nabla=\mathrm{d}\ln T/\mathrm{d}\ln P}\,$ and $\nabla_\mathrm{ad}$ is the adiabatic temperature gradient. The composition of the background medium is pure $^{12}\mathrm{C}$. The system is assumed to be spherically symmetric, with temperature and density at $\,{r=0}\,$ given by $\,{T_0=10^9\ \mathrm{K}}\,$ and $\,{\rho_0=10^7\ \mathrm{g\,cm^{-3}}}$. These thermodynamic conditions are typical of the interiors of, e.g., electron-capture or core-collapse supernova progenitors \citep[e.g.,][]{takahashi2013,muller2020}. To account for effects of electron degeneracy, in these simulations we employ the Helmholtz EoS.

The initial stratification is generated numerically by solving the coupled system of ODEs enforcing hydrostatic equilibrium and an adiabatic temperature gradient,
\begin{align}
\partial_r P &= \rho g_r, \label{eq:p-hse} \\
\partial_r T &= -\frac{T}{H_P}\nabla_\mathrm{ad}, \label{eq:T-hse}
\end{align}
subject to the constraints
\begin{align}
\rho &= \rho_\mathrm{EoS}(P,T,X_{^{12}\mathrm{C}}=1), \\
g_r &= -\frac{P}{\rho H_P},
\end{align}
where $H_P$ is the pressure scale height. Its value is fixed such that $H_P = R_\mathrm{max}/3$, where $R_\mathrm{max} = c_{s,0}\,(0.75\,\mathrm{s}) = 2.8\times10^8\,\mathrm{cm}$ is the radial extent of the domain and $c_{s,0}$ is the sound speed at $r=0$ derived from the EoS. With this choice of parameters, the pressure decreases by a factor of $\approx20$ across the domain.

The mass density $\rho$ is obtained from pressure, temperature, and composition, using a Newton--Raphson root-finding procedure analogous to that described in Sect.~\ref{sec:eos}. Once the value of the mass density is retrieved, the adiabatic temperature gradient $\nabla_\mathrm{ad}$ is evaluated through a forward EoS call,
\begin{equation}
(\rho,T,X_{^{12}\mathrm{C}}) \rightarrow \nabla_\mathrm{ad}.
\end{equation}
For the chosen parameterization, Eq.~(\ref{eq:p-hse}) admits the analytical solution
\begin{equation}
P(r) = P_0 \exp\left(-\frac{r}{H_P}\right).
\end{equation}
The coupled equations governing the stratification are integrated numerically using a fourth-order Runge--Kutta method on a 1D grid with $10^5$ radial nodes. The integration starts from $r=0$ with $T_0$ and $P_0 = P_\mathrm{EoS}(\rho_0, T_0, X_{^{12}\mathrm{C}})$ and proceeds outward. Throughout the resulting stratification, the contribution of electron degeneracy to the total thermal pressure $P$ is approximately $50\%$.

The newly generated stratification is mapped onto a 3D spherical wedge spanning
\[
(r,\theta,\psi)\in(R_\mathrm{max}/10,R_\mathrm{max})\times(\pi/3,2\pi/3)\times(-\pi/6,\pi/6),
\]
where $\theta$ and $\psi$ denote the colatitude and azimuthal angles, respectively. Reflecting boundary conditions are applied at all grid domain boundaries (see Sect.~\ref{sec:bcs}). The isentropic stratification is perturbed at $\,{r=R_\mathrm{max}/3}\,$ by replacing a small fraction of $^{12}\mathrm{C}$ with $^{4}\mathrm{He}$ according to
\begin{equation}
X_{^{4}\mathrm{He}}(\tilde{r})=
\begin{cases}
\Delta X_{^{4}\mathrm{He}}\cos^2\left(\frac{\pi \tilde{r}}{2r_b}\right), & \tilde{r}\le r_b, \\
0, & \tilde{r}>r_b,
\end{cases}
\end{equation}
where $\,{r_b=R_\mathrm{max}/8}\,$ is the radius of the bubble, $\,{\tilde{r}=\sqrt{(x-R_\mathrm{max}/3)^2+y^2+z^2}}$, and $\,{X_{^{12}\mathrm{C}} = 1-X_{^{4}\mathrm{He}}}$. 
Helium is allowed to burn into carbon via the $3\alpha$ reaction,
\begin{equation}
^{4}\mathrm{He}+^{4}\mathrm{He}+^{4}\mathrm{He}\rightarrow {}^{12}\mathrm{C}+\gamma.
\end{equation}
The amplitude of the perturbation is set to ${\Delta X_{^{4}\mathrm{He}}=0.01}$. Inside the bubble, the temperature is recomputed at ${t=0}$ to account for the modified composition while maintaining mechanical equilibrium,
\begin{equation}
T(r\le r_b)=T_\mathrm{EoS}(\rho,P,X_l).
\end{equation}
This operation results in a negative temperature perturbation inside the bubble with an amplitude relative to the surrounding isentropic background of $\,{\approx5\times10^{-3}}$.

Thermal diffusion is enabled and the opacity is assumed to be spatially constant with value $\,{\kappa=3\times10^{-13}\ \mathrm{cm^{2}\, g^{-1}}}$. This value is chosen such that the P\'eclet number
\begin{equation}
\mathrm{Pe} = 3\kappa \rho_{r_b}\frac{u_b R_\mathrm{max}}{c} \approx 0.1,
\end{equation}
implying efficient thermal diffusion over the rising time of the buoyant bubble. Here $\,{u_b\sim3\times10^{6}\ \mathrm{cm\,s^{-1}}}\,$ is the characteristic initial rise velocity of the bubble and $\,{\rho_{r_b}=5\times10^{6}\ \mathrm{g\,cm^{-3}}}\,$ is the mass density at the center of the bubble at $t=0$.

We run 3D, purely hydrodynamical simulations on grids with $32^3$, $64^3$, $128^3$, and $256^3$ cells. In these simulations, deviation well-balancing is used (see Sect.~\ref{sec:wb}). To treat fast thermal diffusion, we use the RKL2 STS method described in Sect.~\ref{sec:sts}. The thermal diffusion operator is applied only to the temperature deviation according to Eq.~(\ref{eq:diffusion_temp}). This approach allows us to simulate the two competing processes that determine the thermal content of the bubble (namely thermal diffusion and nuclear burning) without affecting the background stratification. On the $256^3$ grid, the ratio between the hyperbolic and parabolic CFL time steps is $169$, which results in 26 RKL2 STS substeps. For these simulations, we use the 
PPH reconstruction method.  The $3\alpha$ reaction is coupled to the system of gas-dynamics equations and gravity using Strang splitting, and the nuclear network solver employs the TR-BDF2 implicit time stepper. Overall, the combined numerical methods are expected to achieve second-order accuracy.

Figure~\ref{fig:bb-sli} shows the evolution of the bubble from $\,{t=0}\,$ to $\,{t=35\,\mathrm{s}}\,$ on the $256^3$ grid. Within $\,{t \approx 1\,\mathrm{s}}$, thermal energy is rapidly transported from the hotter surroundings into the cooler bubble, thereby increasing its temperature. This fast transient, alongside the increase in pressure due to heating by $3\alpha$, breaks the initial mechanical equilibrium causing the bubble to expand. The resulting buoyant acceleration remains weak, and the bubble reaches Mach numbers of $\mathcal{M}\sim0.01$ by the end of the simulation. During the buoyant rise of the bubble, the helium mass decreases by roughly a factor of $5\%$ as a result of the action of $3\alpha$. However, the entropy of the bubble remains almost unaltered over time because thermal diffusion acts as a balancing mechanism that efficiently removes excess entropy produced by nuclear burning.  At later times, as the central region of the bubble develops the largest density fluctuations, it accelerates more rapidly than its surroundings, leading to the onset of nonlinear effects and the formation of a vortex ring along the bubble’s sides.

To quantify the numerical convergence of the combined numerical methods, we employ Richardson extrapolation \citep[see, e.g.,][]{press1992}. For three consecutive grids with refinement ratio ${r=2}$, the estimated convergence rate of the numerical solution is computed as 
\begin{equation}\label{eq:richex}
p = \frac{\ln\!\left(\dfrac{||q_3 - q_2||_1}{||q_2 - q_1||_1}\right)}{\ln(r)},
\end{equation}
where $q_m$ denotes the numerical solution obtained on grid $m$. In Eq.~(\ref{eq:richex}) the $|| \cdot ||_1$ operator denotes the $L_1$ error norm, e.g.,
\begin{equation}
  || q_2 - q_1||_1 = \frac{1}{N_r N_\theta N_\psi}\sum_{i=1}^{N_r} \sum_{j=1} ^{N_\theta} \sum_{k=1}^{N_\psi} | q_{2,i,j,k}-q_{1,i,j,k}|.
\end{equation}
The subtraction of one numerical solution from another is performed by restricting the finer grid to the coarser grid through volume averaging of a block of ${2\times2\times2}$ neighboring cells. The convergence rates as functions of time obtained from the $32^3$--$64^3$--$128^3$ and $64^3$--$128^3$--$256^3$ grid combinations are shown in Fig.~\ref{fig:bb-Rich}. In particular, convergence rates are computed for the helium mass fraction, entropy, and temperature deviations from the background state. At early times in the evolution ($\,{t\lesssim16\,\mathrm{s}}\,$), while the bubble remains smooth, the code achieves the expected second-order convergence. At later times, the bubble develops steep gradients in both composition and entropy at the top of the bubble that are resolved by only a few grid cells, as shown in Fig.~\ref{fig:bb-rprofs}. As a result, the observed convergence rate drops to first order. The evolution of the convergence rate is consistent across the different grid combinations and variables considered. 
Moreover, thanks to the well-balanced property of \texttt{PHLEGETHON}, the background stratification remains isentropic to a level of ${\sim10^{-6}}$, with most of the residual error arising from the biquintic interpolation of the Helmholtz EoS table. Overall, this test demonstrates that \texttt{PHLEGETHON} can robustly handle the coupled effects of hydrodynamics, nuclear burning, thermal diffusion, and a realistic equation of state in a strongly stratified medium. 

\begin{figure*}
\centering
\includegraphics[width=1.0\textwidth]{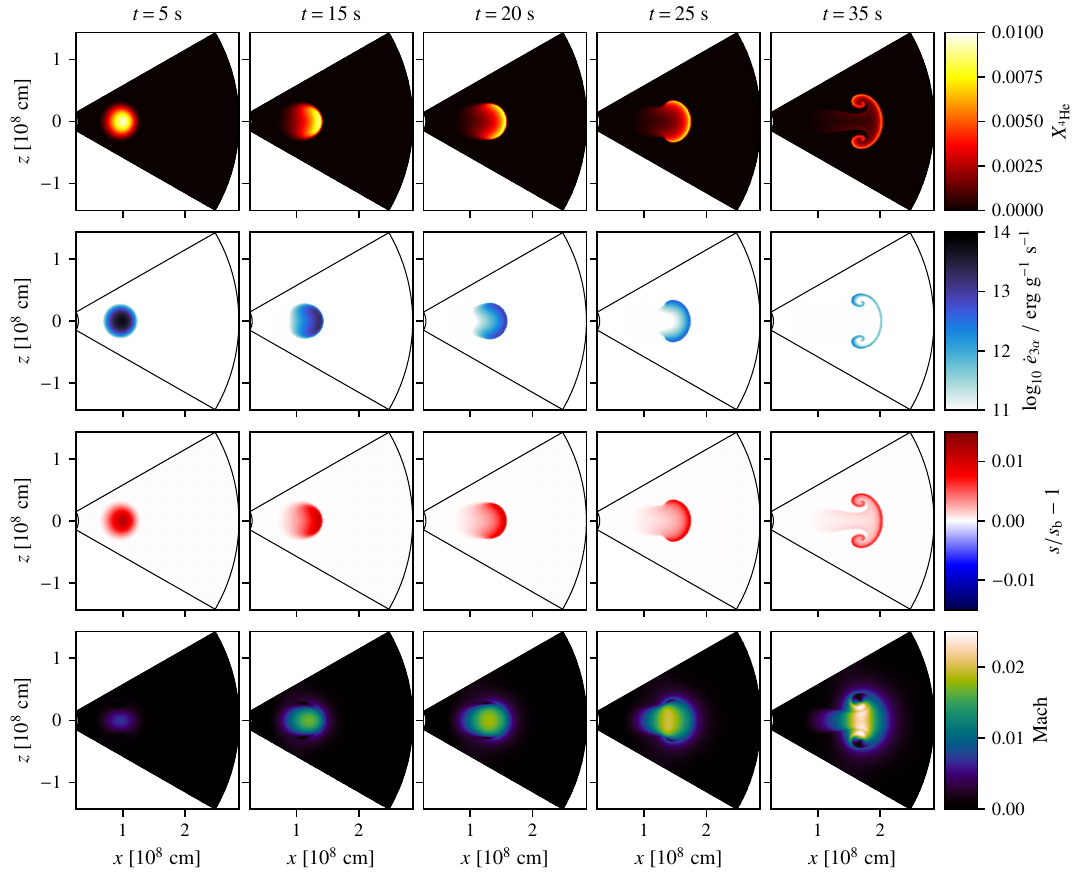}
\caption{Slices at ${\psi=0}$ showing the time evolution of the helium mass fraction, specific energy generation rate from the $3\alpha$ reaction, specific entropy deviations from the background isentropic state, and Mach number in the burning--radiating buoyant bubble problem. The panels show results computed on a $256^3$ spherical wedge grid.}
\label{fig:bb-sli}
\end{figure*}

\begin{figure}
\includegraphics[width=\linewidth]{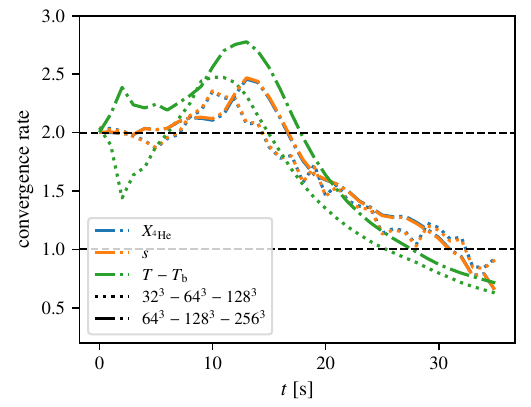}
\caption{Convergence rates as functions of time for the helium mass fraction, specific entropy, and temperature deviation from the hydrostatic background state in the burning--radiating buoyant bubble problem. Dotted and dashed-dotted lines represent results obtained from Richardson extrapolation using the $32^3$, $64^3$, and $128^3$ grids and the $64^3$, $128^3$, and $256^3$ grids, respectively. The two horizontal black dashed lines indicate first- and second-order convergence rates and serve as guides to the eye.}
\label{fig:bb-Rich}
\end{figure}

\begin{figure}
\includegraphics[width=\linewidth]{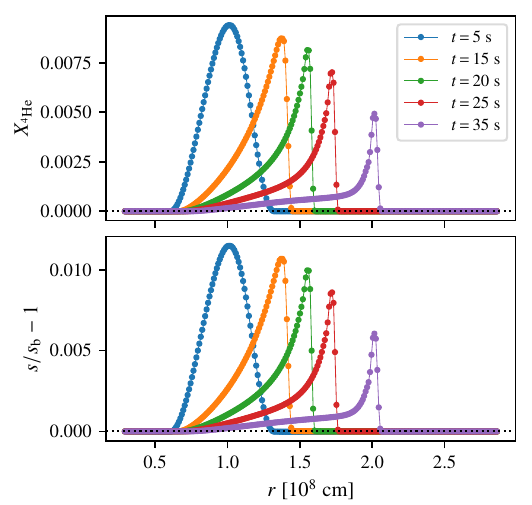}
\caption{Radial profiles of helium mass fraction and deviations of specific entropy from the hydrostatic background state in the burning--radiating buoyant bubble problem, along the ray at $\,{\theta=\pi/2}\,$ and $\,{\psi=0}$.}
\label{fig:bb-rprofs}
\end{figure}

\subsection{Andrassy code-comparison setup}

As a last test, we run simulations of the code-comparison setup of \cite{andrassy2022}. This problem involves turbulent convective motions simulated in plane-parallel geometry, driven by a time-independent heat source placed close to the bottom boundary of the computational domain. The convection interacts with a convective boundary that is initially located at half the vertical domain height. This interaction excites waves and drives mixing processes, causing the convection zone to grow in mass and drift over time in the direction opposite to gravity. This setup provides an excellent test of whether a given combination of numerical schemes can accurately capture a broad range of hydrodynamic phenomena closely related to those occurring in convection zones and radiative regions inside stars. Because the resulting flows are both subsonic and strongly stratified, this problem also represents a good benchmark for testing the well-balanced and low-Mach-number properties of the chosen numerical methods. The outputs of the simulations presented in \cite{andrassy2022} are publicly available\footnote{\url{https://github.com/robert-andrassy/CoCoPy}}, making this test particularly well suited for direct quantitative comparisons with established stellar hydrodynamics codes, including \texttt{SLH}, \texttt{PROMPI}, \texttt{FLASH}, \texttt{PPMSTAR}, and \texttt{MUSIC}. The specific code versions and numerical methods used to generate the reference data are documented in \cite{andrassy2022} and are adopted here for the corresponding comparisons.

The spatial domain spans $(x,y,z) \in (-1,1)\times(1,3)\times(-1,1)$. Gravity points in the negative $y$-direction and is described by a smooth function of the $y$-coordinate. The initial stratification is piecewise polytropic such that in the lower half of the domain the polytropic index is equal to the adiabatic index of the classical gas equation of state employed, $\,{\gamma = 5/3}$, yielding an isentropic stratification. This region smoothly connects to a subadiabatic stratification in the upper half of the domain. A detailed description of the initial conditions can be found in \cite{andrassy2022}.

We run simulations at the intermediate $256^3$ grid resolution used in \cite{andrassy2022} until $\,{t = 2000}\,$ in code units, testing both our PPH and limited fifth-order reconstruction methods. The Riemann solver used in both simulations is LHLLC. Boundary conditions are periodic at the lateral sides of the box and reflective in the direction of gravity.

Figure~\ref{fig:cc-mach2d} shows snapshots of the Mach number in the $z = 0$ plane halfway through the simulation. Overall, the morphology of the flows in both of our simulations is similar to the results obtained by the five hydrodynamic codes considered in the 2022 study. The convection exhibits a relatively wide scale separation, with large-scale buoyant flows driven by the heat source that develop shear instabilities and undergo a turbulent cascade. The resulting Mach numbers in both simulations are similar, with typical values ranging from $0.02$ to $0.08$, consistent with the results reported in \cite{andrassy2022}. Other properties of the system dynamics, such as the location of the upper convective boundary and the characteristic wavelengths of internal gravity waves visible in the upper part of the domain, are qualitatively similar in the two runs. Nonlinear effects give rise to marginal wave breaking, which appears to be captured in both simulations. However, the limited fifth-order reconstruction method produces smaller-scale structures in both the convective and stable regions. This result is consistent with the findings of \cite{leidi2024}, who showed that higher-order reconstruction methods tend to yield finer-scale flow structures in both convection and wave motions.

Similar to the codes tested in \cite{andrassy2022}, \texttt{PHLEGETHON} relies on the implicit large-eddy  simulation (ILES) approach to model viscous dissipation of kinetic energy in turbulent flows. To verify the ILES properties of the methods used in our code, we compute the kinetic energy power spectrum across the $\,{y = 1.7}\,$ plane in the convective layer as a function of the horizontal wavenumber $\,{k_\mathrm{h}=\sqrt{k_x^2+k_y^2}}$. The spectrum is averaged over the time interval $\,{t \in (500,2000)}\,$ to suppress statistical variations arising from the chaotic nature of the turbulent flows. The results are shown in Fig.~\ref{fig:cc-spectra}, where we compare them to those of the other hydrodynamic codes used in the original code-comparison study. Overall, there is excellent agreement with the other codes, especially in the inertial range of the spectrum, where the scaling of kinetic power follows Kolmogorov's ${-5/3}$ law. Differences between codes become more apparent in the viscous range, where the amount of numerical dissipation depends on the specific combination of methods employed. \texttt{PHLEGETHON}'s solutions closely follow that of \texttt{SLH} up to $\,{k_\mathrm{h} = 50}$, beyond which the power resulting from the simulation using PPH decreases significantly relative to \texttt{SLH}. At the Nyquist wavelength, the limited fifth-order reconstruction method yields the least dissipative scheme in the ensemble after \texttt{SLH}, which employs the flux-splitting-based $\mathrm{AUSM}^+$-$\mathrm{up}$ Riemann solver \citep{liou2006} and a semi-discrete adaptation of the piecewise-parabolic method of \cite{colella1984a}. Figure~\ref{fig:cc-spectra} also shows kinetic energy power spectra computed at $\,{y = 2.7}$, which lies in the convectively-stable region. Again, we observe very good agreement with the other codes, with the largest spread occurring near the Nyquist wavelength where numerical dissipative processes dominate the dynamics of the waves.

The test problem includes a passive scalar whose abundance at $t=0$ is zero in the adiabatic layer while smoothly transitioning to unity in the subadiabatic part of the stratification. This passive scalar acts as a tracer for the fluid in the stable region, so it allows the position of the boundary between the convective and stable regions to be tracked over time. This result is obtained by locating the steepest radial gradient in horizontal averages of the scalar. In fact, convection efficiently mixes any scalar abundance that enters the convective layer, so the steepest radial gradient in $X$ naturally develops at the convective boundary. This procedure enables us to quantify how much mass has been entrained into the convective layer. CBM plays a crucial role in stellar structure and evolution \citep{zahn1991}, so it is vital that the numerical methods implemented in the code can capture this process accurately. In Fig.~\ref{fig:cc-ment}, we show the time evolution of the total mass entrained into the convective layer in the two \texttt{PHLEGETHON} simulations, alongside the results obtained by \cite{andrassy2022}. Both of our tested methods yield results consistent with the five hydrodynamic codes considered in the original study. At $\,{t=2000}\,$ in code units, our results show a relative deviation from the reference solution (computed on a $512^3$ grid with the \texttt{PPMSTAR} code) of 5\% for the simulation using the limited fifth-order reconstruction method and and 1\% for the simulation using PPH. Overall, the results presented in this section are encouraging because they confirm that \texttt{PHLEGETHON} can accurately resolve hydrodynamic phenomena relevant to stellar interiors, achieving a level of accuracy comparable to established codes in the field.

\begin{figure}
\centering
\includegraphics[width=\linewidth]{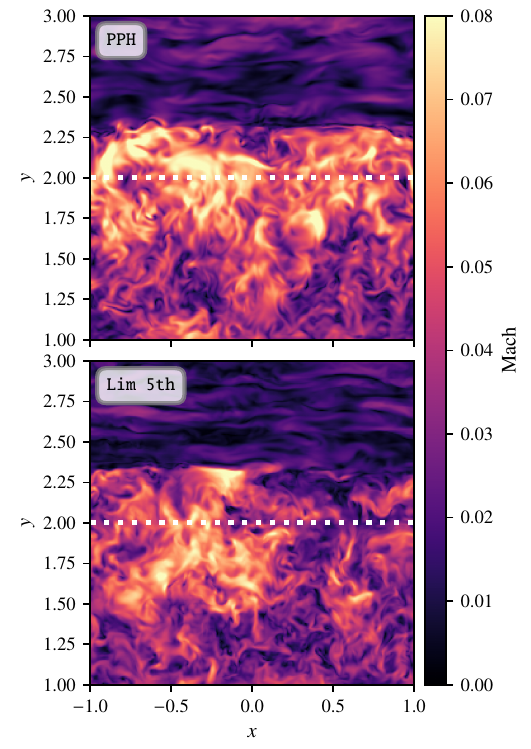}
\caption{Snapshots of the Mach number in the ${z = 0}$ plane in the code-comparison benchmark of \cite{andrassy2022} run with \texttt{PHLEGETHON} on $256^3$ grids at $\,{t=1000}$. The two panels show results obtained with our PPH and limited fifth-order reconstruction methods, respectively. The white dotted lines represent the location of the upper boundary of the adiabatic region at $\,{t=0}$.}
\label{fig:cc-mach2d}
\end{figure}

\begin{figure}
\centering
\includegraphics[width=\linewidth]{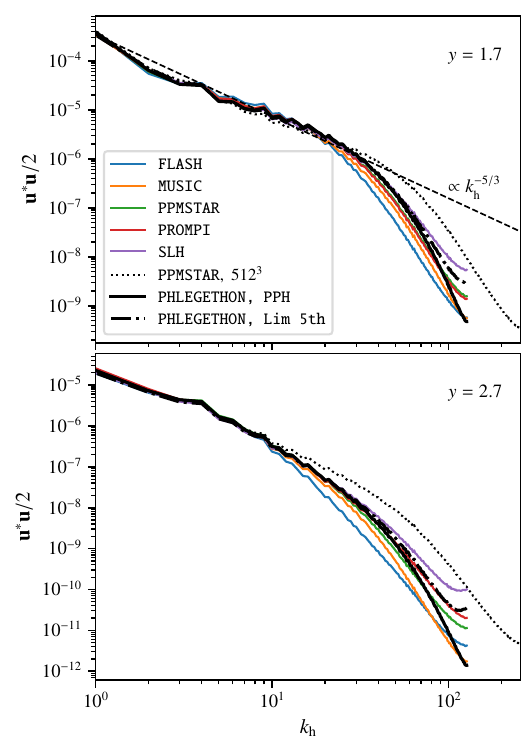}
\caption{Power spectra of specific kinetic energy as functions of the horizontal wavenumber $k_\mathrm{h}$ in the code-comparison test problem of \cite{andrassy2022}, for simulations run with the PPH and limited fifth-order reconstruction methods on $256^3$ grids. The power spectra shown here include those presented in \cite{andrassy2022}, which are publicly available at \url{https://github.com/robert-andrassy/CoCoPy}. The top panel shows spectra computed in the convective layer at $\,{y=1.7}$, while the bottom panel shows spectra computed in the subadiabatic region at $\,{y=2.7}$. All spectra are averaged over the time interval $\,{t \in (500,2000)}\,$ in code units. In the top panel, the black dashed line represents Kolmogorov's $k_\mathrm{h}^{-5/3}$ scaling law.}
\label{fig:cc-spectra}
\end{figure}

\begin{figure}
\centering
\includegraphics[width=\linewidth]{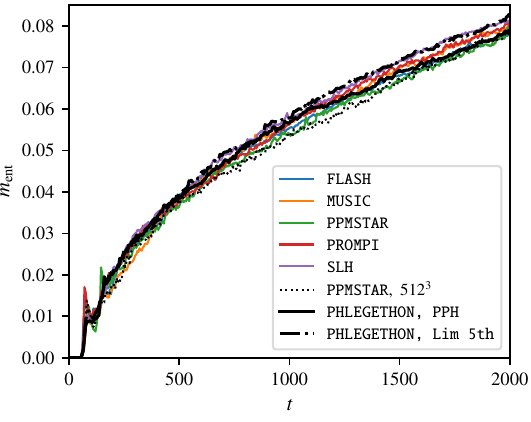}
\caption{Total mass entrained from the upper stable region into the convective layer as a function of time in the code-comparison test problem of \cite{andrassy2022} run on $256^3$ grids. Results obtained with our PPH and limited fifth-order reconstruction methods are shown alongside the outputs from the original code-comparison study.}
\label{fig:cc-ment}
\end{figure}

\end{section}

\begin{section}{Magnetohydrodynamic simulation of a core-collapse supernova progenitor}\label{sec:ccsnp}

In this section, we apply a combination of the numerical methods described in Sect.~\ref{sec:methods} to model an astrophysical scenario. In particular, we consider an extended convective shell in a massive star shortly before the onset of core collapse. This setup involves the interplay of several physical processes, including turbulent convection, small-scale dynamo action, nuclear energy generation, and neutrino cooling. The simulation requires the use of a nuclear reaction network solver, an electron--positron equation of state, constrained transport for evolving the magnetic field while preserving the solenoidal constraint, well-balanced methods to maintain the hydrostatic background stratification, and low-dissipation Riemann solvers to reduce numerical dissipation when modeling the subsonic convective flows. Therefore, the considered setup constitutes a demanding test problem that allows us to assess the reliability of \texttt{PHLEGETHON} in production-quality astrophysical simulations where multiple complex numerical methods are used concurrently. \\
\newline

\subsection{Initial conditions}
We consider, as input model for the MHD simulation, a 1D stellar evolution calculation produced by \cite{whitehead2026} using revision 10398 of the Modules for Experiments in Stellar Astrophysics (\texttt{MESA}) code \citep{Paxton2011,Paxton2013,Paxton2015,Paxton2018,Paxton2019}. \cite{whitehead2026} presents a large grid of low-metallicity (${Z = 0.001}$) massive-star (10--45 $M_\odot$) models, investigating CBM with traditional, \texttt{f0p02}, versus updated, \texttt{SH21}, prescriptions \citep[see][for more details about the \texttt{f0p02} and \texttt{SH21} CBM prescriptions]{whitehead2026}. We choose for our initial conditions the model \texttt{25f0p02} with initial mass 25 $M_\odot$ from grid \texttt{f0p02}, which assumes exponential diffusive CBM with ${f = 0.02}$ above every
upper convective boundary and ${f = 0.004}$ below every lower convective boundary \citep[see][for more details]{whitehead2026}. Our choice is motivated by the fact that this model presents, right before collapse, a large convective shell above the iron core, as a product of the merging of multiple shells of different composition. The size of the convective zone, the presence of vigorous turbulent convective flows, and the multiple nuclear burning reactions occurring inside the shell represent an interesting but complex physical problem. \texttt{PHLEGETHON} incorporates all the numerical methods needed to model these different physical mechanisms, and most importantly their interplay, thus demonstrating the full capabilities of the code. \\
Figure \ref{fig:kip} shows the structure evolution diagram of the stellar evolution model. A zoom-in on this area is presented in the right panel of Fig.~\ref{fig:kip}, showing the time domain covered by our 3D simulation, which corresponds to the final evolution of the convective shell until the time of predicted collapse.

\begin{figure*}
\centering
\footnotesize
\includegraphics[trim={0.33cm -0.25cm 1.25cm 0cm},clip,width=0.56\textwidth]{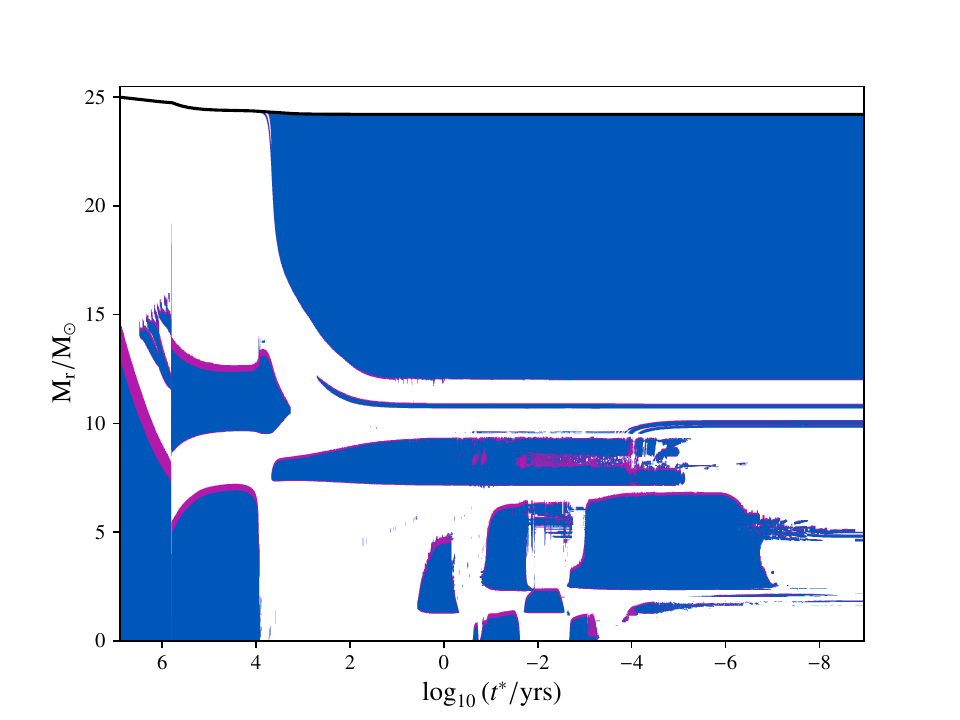}
\includegraphics[trim={0.25cm 0cm 0cm 0cm},clip,width=0.435\textwidth]{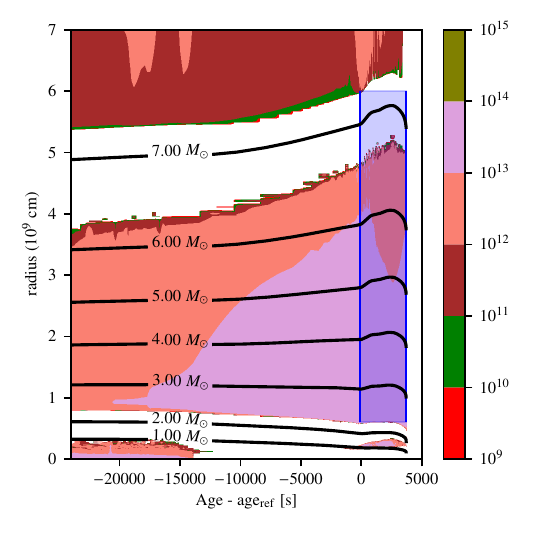}\put(-314,33){\color{red}\Huge\boldmath$\uparrow$}
\caption{{\it Left}: Structure evolution diagram of the \texttt{f0p02} 25 M$_\odot$ 1D \texttt{MESA} model as a function of the time left until the predicted collapse of the star (in years, log scale). Convective zones are drawn in blue, CBM zones in purple. The red arrow indicates the time and shell where initial conditions were extracted. {\it Right:} Zoom-in on the C--Ne--O shell merger, as a function of time in seconds from the start of the 3D simulations. In color scale, the kinetic energy per unit mass ($\mathrm{erg\,g^{-1}}$). The black lines are isomass contours. The vertical blue bars represent the radial and time extent of the 3D simulations.}\label{fig:kip}
\end{figure*}

\subsection{Mapping the stratification onto the 3D grid}

The initial conditions for the 3D simulation are extracted  approximately one hour before the predicted core collapse. The stellar stratification, including mass density, thermal pressure, temperature, chemical composition, and gravitational acceleration, is mapped onto a uniform 3D Cartesian grid with $512^3$ cells in $4\pi$ geometry. While both codes use a similar EoS, to ensure consistency with the EoS used in the \texttt{PHLEGETHON} model, the input stratification is first re-integrated in hydrostatic equilibrium while enforcing the profiles in $T$, $\widebar{A}$, and $\widebar{Z}$ as functions of radius from the \texttt{MESA} model using the Helmholtz EoS. The resulting stratification is shown in Fig.~\ref{fig:ccsnp-ics} for mass density, temperature, thermal pressure, specific entropy, $\widebar{A}$, and degeneracy parameter,
\begin{equation}
    \eta = \frac{\Phi_e}{k_bT},
\end{equation}
where $\Phi_e$ is the electron chemical potential and $k_b$ is Boltzmann's constant. Relative deviations with respect to the input \texttt{MESA} model after the re-integration are at most $2\%$.

The mapped model spans radii from ${0.6\times 10^9\ \mathrm{cm}}$ to ${6\times 10^9\ \mathrm{cm}}$. We apply the inscribed boundaries technique (see Sect.~\ref{sec:bcs}) and place solid spherical boundaries at the inner and outer domain radii. This choice ensures that the lower boundary of the convective shell coincides with the innermost embedded sphere. The computational domain thus includes the nearly isentropic convective shell and a narrow subadiabatic helium-rich layer above it. Within this spatial domain, the temperature reaches a maximum of $2\times 10^9\ \mathrm{K}$ at the base of the convective shell. The pressure and density decrease across the mapped radial domain by factors of 2159 and 457, respectively, corresponding to approximately 7.7 pressure scale heights and 6.1 density scale heights. At the base of the convective shell, electron degeneracy accounts for roughly $7\%$ of the thermal pressure, whereas radiation pressure accounts for approximately $46\%$. Unlike in, e.g., \cite{muller2016} or \cite{yadav2020}, here the lower boundary of the convective shell is held fixed in space and time for simplicity.

The hydrostatic background stratification is seeded with a low-amplitude thermal pressure white noise perturbation (${\delta P/P \sim 10^{-4}}$) to break spherical symmetry. Moreover, at $t=0$, we introduce a weak magnetic seed field which is computed from a vector potential $\mathbf{B} = \boldsymbol{\nabla} \times\mathbf{A}$, with $\mathbf{A}=(A_x,A_y,A_z)$ being defined on a compact support as
\begin{equation}
\begin{aligned}
A_\psi(\tilde{r}) &=
\begin{cases}
A_0 \cos^2\!\Big(\frac{\pi \tilde{r}}{2}\Big), & \tilde{r} \le 1, \\
0, & \tilde{r} > 1,
\end{cases} \\
A_x(\tilde{r},\psi) &=
-A_\psi(\tilde{r}) \sin\psi,
 \\
A_y(\tilde{r},\psi) &=
A_\psi(\tilde{r}) \cos\psi, \\
A_z(\tilde{r},\psi) & = 0.
\end{aligned}
\end{equation}
Here, $\psi$ is the polar angle and 
\begin{equation}
    \tilde{r} = \frac{1}{a_y}\sqrt{(x-x_0)^2+(y-y_0)^2+z^2},
\end{equation}
where
\begin{align}
    x_0 = & \ R_\mathrm{c} \cos(\psi), \\ 
    y_0 = & \ R_\mathrm{c} \sin(\psi).
\end{align} 
This choice of vector potential results in a purely poloidal magnetic seed field. Because rotation is absent in this setup, we expect a small-scale turbulent dynamo to dominate the magnetic field amplification \citep[see, e.g.,][]{batchelor1950,schekochihin2004a}. The strength and topology of the seed field do not significantly affect the dynamo action or the saturated field strength in the nonlinear regime, provided the seed is weak enough not to directly influence the turbulent flows \citep{seta2020}.  Therefore, we set $\,{a_y = 0.6 \times 10^9\ \mathrm{cm}}$, $\,{R_\mathrm{c} = 3.3 \times 10^9\ \mathrm{cm}}$, and $\,{A_0 = 0.6\times 10^{13}\ \mathrm{G\,cm}}$, resulting in a maximum seed field $\,{B_\mathrm{max} \approx 10^4\ \mathrm{G}}$, which is much weaker than the expected convective equipartition field ($\,{\sim 10^{10}\ \mathrm{G}}$).  

\subsection{Methods}
Gravity is held fixed throughout the simulation, and we employ the deviation method (see Sect.~\ref{sec:wb}) to preserve the hydrostatic equilibrium of the background stratification. The Helmholtz equation of state with Coulomb corrections is used (see Sect.~\ref{sec:eos}). Spatial reconstruction is performed using the PPH method and the Riemann solver is LHLLD (see Sects.~\ref{sec:reconstruction} and \ref{sec:riemann}). Auxiliary indices $\gamma_e$ and $\gamma_c$ are reconstructed following the procedure described in Sect.~\ref{sec:reconstruction}. For this simulation, diffusive heat transport is not considered. 

Nuclear burning is modeled with a 12-species C-, Ne-, and O-burning network, including $\mathrm{p}$, $^{4}\mathrm{He}$, $^{12}\mathrm{C}$, $^{16}\mathrm{O}$, $^{20}\mathrm{Ne}$, $^{23}\mathrm{Na}$, $^{24}\mathrm{Mg}$, $^{28}\mathrm{Si}$, $^{31}\mathrm{P}$, $^{32}\mathrm{S}$, $^{36}\mathrm{Ar}$, and $^{40}\mathrm{Ca}$. The initial composition inside the convective shell consists of $60-80\%$ $^{16}\mathrm{O}$, $9\%$ $^{24}\mathrm{Mg}$, ${6-10\%}$ $^{28}\mathrm{Si}$, ${2-5\%}$ $^{20}\mathrm{Ne}$, with minor contributions (${\approx 1\%}$) from $^{32}\mathrm{S}$ and $^{12}\mathrm{C}$. In the subadiabatic layer above the convective shell, the mass fraction of $^{4}\mathrm{He}$ increases outward to about $60\%$, while that of $^{12}\mathrm{C}$ is approximately $20\%$ (see also Fig.~\ref{fig:ccsnp-Xprofs}). The network includes 
\begin{equation}
\begin{array}{lll}
3\alpha(\gamma)^{12}\mathrm{C}, &
^{12}\mathrm{C}(\alpha,\gamma)^{16}\mathrm{O}, &
^{16}\mathrm{O}(\alpha,\gamma)^{20}\mathrm{Ne}, \\
^{20}\mathrm{Ne}(\alpha,\gamma)^{24}\mathrm{Mg}, &
^{23}\mathrm{Na}(\mathrm{p},\alpha)^{20}\mathrm{Ne}, &
^{23}\mathrm{Na}(\mathrm{p},\gamma)^{24}\mathrm{Mg}, \\
^{24}\mathrm{Mg}(\alpha,\gamma)^{28}\mathrm{Si}, &
^{28}\mathrm{Si}(\alpha,\gamma)^{32}\mathrm{S}, &
^{31}\mathrm{P}(\mathrm{p},\gamma)^{32}\mathrm{S}, \\
^{31}\mathrm{P}(\mathrm{p},\alpha)^{28}\mathrm{Si}, &
^{32}\mathrm{S}(\alpha,\gamma)^{36}\mathrm{Ar}, &
^{36}\mathrm{Ar}(\alpha,\gamma)^{40}\mathrm{Ca}, \\
^{12}\mathrm{C}(^{12}\mathrm{C},\mathrm{p})^{23}\mathrm{Na}, &
^{12}\mathrm{C}(^{12}\mathrm{C},\alpha)^{20}\mathrm{Ne}, &
^{16}\mathrm{O}(^{16}\mathrm{O},\mathrm{p})^{31}\mathrm{P}, \\
^{16}\mathrm{O}(^{16}\mathrm{O},\alpha)^{28}\mathrm{Si}, &
^{16}\mathrm{O}(^{12}\mathrm{C},\alpha)^{24}\mathrm{Mg} &
^{20}\mathrm{Ne}(^{12}\mathrm{C},\mathrm{p})^{31}\mathrm{P} \\
^{20}\mathrm{Ne}(^{12}\mathrm{C},\alpha)^{28}\mathrm{Si},  &
\end{array}
\end{equation}
and their reverse, for a total of 38 reactions.
Nuclear burning is coupled to the MHD equations using Godunov splitting and solved with the implicit backward-Euler time integration scheme. Thermal neutrino losses are also included (see Sect.~\ref{sec:nuclear-network}). 

\begin{figure}
\includegraphics[width=\linewidth]{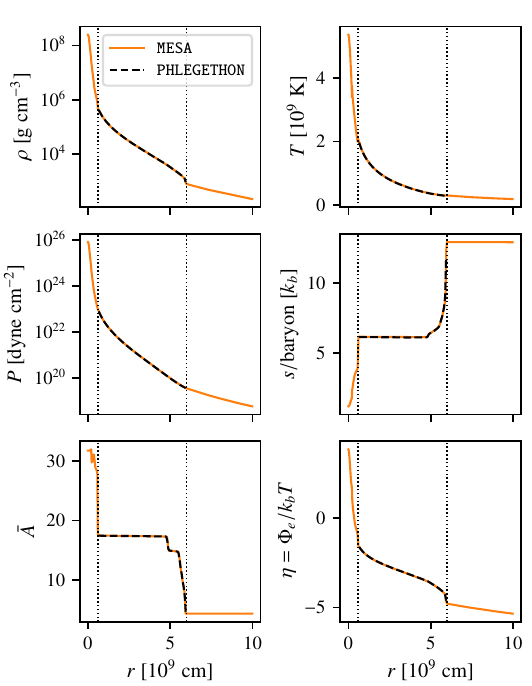}
\caption{Initial stratification used for the simulation of the core-collapse supernova progenitor presented in Sect.~\ref{sec:ccsnp}. The six panels show mass density, temperature, thermal pressure, specific entropy per baryon, mean mass number ($\widebar{A}$) and electron degeneracy parameter ($\eta$). The orange solid line corresponds to the 1D \texttt{MESA} model, while the black dashed line shows the re-integrated stratification, which is used as initial conditions for the 3D \texttt{PHLEGETHON} simulation. The vertical black dotted lines denote the boundaries of the spatial domain of the \texttt{PHLEGETHON} model.}
\label{fig:ccsnp-ics}
\end{figure}

\subsection{3D magnetohydrodynamic simulation}

The initial perturbations, together with nuclear energy release, drives buoyant bubbles of higher entropy relative to the background. Within $\,{\sim 200\ \mathrm{s}}\,$ of simulation time, these bubbles develop into fully turbulent convective motions, achieving Mach numbers in the range $\,{0.1 \lesssim \mathcal{M} \lesssim 0.3}$. The convective plumes collide with the boundary of the neighboring subadiabatic layer, overturning and triggering mixing processes that entrain material into the convective shell. Turbulent eddies stretch the initial magnetic field at spatial scales close to viscous range, initiating a small-scale turbulent dynamo. The time evolution of the total kinetic and magnetic energies in the convective shell is shown in Fig.~\ref{fig:ccsnp-tprof}. Within the first $1000$~s, convective flows achieve a root-mean-square speed $u_\mathrm{rms}$ of ${4\times10^7\ \mathrm{cm\ s^{-1}}}$. The corresponding convective turnover timescale, defined as
\begin{equation}
    \tau_\mathrm{conv} = \frac{2L}{u_\mathrm{rms}},
\end{equation}
is $\,{\tau_\mathrm{conv} \approx 250\ \mathrm{s}}$, where $\,{L \approx 5\times 10^9\ \mathrm{cm}}\,$ is the radial extent of the convective shell. In the first eight convective turnovers, the magnetic energy grows exponentially by more than ten orders of magnitude relative to the initial seed value. The kinetic energy experiences a burst within the first $500\ \mathrm{s}$ before gradually decreasing, tracking the decline in nuclear luminosity that drives convection.  The setup is evolved until ${t=4000\ \mathrm{s}}$, time at which the 1D stellar model predicts the onset of core-collapse.

At $\,{t \approx 1500\ \mathrm{s}}$, the turbulent dynamo saturates. The Lorentz force begins to feed back on the convective flows, suppressing small-scale structures, while the stretching of magnetic field lines is balanced by numerical Ohmic dissipation, thus achieving a statistical steady state configuration. The resulting magnetic field strength reaches values of $\,{\sim 10^9}\,$--$10^{10}\ \mathrm{G}$, corresponding to $10$--$30\%$ of equipartition with the local kinetic energy. Figure~\ref{fig:ccsnp-spectra} shows the power spectra of kinetic and magnetic energy as a function of angular wavenumber $\ell$ extracted at $\,{r = 3\times10^9\ \mathrm{cm}}$. The spectra are averaged over $\,{t \in (2000,4000)\ \mathrm{s}}\,$ to suppress statistical fluctuations due to the chaotic nature of the turbulent flows, covering the nonlinear, saturated phase of the dynamo. For comparison, results from a purely hydrodynamical simulation run on the same grid are also shown. Both the hydrodynamic and MHD simulations show a narrow inertial range (${3 \lesssim \ell \lesssim 10}$) where the kinetic energy spectrum closely follows the Kolmogorov scaling law ($\ell^{-5/3}$). Such scaling is often seen in simulations of late convective shells of core-collapse supernova progenitors \citep[e.g.,][]{yadav2020,fields2020,varma2020}. The hydrodynamic simulation also exhibits a power excess across a wide range of angular wavenumbers between the inertial and viscous ranges. This phenomenon has been observed in numerous simulations of turbulent flows and is commonly referred to in the literature as the `bottleneck effect' \citep[see, e.g.,][]{falkovich1994,dobler2003,andrassy2022}. Although several hypotheses have been proposed regarding the underlying mechanism responsible for this power excess, including non-local energy transport \citep{minnini2008}, incomplete thermalization \citep{frisch2008}, and backscatter from smaller scales \citep{johnson2021}, its origin remains unclear. 

Consistent with previous MHD studies of analogous convective shells of massive stars \citep[e.g.,][]{varma2020,leidi2023}, magnetic power accumulates near the bottom of the inertial range and declines rapidly at the resistive scale. At large scales ($\,{\ell \lesssim 30}\,$), the magnetic field is subdominant, while at smaller scales ($\,{\ell \gtrsim 30}\,$), it reaches superequipartition strenghts down to the Nyquist wavenumber. The kinetic energy spectrum exhibits a prominent peak at $\,{\ell = 2}\,$ (quadrupole), and deviates from the hydrodynamical case around $\,{\ell \approx 10}\,$, where the turbulent dynamo starts to efficiently amplify magnetic energy at the expense of the kinetic energy content of the convection. The quadrupolar behavior can be understood as a consequence of the inner boundary conditions, which inhibit radial convective motions near the center and thereby suppress the largest-scale overturning modes. In contrast, multidimensional simulations of core convection in massive stars tend to exhibit significant power in large-scale, dipole-like flow components \citep[e.g.,][]{woodward2019,herwig2023,lecoanet2023,blouin2023,blouin2024}. Although in reality the lower boundary of the convective shell is not solid, simulations of late burning shells in massive stars show that the buoyant response of the stable layer beneath the convective boundary is strong enough to prevent significant entrainment processes \citep[see also][]{meakin2007,jones2017,varma2020,rizzuti2024}.

In the ILES approach employed by \texttt{PHLEGETHON}, the fluid and magnetic Reynolds numbers cannot be prescribed independently. Rather, their effective values are determined by the dissipative properties of the numerical methods used in the simulation, including the spatial reconstruction scheme, the Riemann solver, and the method used to preserve the solenoidal constraint on the magnetic field \citep[e.g.,][]{kritsuk2011}. The same applies to the magnetic Prandtl number, which is set by the ratio of the viscous and resistive coefficients. For commonly used ILES methods, the effective magnetic Prandtl number is often of order unity \citep{shivakumar2025}, whereas stellar interiors can exhibit magnetic Prandtl numbers spanning many orders of magnitude, from approximately $10^{-4}$ to $10^{8}$, depending on the thermodynamic conditions and location within the star \citep{jermyn2022}. Turbulent dynamos are particularly sensitive to these parameters because magnetic-field amplification is fundamentally a small-scale process \citep{schekochihin2004a,tobias2021}. Previous ILES studies of turbulent dynamos in stellar convective layers have demonstrated that both the dynamo growth rate and the saturated magnetic-field strength tend to increase with increasing the effective resolution of the turbulent flows \citep{rempel2014,rempel2023,leidi2023}. Consequently, the quantitative saturation level obtained here should not be interpreted as a resolution-independent prediction of the physical magnetic-field strength. Instead, it should be understood as the result obtained for the numerical dissipation and effective dimensionless parameters associated with the numerical method and resolution employed here. Nevertheless, the saturated magnetic-field strength obtained in our study is consistent with previous ILES simulations of turbulent dynamos in stellar interiors, which typically find magnetic energies at a 10-50\% fraction of the equipartition value \citep{hotta2016,varma2020,leidi2023}. Consistent with these previous studies, we find that the coarser $256^3$ simulation exhibits a reduced mean dynamo growth rate and a lower saturated magnetic-field strength, reaching approximately $70\%$ of the strength observed in the nominal-resolution simulation (see Fig.~\ref{fig:ccsnp-tprof}).

2D slices in the equatorial (${z=0}$) plane are shown in Fig.~\ref{fig:ccsnp-slices}, displaying the $^{20}\mathrm{Ne}$ mass fraction, the nuclear energy release by alpha-capture on oxygen, radial velocity, magnetic field strength in the saturated stage of the turbulent dynamo, the Mach number, and specific entropy fluctuations at $\,{t = 2400\ \mathrm{s}}$. The convective flow exhibits a clear scale separation, with small-scale turbulence superimposed on large-scale overturning plumes driven by nuclear energy release at the base of the shell. At this stage, the root-mean-square convective velocity has dropped to $\,{u_\mathrm{rms} \approx 2\times 10^7\ \mathrm{cm\ s^{-1}}}$, which leads to a convective turnover time of $\,{\approx500}\,$ s. The average magnetic field strength decreases with radius, reflecting the scaling of the equipartition field with the square root of the local density, which declines outward, while the root-mean-square velocity remains nearly uniform. In the convective shell, root-mean-square fluctuations of both mass density and specific entropy are on the order of $1\%$ compared with the background stratification.

The nuclear energy generation rate from $^{16}\mathrm{O}(\alpha,\gamma)^{20}\mathrm{Ne}$ shown in Fig.~\ref{fig:ccsnp-slices} highlights the entrainment of material from the subadiabatic layer above into the convection zone. The $\alpha$-capture reaction exhibits a double-peaked distribution. One peak is located at the base of the shell, where temperatures are highest. There, photodisintegration of $^{20}\mathrm{Ne}$ produces a $10^{-8}$ mass fraction abundance of $^{4}\mathrm{He}$. Part of those $\alpha$ particles gets captured on $^{16}\mathrm{O}$ to generate $^{20}\mathrm{Ne}$, while the rest undergoes $\alpha$ captures on $^{20}\mathrm{Ne}$, $^{24}\mathrm{Mg}$, and $^{28}\mathrm{Si}$, allowing the burning phase to progress. The second peak lies within the convective region, and it is mostly driven by the entrainment of helium-rich material from the upper stable region, which provides additional $\alpha$ particles for the neon burning. This behavior is sometimes known as `convective-reactive', and it can predict the occurrence of multiple burning episodes within the same convective region, often leading to strong gradients and asymmetries in chemical composition \citep[see][]{moczak2018,yadav2020,rizzuti2024}. Because the entrainment occurs through localized plumes that scrape material from the outer stable layer, rather than uniformly around the convective boundary, the resulting enhancement of the $^{16}\mathrm{O}(\alpha,\gamma)^{20}\mathrm{Ne}$ energy-generation rate is intrinsically asymmetric in the azimuthal direction, as visible in the upper-right panel of Fig.~\ref{fig:ccsnp-slices}.

This two-regime burning is also evident in Fig.~\ref{fig:ccsnp-edot}, which shows horizontally averaged nuclear energy generation rates per unit mass as functions of radius for several reactions at $\,{t=2400\ \mathrm{s}}$. The energetics at the base of the convective shell are dominated by reactions associated with carbon and neon burning, together with $\alpha$-capture reactions on $^{24}\mathrm{Mg}$ and $^{28}\mathrm{Si}$. In contrast, reactions associated with oxygen burning contribute only subdominantly to the total energy generation.

By the end of the simulation at $\,{t = 4000\ \mathrm{s}}$, mass fractions are largely homogenized by convective mixing (see Fig.~\ref{fig:ccsnp-Xprofs}), with the exception of $^{4}\mathrm{He}$ which shows important radial gradients in chemical composition as a result of the convective-reactive behavior. Significant entrainment provides important amounts of extra helium and carbon into the convective region and expands the outer boundary of the convective shell nearly to the domain boundary. While the total partial masses of most species remain largely unchanged, $^{20}\mathrm{Ne}$ is depleted by $\sim 50\%$.  

To interpret the 3D simulation results in a reduced 1D framework, we employ implicit large-eddy simulation Reynolds averaging (RA-ILES; \citealt{meakin2007,viallet2013,moczak2018,arnett2019}). In this approach, variables are decomposed into mean and fluctuating components using both Reynolds and Favre averages, denoted $\,{\langle \cdot \rangle}\,$ and $\,{\langle \widetilde{\cdot} \rangle}$, respectively. For a generic quantity $q$, the Reynolds decomposition is $\,{q = \langle q \rangle + q'}$, where the bracket operator $\langle \cdot \rangle$ denotes the time average of the horizontal (spherical-shell) average over the full solid angle $4\pi$. The Favre decomposition for density-weighted quantities is $\,{q = \langle \widetilde{q} \rangle + q''}$, where $q'$ and $q''$ are the fluctuations, and the Favre average is defined as $\,{\langle \widetilde{q} \rangle = \langle \rho q \rangle / \langle \rho \rangle}$. This formalism allows us to compute turbulent fluxes and dissipation terms directly from the simulation data, providing insight into the transport of entropy, composition, and energy by turbulent motions.

Figure~\ref{fig:ccsnp-rans} shows RA-ILES results for the entropy evolution equation and $^{4}\mathrm{He}$ evolution, which can be written as
\begin{align} \langle \rho \rangle D_t \langle \widetilde{s} \rangle + \nabla_r \langle \rho \rangle \langle \widetilde{s'' u_r''} \rangle &= \langle \rho \rangle \widetilde{\Bigg\langle \frac{\dot{e}_\mathrm{N}}{T} \Bigg\rangle} + \langle \rho \rangle\widetilde{ \Bigg\langle \frac{\dot{e}_\nu}{T} \Bigg\rangle} + \langle \rho \rangle \widetilde{\Bigg\langle \frac{\epsilon_\mathrm{d}}{T} \Bigg\rangle}, \\[1ex] \langle \rho \rangle D_t \langle \widetilde{X} \rangle + \nabla_r \langle \rho \rangle \langle \widetilde{X'' u_r''} \rangle &= \langle \rho \rangle \langle \widetilde{\dot{X}_\mathrm{N}} \rangle, \end{align}
where the operator
\begin{equation}
    \nabla_r  = \frac{1}{r^2} \frac{\partial}{\partial r} \left( r^2 \cdot \right).
\end{equation}
Here, $u_r$ is the radial component of the velocity and $X$ represents the mass fraction of $^{4}\mathrm{He}$. In the entropy equation, $\dot{e}_\mathrm{N}$ and $\dot{e}_\nu$ are the nuclear and neutrino energy generation/loss rates per unit mass, respectively, while $\epsilon_\mathrm{d}$ represents the effective dissipation due to numerical viscosity and resistivity. In the composition equation, $\dot{X}_\mathrm{N}$ is the net nuclear production rate of helium. The divergence terms, $\nabla_r \langle \rho \rangle \langle \widetilde{ s'' u_r''} \rangle$ and $\nabla_r \langle \rho \rangle \langle \widetilde{X'' u_r''} \rangle$, describe the turbulent transport of entropy and composition by convective motions.

Most of the entropy generated at the base of the convective shell is transported outward by turbulent convection. Nuclear luminosity increases by $\,{\sim 10\%}\,$ near $\,{3\times10^9\ \mathrm{cm}}\,$ due to $\alpha$-capture reactions induced by the entrainment of helium from the subadiabatic layer. The convective entropy flux is positive throughout the shell, becoming negative at the boundary with the subadiabatic region, indicating entrainment of entropy in addition to composition. The contribution of neutrino cooling is approximately $5\%$ of the nuclear entropy luminosity, which is itself comparable to the entropy luminosity associated with viscous and resistive dissipation. The $^{4}\mathrm{He}$ profiles show that in the upper half of the convective shell, abundance changes are dominated by inward turbulent fluxes. At radii between $\,{1.4\times10^9\ \mathrm{cm}}\,$ $4\times10^9\ \mathrm{cm}$, $\alpha$-capture reactions balance inward turbulent transport. At the base, the small imbalance between photodisintegration of $^{20}\mathrm{Ne}$ and $\alpha$-capture on $^{16}\mathrm{O}$ is compensated by outward turbulent fluxes.

The presented 3D MHD simulation demonstrates that \texttt{PHLEGETHON} is capable of robustly modeling the complex interplay of turbulent convection, nuclear burning, and magnetic field amplification in realistic stellar environments. These results highlight the ability of the numerical methods implemented in \texttt{PHLEGETHON} to handle highly nonlinear, multi-physics problems relevant to late-stage stellar evolution. Moreover, this simulation exhibits features commonly observed in previous studies of this type, including Kolmogorov-like turbulence, the dominance of large-scale flow patterns, and super-equipartition field strengths in the viscous range. These results further support the conclusion that the code can reliably simulate the physics of core-collapse supernova progenitors, in line with other established codes in the field.

\begin{figure}
\includegraphics[width=\linewidth]{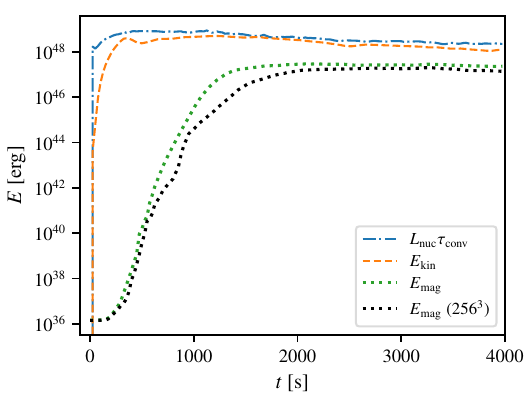}
\caption{Time evolution of the volume-integrated kinetic (orange dashed line) and magnetic energy (green dotted line) in the convective zone of the simulation of the core-collapse supernova progenitor star (see Sect.~\ref{sec:ccsnp}). The product of the nuclear luminosity and the convective turnover timescale, $\,{\tau_\mathrm{conv}\approx250\ \mathrm{s}}\,$ (computed within the first 1000 s) is also shown (blue dash-dotted line). The black dotted line denotes the volume integrated magnetic energy in a simulation run with $256^3$ cells.}
\label{fig:ccsnp-tprof}
\end{figure}

\begin{figure}
\includegraphics[width=\linewidth]{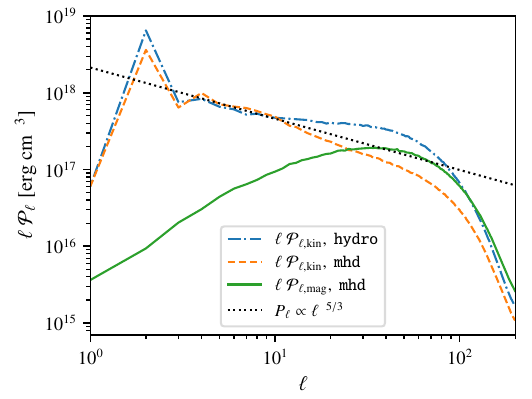}
\caption{Kinetic (orange dashed line) and magnetic (green solid line) power spectra as functions of the angular wavenumber $\ell$ in the simulation of the core-collapse supernova progenitor discussed in Sect.~\ref{sec:ccsnp}. The spectra are averaged over the time interval ${t \in (2000,4000)\ \mathrm{s}}$ and multiplied by $\ell$ to better highlight the distribution of power across scales. For reference, the kinetic power spectrum from a purely hydrodynamical simulation (blue dash-dotted line) is also shown. All curves are evaluated at a radius of $\,{3\times 10^9\ \mathrm{cm}}$. The black dashed line represents Kolmogorov $\ell^{-5/3}$ scaling law.}
\label{fig:ccsnp-spectra}
\end{figure}

\begin{figure*}
\centering
\includegraphics[width=0.95\textwidth]{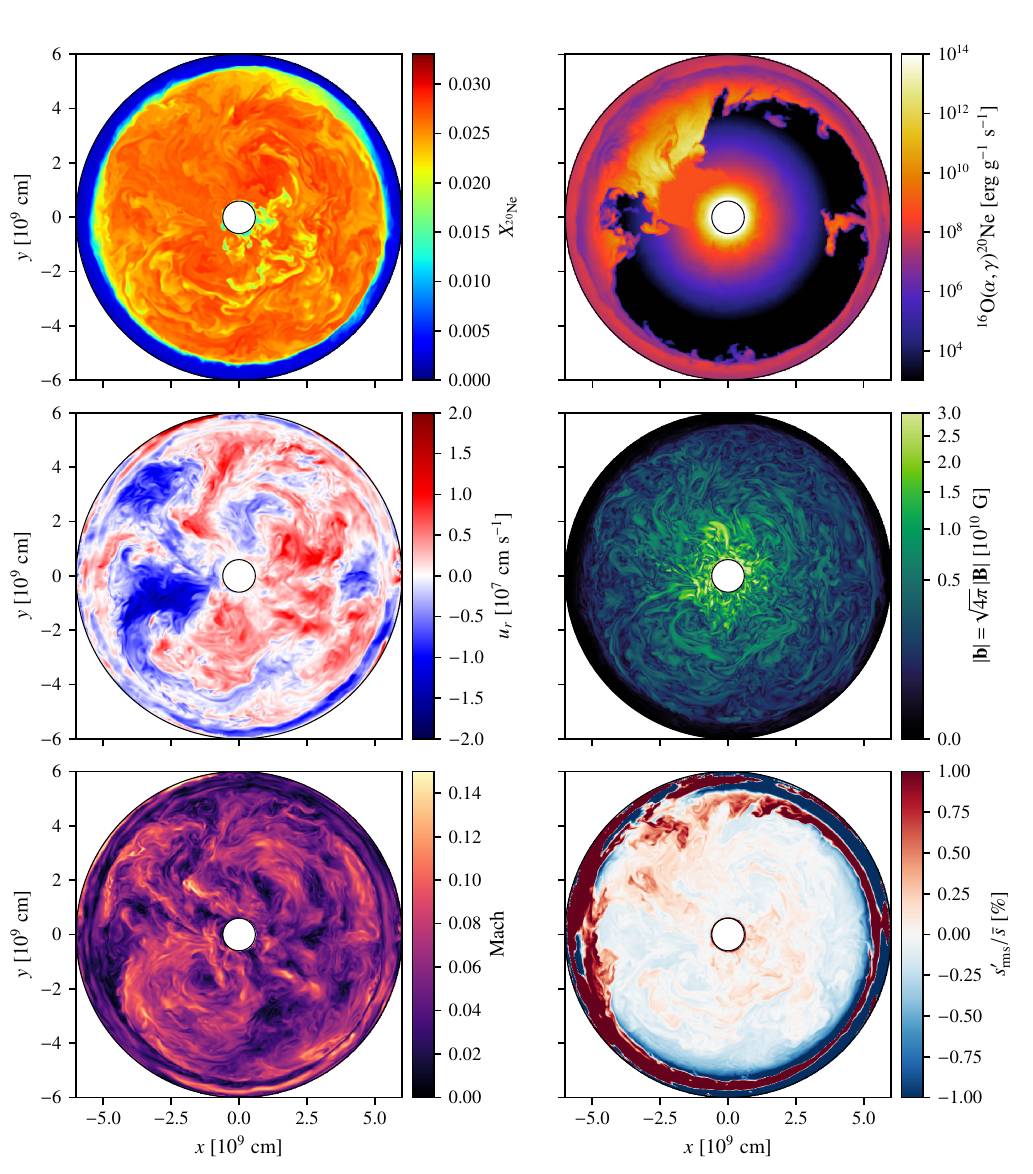}
\caption{2D slices in the equatorial ($\,{z=0}\,$) plane from the simulation of the convective shell presented in Sect.~\ref{sec:ccsnp}. The six panels show the mass fraction of $^{20}\mathrm{Ne}$ (upper left), the nuclear energy generation rate per unit mass from $\alpha$-capture on $^{16}\mathrm{O}$ (upper right), radial velocity (center left), the magnetic field strength (center right), Mach number (lower left), and relative root-mean-square fluctuations in specific entropy (lower right). All panels are extracted at $\,{t = 2400\ \mathrm{s}}$. To save disk space, the energy generation rate from nuclear reactions (including $\alpha$-capture on $^{16}\mathrm{O}$) is recorded in a separate output channel, where neighboring $\,{2\times2\times2}\,$ cell blocks are first averaged before writing the output to disk. Movie available at \url{https://youtu.be/zIzC-aaGXF4}.}
\label{fig:ccsnp-slices}
\end{figure*}

\begin{figure}
\includegraphics[width=\linewidth]{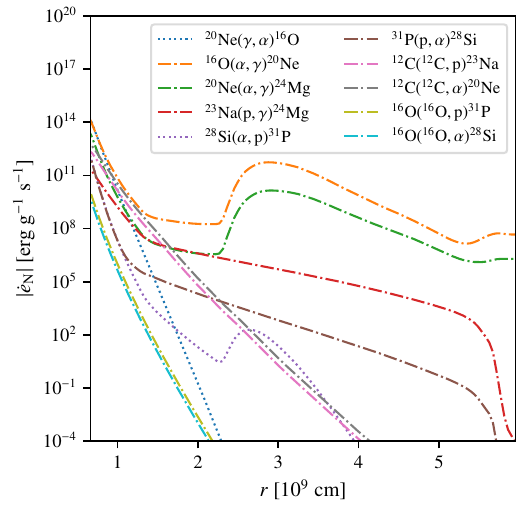}
\caption{Radial profiles (horizontal averages) of the nuclear energy generation rate per unit mass for several reactions in the simulation of the core-collapse supernova progenitor star. Dash-dotted lines indicate positive energy generation, while dotted lines indicate energy losses. All curves are extracted at $\,{t =2400\ \mathrm{s}}$.}
\label{fig:ccsnp-edot}
\end{figure}

\begin{figure}
\includegraphics[width=\linewidth]{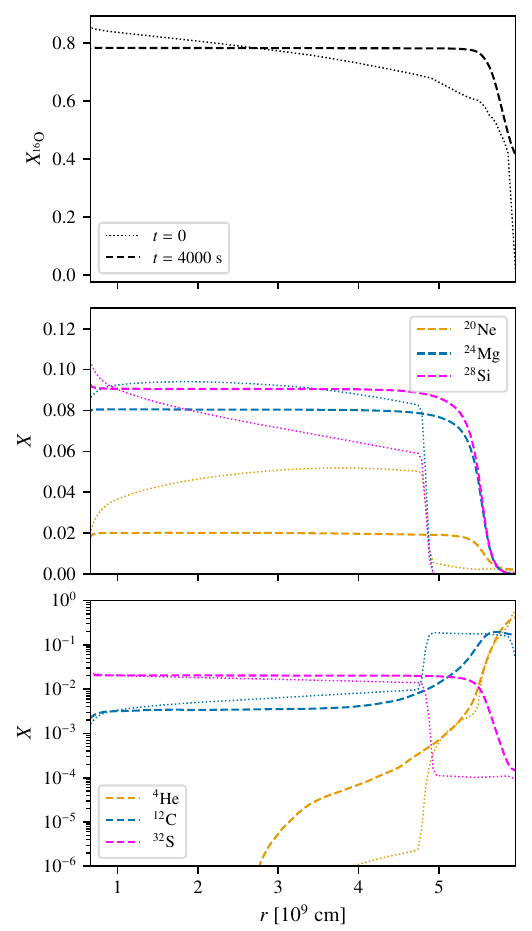}
\caption{Reynolds-averaged mass fractions as functions of radius for the simulation of the core-collapse supernova progenitor. Dotted lines show the initial profiles at $\,{t = 0}$, while dashed lines correspond to the profiles at \mbox{$t = 4000\ \mathrm{s}$}.}
\label{fig:ccsnp-Xprofs}
\end{figure}

\begin{figure}
\includegraphics[width=\linewidth]{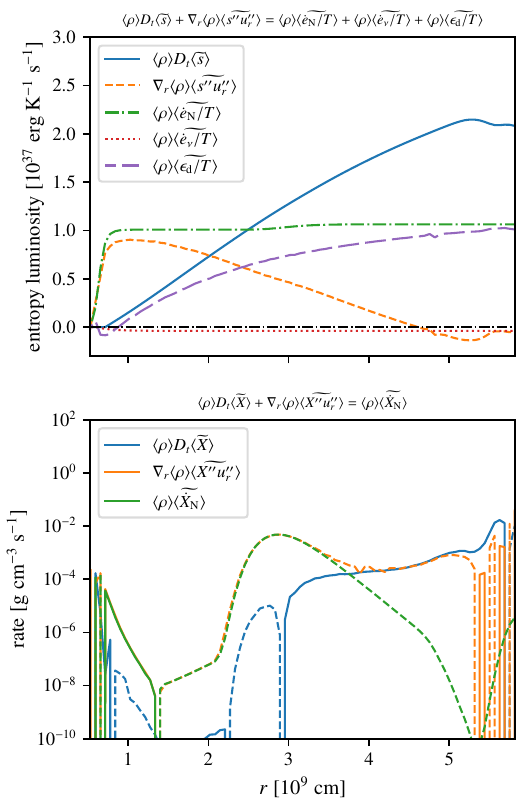}
\caption{Reynolds-averaged terms in the entropy and $^{4}\mathrm{He}$ evolution equations. All quantities are averaged over the time interval ${t \in (2000,4000)\ \mathrm{s}}$. Reynolds and Favre averages are denoted by $\langle \cdot \rangle$ and $\langle \widetilde{\cdot} \rangle$, respectively \citep[see also][]{viallet2013}. In the bottom panel, solid lines indicate positive contributions, while dashed lines indicate negative contributions. For the entropy equation, the individual terms are integrated over volume to yield cumulative profiles.}
\label{fig:ccsnp-rans}
\end{figure}

\end{section}

\begin{section}{Conclusions}\label{sec:conclusions}

In this paper, we have presented \texttt{PHLEGETHON}, a finite-volume Eulerian code optimized for applications in stellar magnetohydrodynamics. The code employs a rich suite of numerical methods that allow it to simulate a broad range of dynamical conditions, from low-Mach-number flows in steeply stratified media to supersonic regimes. This versatility is achieved through the use of low-dissipation Riemann solvers, which significantly reduce numerical dissipation in subsonic flows while remaining robust at high Mach numbers, and a well-balanced discretization, which is essential for accurately modeling slow flows in steep stratifications. Several reconstruction methods are available, allowing the user to prioritize robustness in highly dynamic flows or accuracy in smooth regions.  

\texttt{PHLEGETHON} supports arbitrary nuclear reaction networks and equations of state tailored for stellar plasmas, including the effects of electron degeneracy, pair production, and partial ionization. Stiff diffusive processes, such as thermal conduction and radiative diffusion, can be efficiently integrated using super-time-stepping, circumventing the restrictive parabolic CFL stability criterion. A staggered formulation of constrained transport ensures that the magnetic field remains divergence-free to round-off error. A gravitational Poisson solver is available when the gravitational potential needs to be explicitly evolved in time. 

\texttt{PHLEGETHON} runs on CPUs with MPI-based parallelization and has been shown to scale efficiently up to $32768$ cores on the Tier-2 HoreKa HPC system, making it suitable for large-scale, production-quality simulations. Extensive verification of the methods implemented in the code has been performed in Sect.~\ref{sec:verification} using standard benchmark tests, including 1D Brio--Wu shock tubes, 2D magnetized vortices, and the diffusion of a temperature pulse, which confirm second-order convergence for most combinations of numerical methods. More advanced setups, including convective flows driven by an energy source such as the 3D code-comparison test of \cite{andrassy2022}, further demonstrate that \texttt{PHLEGETHON} can accurately capture convective and turbulent dynamics in stratified, astrophysically relevant environments. 

To illustrate that \texttt{PHLEGETHON} can robustly handle production-quality astrophysical setups, we ran 3D MHD simulations of a convective shell in a 25 $M_\odot$ core-collapse supernova progenitor star in Sect.~\ref{sec:ccsnp}. This model contains all the ingredients necessary for advanced applications to stellar interiors, including strong stratification, nuclear burning networks, neutrino cooling, a complex equation of state, and magnetic fields. These simulations capture small-scale turbulent dynamos, convective entrainment, and detailed nuclear burning processes, and demonstrate that the code can be used reliably for production-quality simulations of complex, multi-physics stellar environments.

The present version of \texttt{PHLEGETHON} does not include explicit viscous or resistive dissipation, and all dissipation is therefore provided by the numerical scheme. This choice reflects the extremely large Reynolds and magnetic Reynolds numbers characteristic of stellar interiors, for which the physical viscous and resistive dissipation scales are generally far below the resolution attainable in global simulations. An ILES approach is therefore appropriate for the primary target applications of \texttt{PHLEGETHON}. The resulting effective viscous and resistive dissipation coefficients are, however, dependent on the resolution and numerical scheme and cannot be set a priori. Several methods have been proposed to quantify these effective coefficients in ILES \citep[e.g.,][]{rembiasz2017,shivakumar2025}. Such estimates can characterize the dissipation generated by a given numerical scheme, but they do not provide independent control over the viscous and resistive dissipation scales or their ratio, i.e., the magnetic Prandtl number. This limitation is particularly relevant for controlled studies of turbulent dynamos, for which the dependence on the Reynolds number, magnetic Reynolds number, and magnetic Prandtl number is of central importance \citep[e.g.,][]{iskakov2007,brandenburg2012,kapyla2018}. The inclusion of explicit viscosity and Ohmic resistivity is therefore planned for future versions of the code.

Overall, \texttt{PHLEGETHON} provides a versatile platform for simulating stellar flows and MHD phenomena across multiple evolutionary stages and stellar types. It allows researchers to investigate the interplay between turbulence, magnetic fields, nuclear burning, diffusive energy transport, and rotation, supporting studies of the complex dynamics of stellar interiors. \texttt{PHLEGETHON} has been made publicly available, enabling the community to apply it to a wide range of astrophysical problems, benchmark its performance, and further extend its capabilities. For example, efforts are currently underway to port the code to GPUs. 

Future applications of the code will focus on open questions in stellar astrophysics. These include the impact of magnetic fields on convective boundary mixing in massive main-sequence stars, which is crucial for determining the size of the helium core and subsequent evolutionary stages. The code will also be used to study the excitation of internal waves, their coupling to magnetic fields, the resulting asteroseismic signals in low- and high-mass stars, and the transport of angular momentum and chemical species by waves. Finally, we will investigate sources of asymmetries in core-collapse and electron-capture supernova progenitors, thereby aiding our understanding of the explosion mechanisms in both cases.
\end{section}

\section*{Acknowledgments}
The authors are grateful to Mattia Lupi (University of Padova) for insightful comments and helpful discussions on geometric multigrid preconditioning. We acknowledge support by the Klaus Tschira Foundation. RA, DG, PSHC and FKR acknowledge funding by the European Union (ERC, ExCEED, project number 101096243). Views and opinions expressed are however those of the authors only and do not necessarily reflect those of the European Union or the European Research Council Executive Agency. Neither the European Union nor the granting authority can be held responsible for them. FR is a fellow of the Alexander von Humboldt Foundation, and acknowledges support by the INAF Mini grant 2024, ‘GALoMS -- Galactic Archaeology for Low Mass Stars’ (1.05.24.07.02). PC acknowledges support by the Deutsche Forschungsgemeinschaft (DFG, German Research Foundation) -- Project-ID MA 4248/3-1. KV is a fellow of the International Max Planck Research School for Astronomy and Cosmic Physics at the University of Heidelberg (IMPRS-HD) and acknowledges financial support from IMPRS-HD. PVFE was supported by the U.S. Department of Energy through the Los Alamos National Laboratory (LANL). LANL is operated by Triad National Security, LLC, for the National Nuclear Security Administration of the U.S. Department of Energy (Contract No. 89233218CNA000001). This work has been assigned a document release number LA-UR-26-22807. The authors gratefully acknowledge the computing time provided on the high-performance computer HoreKa by the National High-Performance Computing Center at KIT (NHR@KIT). This center is jointly supported by the Federal Ministry of Education and Research and the Ministry of Science, Research and the Arts of Baden-Württemberg, as part of the National High-Performance Computing (NHR) joint funding program (https://www.nhr-verein.de/en/our-partners). HoreKa is partly funded by the German Research Foundation (DFG). 
RH acknowledges support from the World Premier International Research Centre Initiative (WPI Initiative), MEXT, Japan, the IReNA AccelNet Network of Networks (NSF, Grant No. OISE-1927130),  {the Wolfson Foundation that part-funded the greenHPC facility at Keele and a 2025 Klaus Tschira Gast professorship from HITS.
The authors used ChatGPT (GPT-5.2, OpenAI) to assist with grammar and language polishing. All scientific content, analysis, and conclusions were written and verified by the authors.

\bibliographystyle{yahapj}
\bibliography{main}

\end{document}